\documentclass{aa}
\def\vec#1{\ensuremath{\mathchoice{\mbox{\boldmath$\displaystyle#1$}}
{\mbox{\boldmath$\textstyle#1$}}
{\mbox{\boldmath$\scriptstyle#1$}}
{\mbox{\boldmath$\scriptscriptstyle#1$}}
}}
\begin{document}
\title{``Atlas'' of numerical solutions for star-disk magnetospheric interaction}
\titlerunning{``Atlas'' of numerical solutions for SDMI}
\authorrunning{M. \v{C}emelji\'{c}}
\author{M. \v{C}emelji\'{c}
 }
\offprints{M. \v{C}emelji\'{c}}
\institute{Nicolaus Copernicus Astronomical Center, Bartycka 18, 00-716
Warsaw, Poland
\email{miki@camk.edu.pl}
}
\date{Received ??; accepted ??}
\abstract
{}
{I report results in numerical simulations of star-disk magnetospheric
interaction. A thin accretion disk with corona above a rotating stellar
surface is simulated in a parameter study, to find trends in the angular
momentum flux. The results are presented in the case of Young Stellar
Objects, but they can be rescaled to other objects with similar geometry.
}
{In the performed resistive and viscous magneto-hydrodynamic simulations,
a quasi-stationary state is reached in the cases with different parameters.
Angular momentum fluxes in the different components of the flow are
computed, to compare the results.
}
{Results in the simulations are presented with the matter density
distribution and magnetic field geometry displayed in an ``Atlas''
of solutions. The torque exerted on the star is computed, together
with the angular momentum flux loaded into outflow, in the cases
when a conical outflow is formed. In the studied part of the parameter
space, I find trends in both components of the flow.
}
   {}


\maketitle

\section{Introduction}
An analytical solution for viscous accretion disk has been given in
\cite{ss73}, few years after the first numerical solution by \cite{pb68}.
As in many numerical and analytical models which followed, the disk structure
was derived separately in radial and vertical direction. The radial structure
was computed by equations averaged over the disk thickness, and the vertical
structure from the hydrostatic equilibrium in the vertical direction. In
some models, the radiative transfer was also taken into account.

Height-averaged computations were shown to be insufficient in the analytical
work by \cite{urp84}, and also in many numerical works, starting with
\cite{kley92}. In all of them, a backflow in the disk midplane occurs,
thought to be of thermal origin.

A thin accretion disk analytical solution in three dimensions has been given
in \cite[hereafter KK00]{KK00}. They derived the equations of polytropic,
viscous hydrodynamical accretion disk, using the Taylor expansion
in the small parameter $\epsilon=H/R$, the disk aspect ratio. In spite of
neglecting the thermal effects, backflow is still present with the values of
the viscosity parameter $\alpha_{\rm v}<0.685$. Such a disk accretes only at
the higher latitudes, closer to the surface, with the backflow in the
mid-plane occurring at a particular distance from the star. Inside this
distance, the matter is always flowing towards the star. If
$\alpha_{\rm v}>0.685$, the disk accretes across the entire height.

With the stellar magnetic field and disk resistivity added in the model,
angular momentum transfer depends on the details of star-disk magnetospheric
interaction. As first shown in \cite{ghl79a,ghl79b}, to correctly
describe this interaction, it is not enough to consider only the
disk. The rotating stellar surface and corona have to be included
in the model, to include the interaction with the region in the disk below,
and beyond the corotation radius. Pioneering such simulations were
\cite{R09,R13} and \cite{zf09,zf13}. As mentioned in
\cite[hereafter ZF09]{zf09}, for closing on the answer to stellar angular
momentum problem, the different regimes of magnetospheric interaction should
be probed, through the exploration of the parameter space. Here I follow this
suggestion.

In \S 2 I demonstrate the difference in geometry in the results with
different parameters, obtained from a set of simulations sweeping through
the parameter space. In \S 3 the results with different parameters
in the simulations are compared, and trends illustrated, with conclusions
listed in \S 4. A short but detailed technical exposition of the code
set-up, and an ``Atlas'' of quasi-stationary states in all the simulations
are presented in Appendix.

\section{Simulations of star-disk magnetospheric interaction}
\begin{table}
\caption{Parameter space in the study presented in ``Atlas'':
the stellar angular velocity $\Omega_\star$, stellar dipole
magnetic field strength $B_\star$, and the magnetic Prandtl
number $P_{\rm m}$-for which the also given the values of corresponding
resistivity parameter $\alpha_{\rm m}$. Shown are the stellar rotation
period and corotation radius in young stellar object (YSO) cases.
}
\label{params} 
\centering                          
\begin{tabular}{ c c c | c c c }        
\hline    
$\Omega_\star/\Omega_{\rm br}$ & $B_\star$(G) & $P_{\rm m}$ &
$\alpha_{\rm m}$ & $P_\star$(days) & $R_{\rm cor}(R_\star)$ \\
\hline\hline
0.05 & 250 & 6.7 & 0.1 & 9.2 & 7.37 \\
0.1 & 500 & 1.67 & 0.4 & 4.6 & 4.64 \\
0.15 & 750 & 0.95 & 0.7 & 3.1 & 3.54 \\
0.2 & 1000 & 0.67 & 1.0 & 2.3 & 2.92 \\
\hline\hline     
\end{tabular}
\end{table}
\begin{figure*}
\includegraphics[width=\columnwidth,height=0.6\columnwidth]{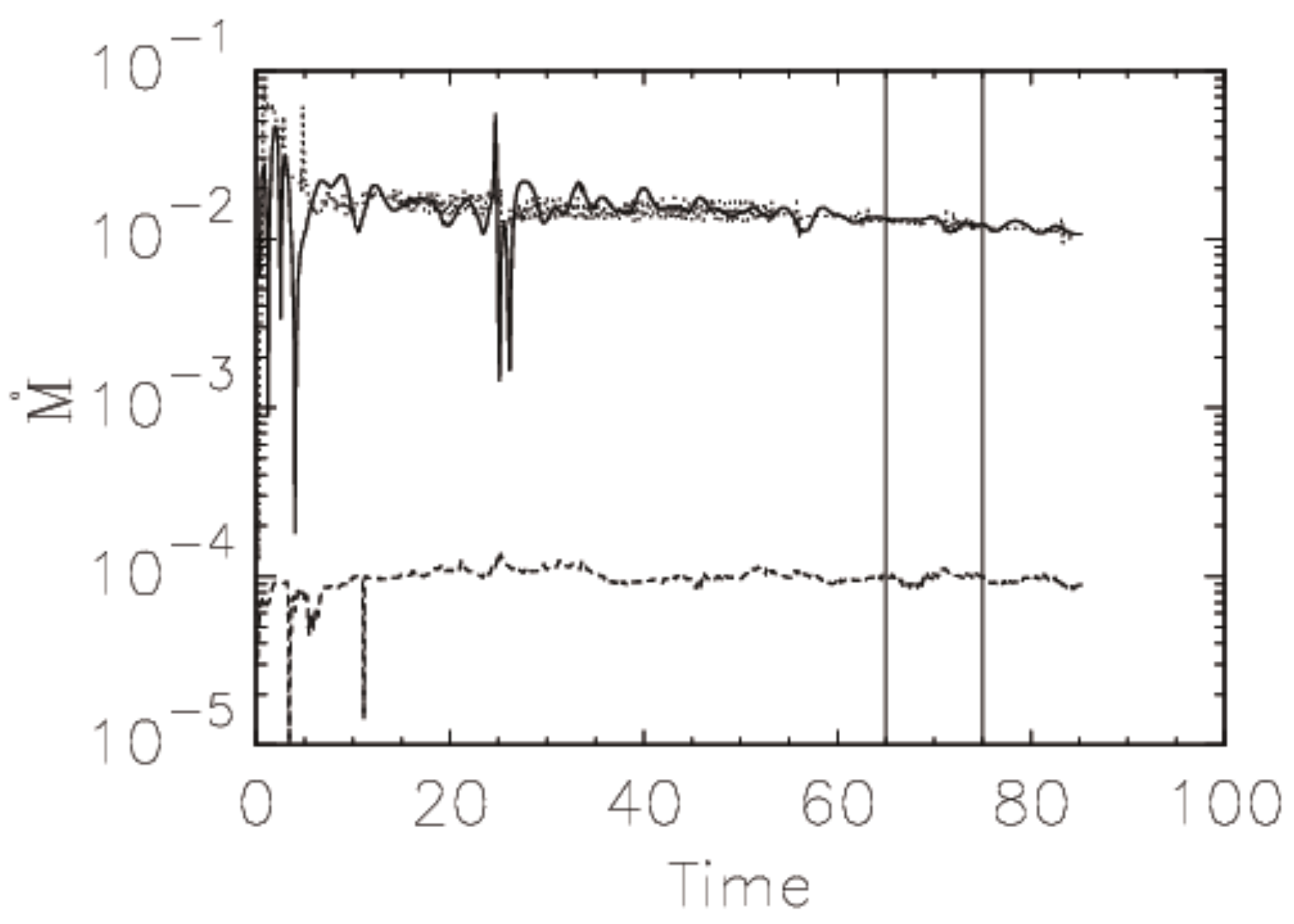}
\includegraphics[width=\columnwidth,height=0.6\columnwidth]{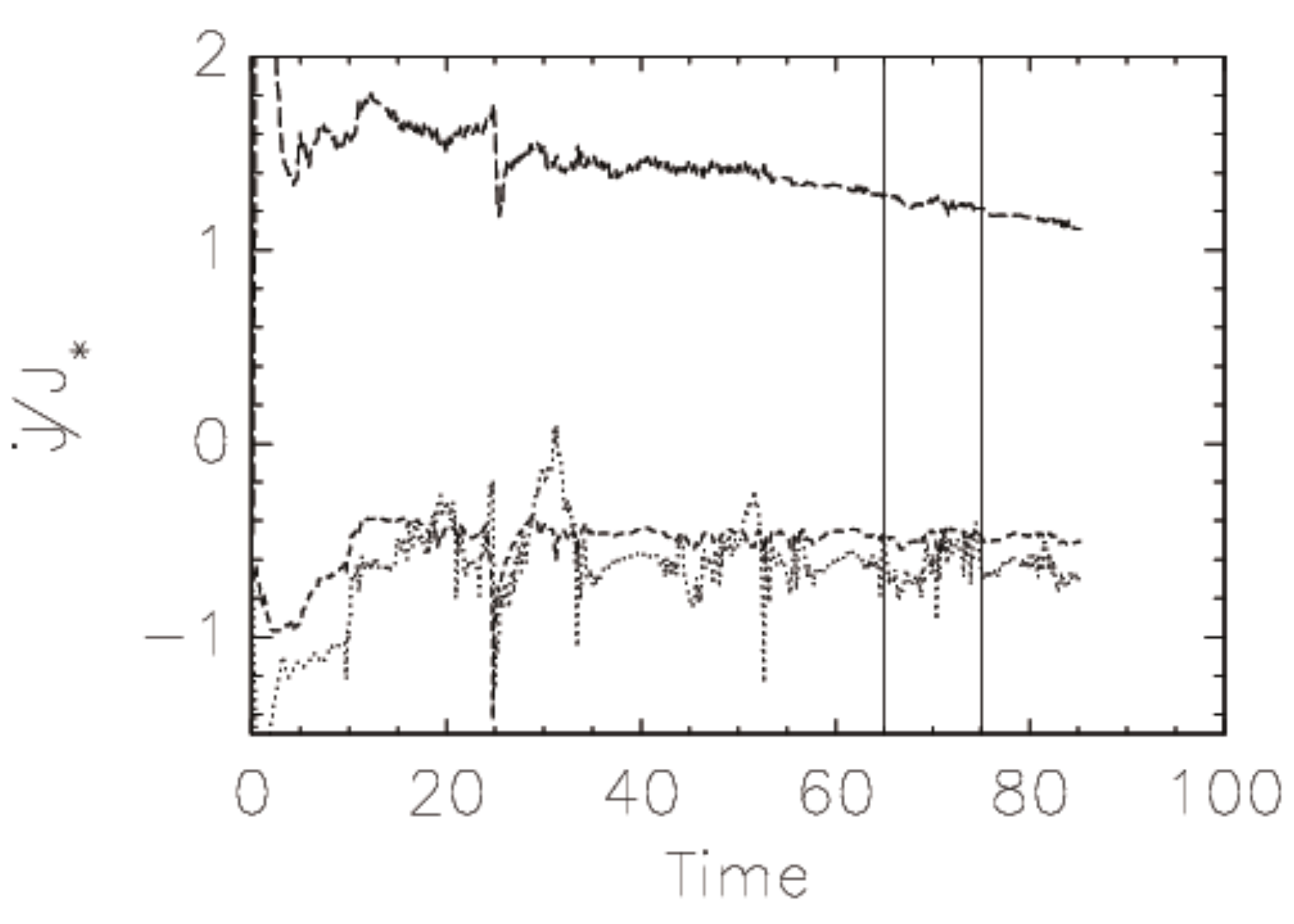}
\caption{Fluxes in the result in a case with
$\Omega_\star=0.1\Omega_{\rm br}$, $B_\star=0.5$~kG, $\alpha_{\rm m}$=1.
In the left panel, with solid, dotted and dashed lines is shown the
mass flux across the disk at R=12$R_\star$, onto the star and into the
stellar wind, respectively. In the right panel, angular momentum flux of
the matter in-falling onto the star from the distances below and beyond
the corotation radius is shown with the long and short dashed lines,
respectively, and in the stellar wind with the dotted line. Vertical
solid lines mark the interval in time, in which is taken an average
for computation of the quasi-stationary state.
}
\label{fig:sols2}
\end{figure*}
\begin{figure}
\includegraphics[width=\columnwidth]{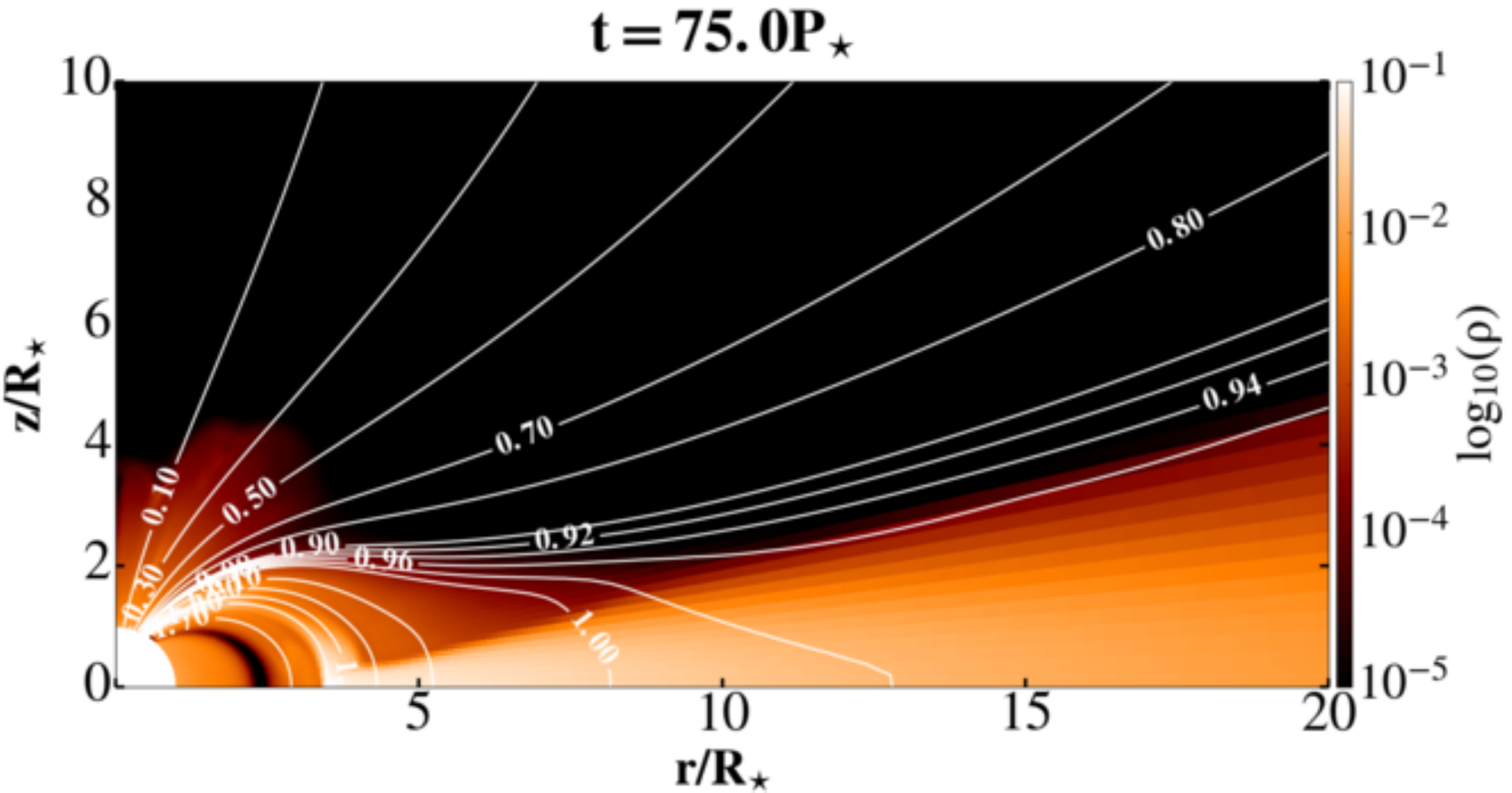}
\includegraphics[width=\columnwidth]{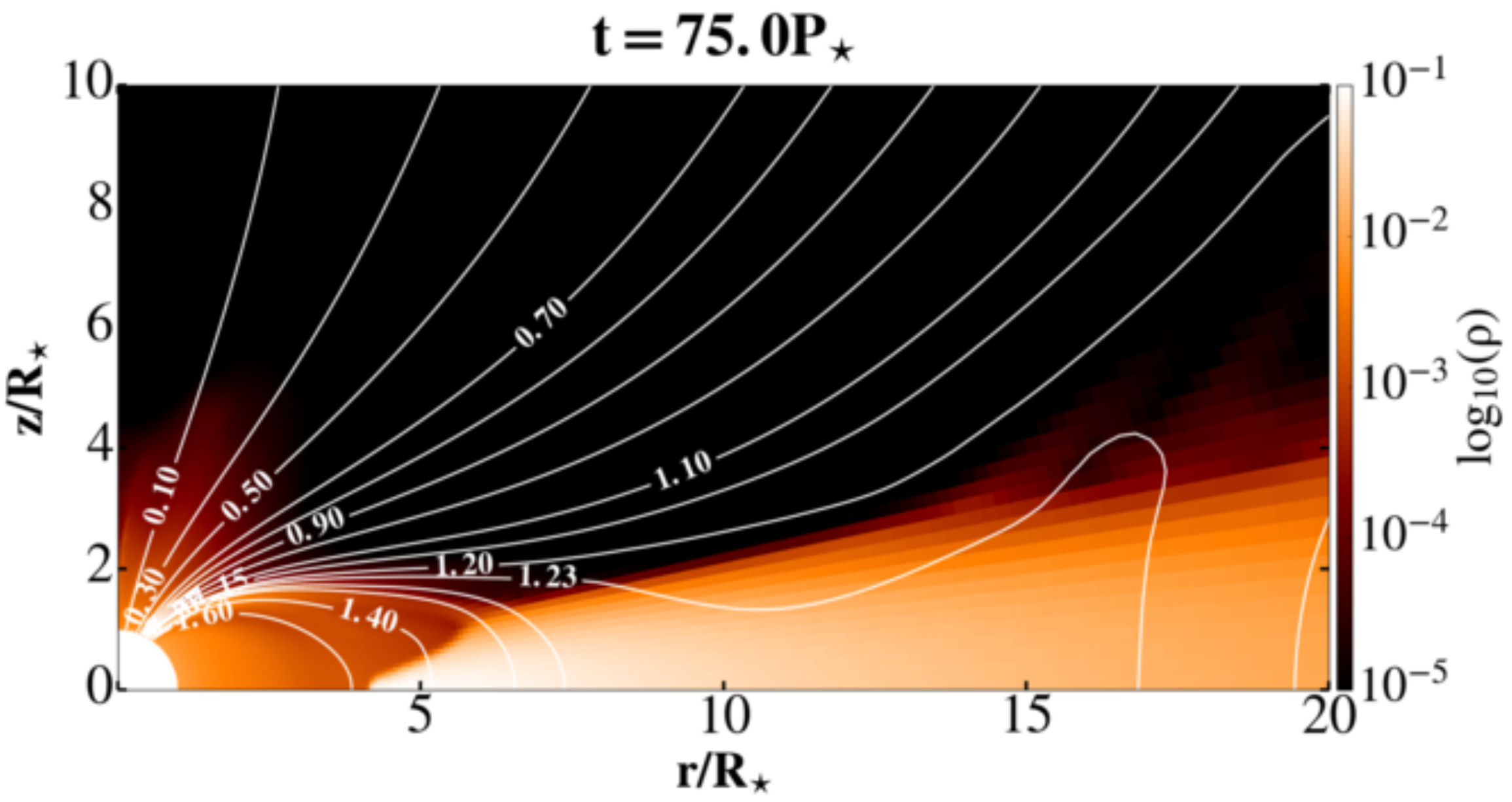}
\includegraphics[width=\columnwidth]{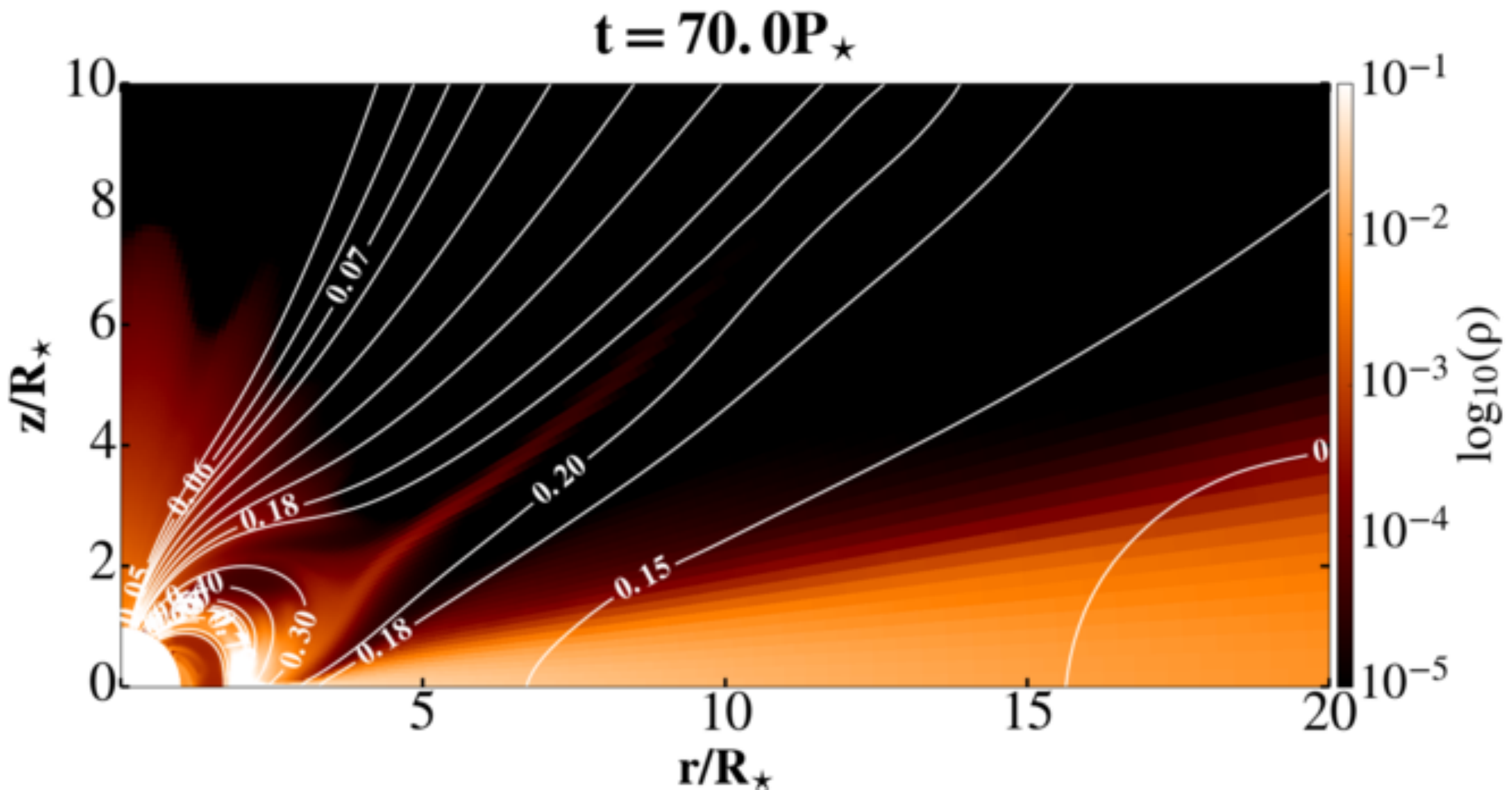}
\caption{Three different cases of geometry in the results. In the top and middle
panels are shown the matter density and a sample of magnetic field lines,
with the stellar magnetic field $B_\star$=1~kG and the resistivity $\alpha_{\rm
m}=1$, in the cases with the stellar rotation rates $\Omega_\star$=0.1$\Omega_{\rm br}$
(top panel) where a stable accretion column is formed) and
$\Omega_\star$=0.15$\Omega_{\rm br}$ (middle panel) in which the faster stellar
rotation prevents the accretion column formation. In the bottom panel is shown
the third case, with stellar magnetic field $B_\star$=0.5~kG, resistivity
$\alpha_{\rm m}=0.1$ and the rotation rate
$\Omega_\star$=0.1$\Omega_{\rm br}$, in which is formed a conical outflow.
}
\label{fig:sols1}
\end{figure}
\begin{figure*}
\includegraphics[width=\columnwidth,height=0.6\columnwidth]{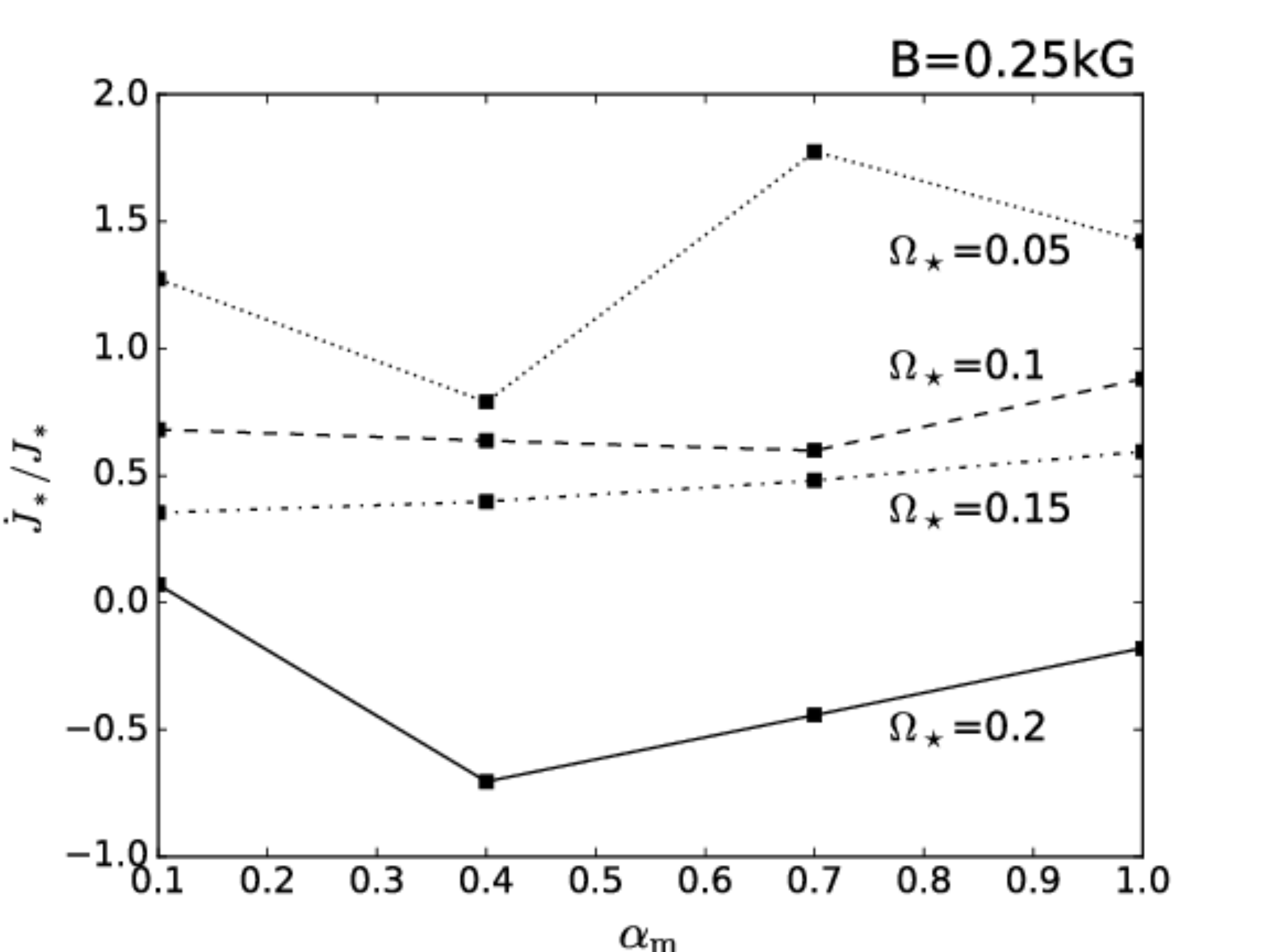}
\includegraphics[width=\columnwidth,height=0.6\columnwidth]{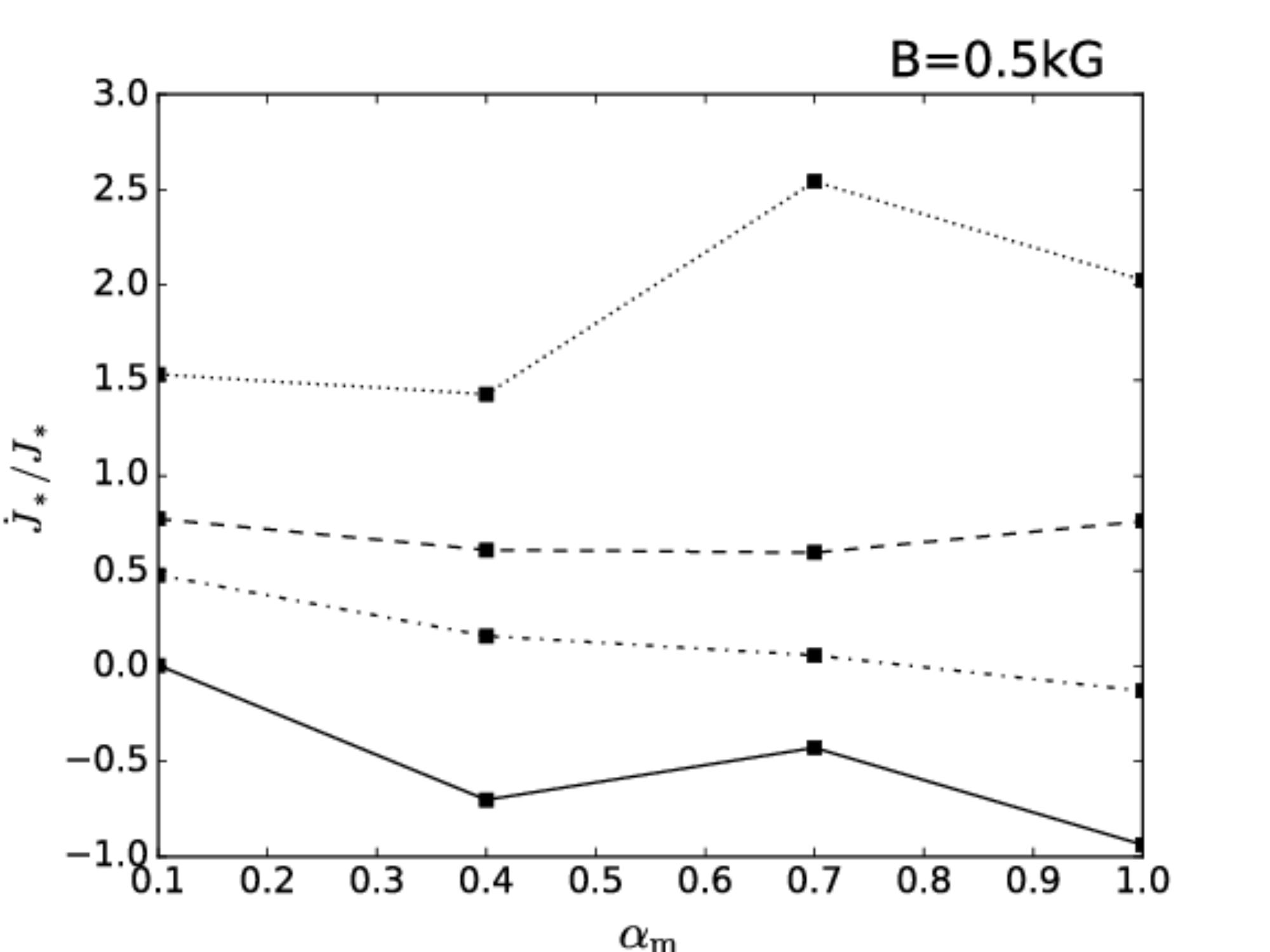}
\includegraphics[width=\columnwidth,height=0.6\columnwidth]{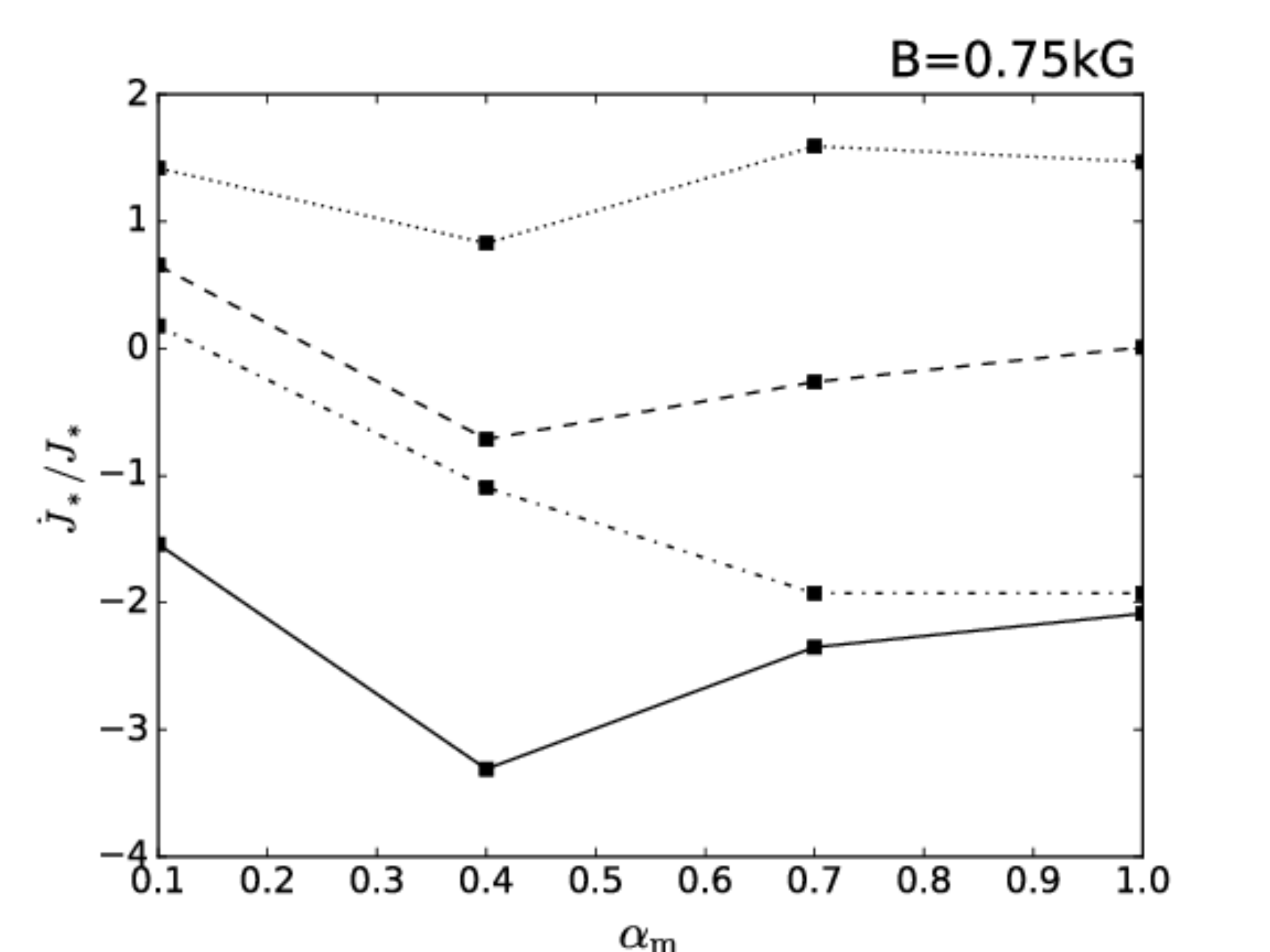}
\includegraphics[width=\columnwidth,height=0.6\columnwidth]{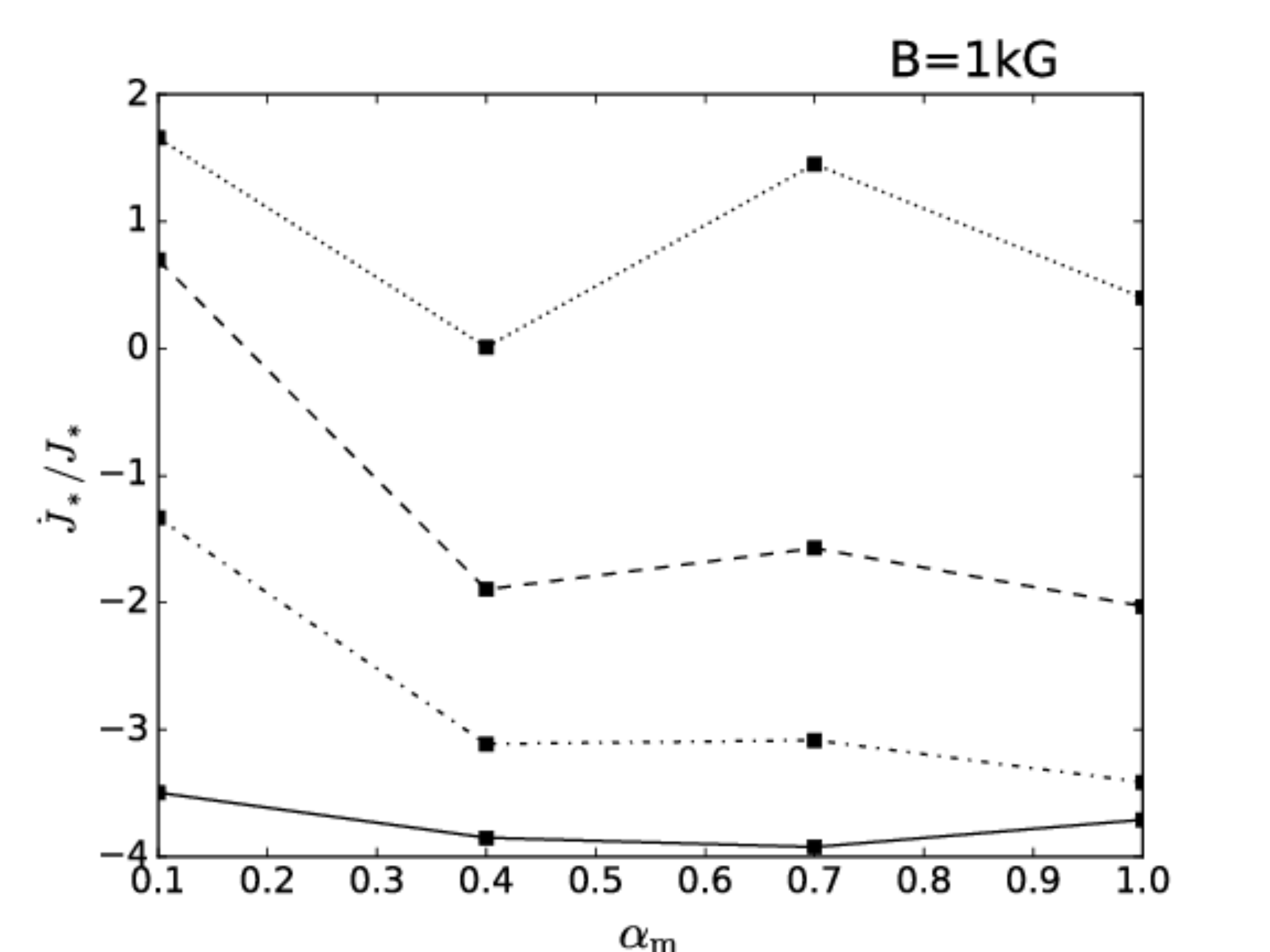}
\caption{Average angular momentum flux transported onto the stellar surface
by the matter in-falling from the disk onto the star through the accretion
column. In each panel is shown a set of solutions with one stellar magnetic
field strength, varying the stellar rotation rate and resistivity. Results
with $\Omega_\star/\Omega_{\rm br}=$~0.05, 0.1, 0.15 and 0.2 are shown with
the dotted, dashed, dash-dot-dot and solid lines, respectively, in the units
of stellar angular momentum expressed in
$J_{\star 0}=\rho_{d0}R_\star^4V_{\rm K\star}$. Positive flux spins-up the
star, negative slows it down. With the increase in stellar rotation rate,
spin-up of the star by the infalling matter decreases, eventually switching
to the spin-down.
}
\label{jstar}
\end{figure*}
\begin{figure}
\includegraphics[width=\columnwidth,height=0.6\columnwidth]{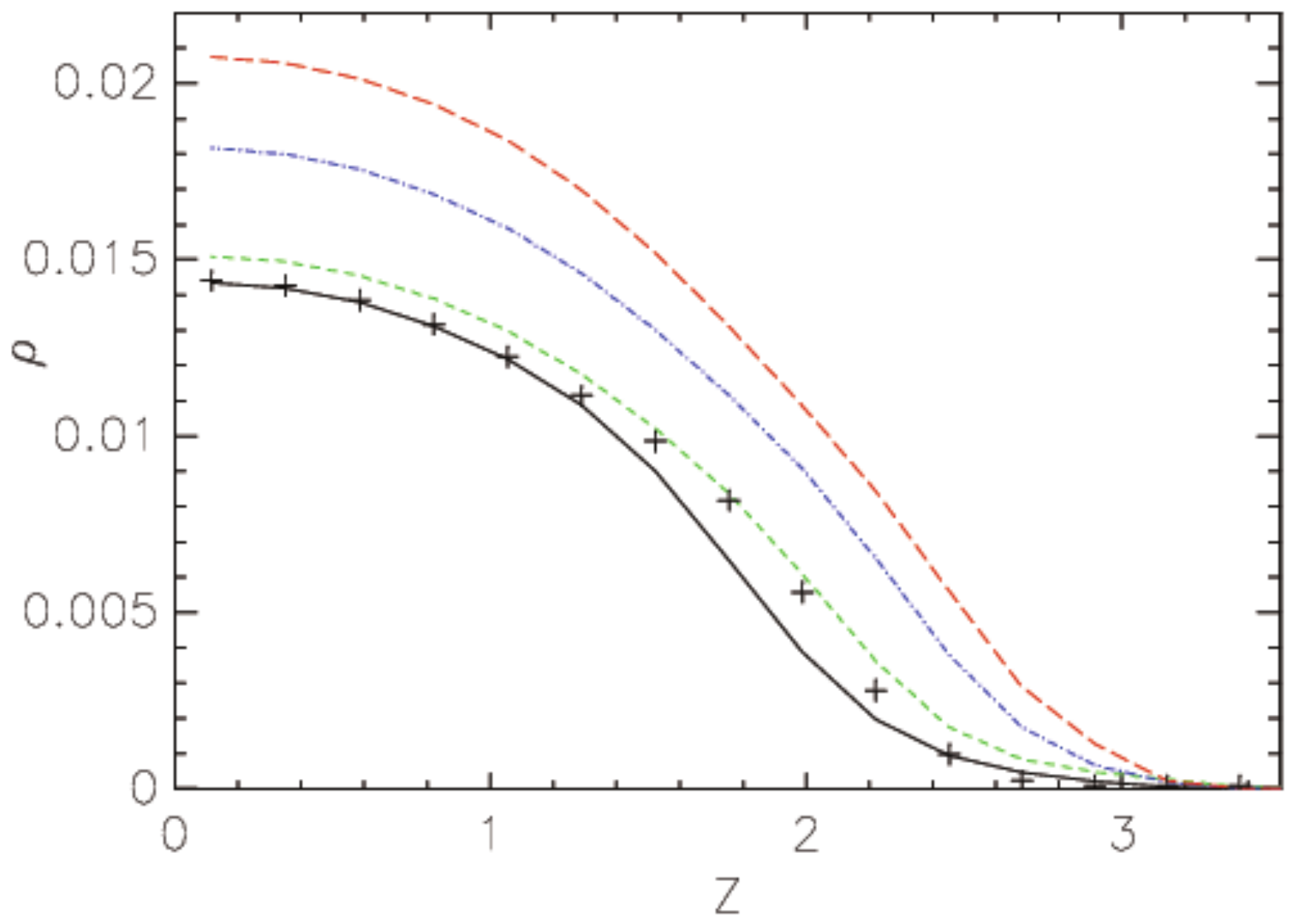}
\caption{
Disk density in the simulations, measured along the disk height at
R=15R$_\star$. With solid (black), short-dashed (green), dash-dotted
(blue) and long-dashed (red) lines are shown results in the cases
with $B_\star$=0.25, 0.5, 0.75 and 1kG, respectively. There is a trend
in density, with the increasing stellar field. Result in the simulations
without magnetic field is depicted in plus symbols.
}
\label{dens}
\end{figure}
\begin{figure}
\includegraphics[width=\columnwidth,height=0.6\columnwidth]{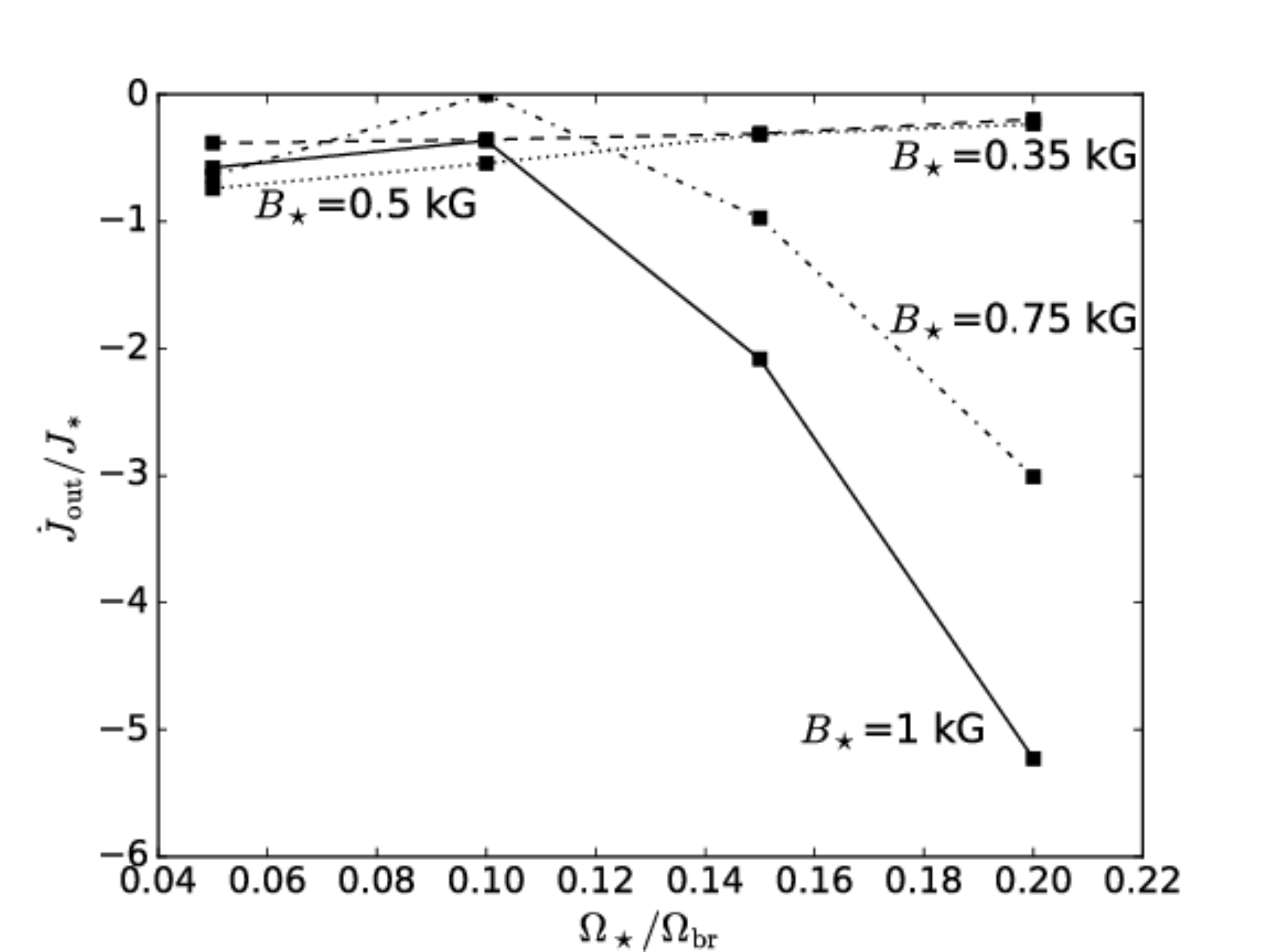}
\caption{Average angular momentum flux in the outflow, which forms
in the cases with $\alpha_{\rm m}$=0.1. It is computed at R=12R$_\star$,
in the cases with different stellar rotation rates. Normalization is to
the stellar angular momentum expressed in
$J_{\star 0}=\rho_{d0}R_\star^4V_{\rm K\star}$. With dotted, dash-dotted,
dashed and solid lines are shown the fluxes in the $B_\star$=0.25, 0.5,
0.75 and 1~kG cases, respectively.
}
\label{jme}
\end{figure}
Using the {\sc pluto} code (v.4.1) \citep{m07,m12}, I perform simulations
of star-disk magnetospheric interaction (SDMI). Initial conditions in
the disk and corona, and boundary conditions at the edges of the
computational domain closely follow ZF09. See Appendix A for the details
of numerical set-up.

Simulations span over 64 points in the parameter space (see
Table~\ref{params}), by varying stellar angular velocity $\Omega_\star$,
expressed in the units of stellar breakup angular velocity
$\Omega_{\rm br}=\sqrt{GM_\star/R_\star^3}$, stellar magnetic field
strength $B_\star$, and the magnetic Prandtl number
\begin{equation}
P_{\rm m}=\frac{2}{3}\frac{\alpha_{\rm v}}{\alpha_{\rm m}},
\end{equation}
where $\alpha_{\rm m}$ is the resistive parameter. In the simulations
presented here, I fixed viscosity parameter to $\alpha_{\rm v}=1$.

To compare the results in the different points in the parameter space,
a quasi-stationary state in each of the simulations is found. I rely on
two measurements: of the mass flux $\dot{M}$ and the angular momentum
flux $\dot{J}$.

The fluxes are computed by integrating 
\begin{equation}
\dot{M}=\int_{\rm S}\rho\vec{v}_{\rm p}\cdot d\vec{S},
\ \dot{J}=\int_{\rm S}\left( r\rho v_\varphi
\vec{v}_{\rm p}-\frac{rB_\varphi\vec{B}_{\rm p}}{4\pi}\right) d\vec{S}\ ,\nonumber
\end{equation}
over the corresponding surface $\vec{S}$ in the different parts of the flow.

Computed are the mass and angular momentum fluxes onto the star, into the
stellar wind, and across the disk height in the middle part of the disk,
at $R_{\rm d}$=12R$_\star$. In the cases in which an outflow forms,
the fluxes loaded into it are also computed at this distance. In the
angular momentum flux onto the star, the part coming from beyond, and below
the corotation radius $R_{\rm cor}=(GM_\star/R_\star^2)^{1/3}$ are computed
separately. A sign convention is such that a positive angular momentum flux
increases the total in the system, and a negative decreases it. In both
fluxes, the equatorial symmetry is taken into account so that fluxes are
computed in a complete meridional plane. 

A typical example, computed in one point of the parameter space in the
simulations, is presented in Fig.~\ref{fig:sols2}. It shows that after
relaxation from the initial and boundary conditions (lasting typically 10-50
stellar rotations), the quasi-stationary state is reached. Oscillations in
the simulations, still present into the quasi-stationary state, are
smoothed-out by averaging over some characteristic interval-typically an
average is taken over ten stellar rotations. In the example considered here,
the averaging interval is from 65 to 75 stellar rotations.

In each of the 64 simulations, such an interval is chosen in
which both the mass and angular momentum fluxes are not varying much. Then
an average value is found of the angular momentum flux through the various
components in the flow during that time interval.

In Appendix B is provided an ``Atlas'' of results in performed numerical
simulations. In each of the cases shown is a snapshot of matter density
in the quasi-stationary state, with a sample of magnetic field lines.

Three geometries in the solutions in ``Atlas'' can be distinguished, shown
in Fig.~\ref{fig:sols1}:

$\bullet$ Disk connected with the star by an accretion column,

$\bullet$ Disk pushed away from the star, without an accretion column,

$\bullet$ Conical outflow above the accretion column connecting the star
and the disk.

Information about the geometry and reach of the stellar magnetic field in the
disk is helpful in choosing the physical parameters in the model for the
post-processing or further analysis of the results.

\section{Trends in the angular momentum flux}
After indicating the quasi-stationary states in simulations, I proceed
to the comparison of solutions presented in ``Atlas''.

In the first example, in Fig.~\ref{dens} is shown a comparison of density in
the middle part of the disk in the simulations with the increasing stellar
magnetic field. The density increases with the increasing magnetic field.
This trend will contribute to the trends in mass and angular momentum fluxes.

In the second example, compared are simulations with the increasing stellar
rotation rate, in which the stellar magnetic field remains unchanged.
Angular momentum flux from the accretion disk loads into various
components of the flow: onto the star through the accretion column,
into the stellar wind and into the conical outflow, in the cases when it is
formed.

By comparing the solutions, I find a trend in the angular momentum flux,
shown in Fig.~\ref{jstar}. With the increase in the stellar rotation rate,
a spin-up of the star by the infalling matter decreases and eventually
switches to a spin-down. A similar outcome is obtained in each of the cases
with different stellar magnetic field strengths. Such a trend is not
surprising since for slowly rotating stars the torque exerted on the
star will depend mostly on the stellar magnetic field, at each value of the
resistivity parameter. The trend will change only after a change in geometry
of the flow.

The third example of a trend in the results is in the cases with
$\alpha_{\rm m}$=0.1, in which a conical outflow is formed. The angular
momentum flux expelled from the system with such an outflow in the cases
with different magnetic field strength is shown in Fig.~\ref{jme}.

The momentum flux in the outflow does not change much, except in the
cases with faster stellar rotation {\em and} large magnetic field, when
the outflow takes away much more, for an order of magnitude, of the
angular momentum from the system. The magnetospheric ejection increases
with the stronger magnetic field and faster rotation.

Trends in the presented examples can be compared with the results in other
models and simulations when they become available. If found robust, such
trends should be compared with the observations and incorporated in the
star formation models.

\section{Conclusions}
In the numerical simulations of star-disk magnetospheric interaction (SDMI),
I investigate angular momentum flux in the system. In a parameter study,
stellar rotation rate, magnetic field, and the disk resistivity are varied,
to obtain a suite of the quasi-stationary solutions. I compute the angular
momentum flux in different components of the flow to compare results in the
cases with different stellar magnetic field strengths.

Discussing the results, in the first example I find a trend in the matter
density along a vertical line in a mid-part of the disk. The density
increases with the increasing stellar magnetic field in the simulation.

The second result is a trend in angular momentum flux onto the star, with
the increasing stellar rotation rate. I find that with the faster stellar
rotation, spin-up of the star decreases, and eventually switches to a
spin-down. 

The third find from the performed simulations is that in the cases with
$\alpha_{\rm m}$=0.1 and a faster rotating star, the angular momentum flux
loaded into the conical outflow increases significantly, with the large
values of the stellar magnetic field. The increase is for an order of
magnitude. In the cases with a small stellar magnetic field, there is no
change in a load of angular momentum in the outflow.

The results apply in disks around young stellar objects (YSOs) and can be
rescaled to disks in a compact binary. I provide a table with scaling
factors for different cases.

I presented here the results with a slowly rotating star. In the
simulations with a stellar rotation faster than 20\% of the stellar breakup
rotation rate, two additional kinds of a solution are obtained, similar to
\cite{R09}: with a fast axial outflow, and with both the conical and axial
outflow. I leave them for a separate study, together with the solutions with
$\alpha_{\rm v}<$0.685, which show a backflow in the initial disk.

\section*{Acknowledgements}
M\v{C} developed the set-up of star-disk simulations while in CEA, Saclay,
France, under the ANR Toupies grant with A.S. Brun. Work in NCAC Warsaw is
funded by a Polish NCN grant no. 2013/08/A/ST9/00795, and a collaboration
with Croatian STARDUST project through HRZZ grant IP-2014-09-8656 is
acknowledged. I thank IDRIS (Turing cluster) in Orsay, France, ASIAA/TIARA
(PL and XL clusters) in Taipei, Taiwan and NCAC (PSK cluster) in Warsaw,
Poland, for access to Linux computer clusters used for the high-performance
computations. The {\sc pluto} team is thanked for the possibility to use the
code, in particular C. Zanni for help with the code modifications. V.
Parthasarathy and F. Bartoli\'{c} are acknowledged for developing the Python
scripts for visualization, N. Bessolaz for the initial version of the set-up,
and M. Flock for useful discussions about the code.





\appendix


\section{Numerical simulations of star-disk magnetospheric interaction}

\begin{figure}
\includegraphics[width=\columnwidth]{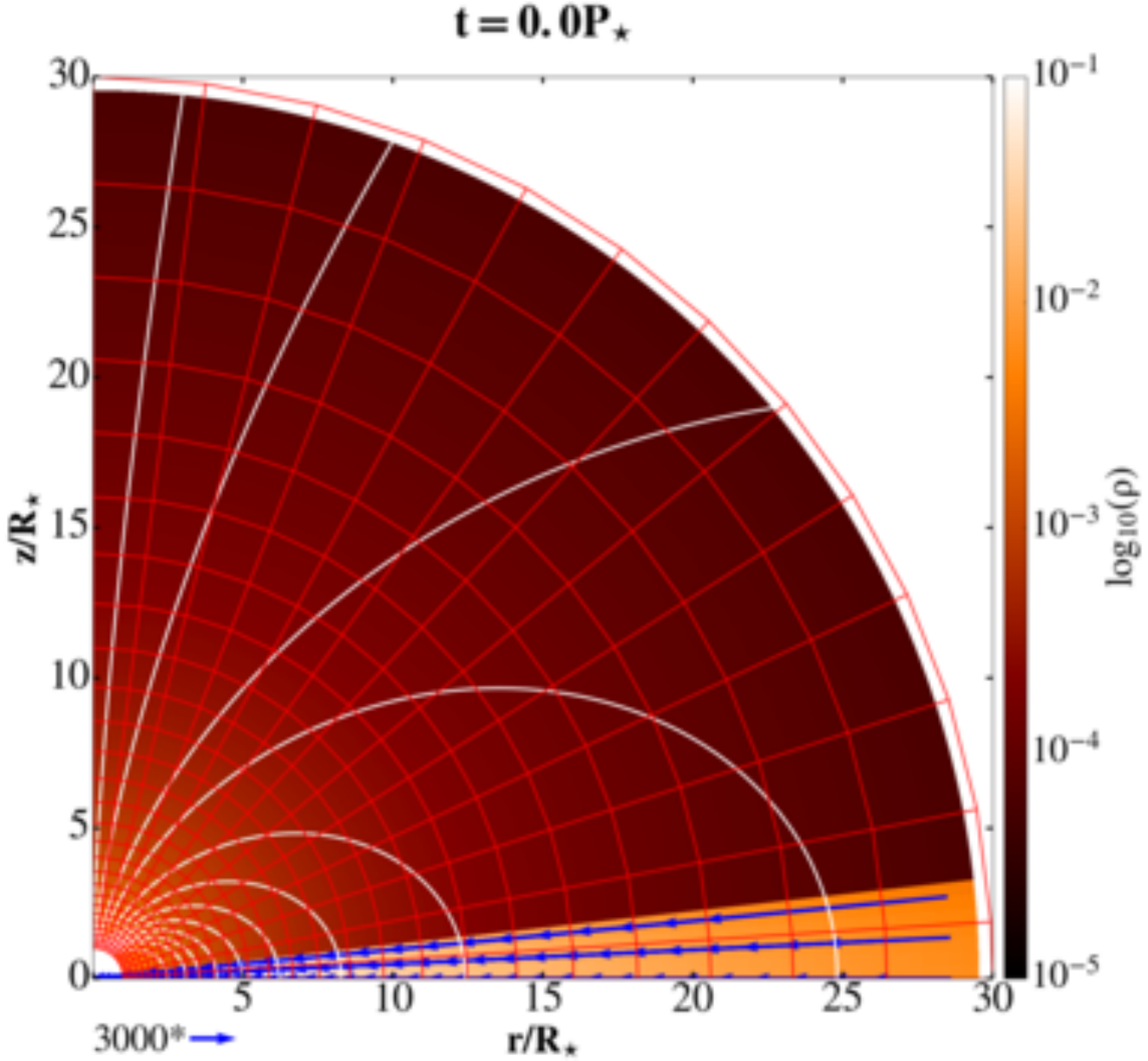}
\caption{The initial density distribution of matter in simulations, with a
sample of the initial poloidal magnetic field lines shown with the solid white
lines. The vectors show the initial velocity distribution in the disk, with
the unit vector length measured in Keplerian velocity units indicated below
the figure. The computational grid is shown in 8x8 blocks of cells.
}
\label{fig:initialcond}
\end{figure}
Star-disk magnetospheric interaction (SDMI) determines the angular momentum
transfer in a star-disk system. Simulations including SDMI have been reported
in works by \cite{R09,R13} with the code which is not publicly available, and
ZF09 and \cite{zf13} with the publicly available code {\sc pluto} (v.3).
Following ZF09, I perform simulations with the updated version of the
{\sc pluto} code (v.4.1) \citep{m07,m12}.

I present a short version of the set-up, amended to facilitate
reproducibility{\footnote{Author is aware of several unsuccessful attempts
in repeating the ZF09 set-up by experienced researchers and students during the
last decade. To my knowledge, the first successful following has been
reported in \cite{cpk17}.}.

The equations solved in the magneto-hydrodynamic (MHD) module of the {\sc pluto}
code are, in the cgs system of units:
\begin{eqnarray}
\frac{\partial\rho}{\partial t}+\nabla\cdot(\rho\vec{\rm v}) =0\\
\frac{\partial\rho\vec{\rm v}}{\partial t}+\nabla\cdot\left[\rho\vec{\rm v}
\vec{\rm v}+\left(P+\frac{\vec{\rm B}\vec{\rm B}}{8\pi}\right)
\vec{\rm I}-\frac{\vec{\rm B}\vec{\rm B}}{4\pi}-\vec{\tau}\right]=\rho\vec{g}\\
\frac{\partial E}{\partial t}+
\nabla\cdot\left[\left(E+P+\frac{\vec{B}\vec{B}}{8\pi}\right)\vec{v}\right] \\
+\nabla\cdot\left[\underbrace{\eta_{\rm m}\vec{J}\times \vec{B}/4\pi - \vec{v}\cdot\vec{\tau}}_{\rm heating\ terms}\right]=
\rho\vec{g}\cdot\vec{v}-\underbrace{{\Lambda}}_{\rm cooling} \\
\frac{\partial\vec{\rm B}}{\partial t}+\nabla\times(\vec{\rm B}
\times\vec{\rm v}+\eta_{\rm m}\vec{J})=0 
\end{eqnarray}
where the symbols have their usual meaning: $\rho$ and $\vec{v}$ are the
matter density and velocity vector, P is the pressure, $\vec{B}$ is the
magnetic field and $\eta_{\rm m}$ and $\vec{\tau}$ represent the
resistivity and the viscous stress tensor, respectively. $\Lambda$ is the
cooling source term, the gravity acceleration is
$\vec{g}=-\nabla\Phi_{\rm g}$, where the gravitational potential of the star
with mass $M_\star$ is equal to $\Phi_{\rm g}=-GM_\star/R$. Then
$g_{\rm R}=-1.0/R^2$ in the code units.

Simulations are performed using the second-order piecewise linear
reconstruction, with a Van Leer limiter in density and magnetic field and
a minmod limiter in the pressure and velocity. To enhance stability, in the
subroutine \verb|plm_states|, the default is set to Van Leer instead of the
less diffusive monotonized central difference limiter. An approximate Roe
solver (hlld in the {\sc pluto} options) is used, with a modification in the
\verb|flag_shock| subroutine: flags are set to switch to more diffusive hll
solver if the internal energy is less than 1\% of the total energy, instead
of switching in the presence of shocks. The second-order time-stepping (RK2)
is employed, and $\nabla\cdot\vec{B}=0$ is maintained by the Constrained
Transport. The magnetic field is evolved with the split-field option, so
that only changes from the initial stellar magnetic field are evolved in
time \citep{tan94,pow99}. In version 4.1 of {\sc pluto} the Constrained
Transport is by default disabled to work with the background field in the
resistive MHD-it is probably an artifact from the older version. To enable
the background field option, the constraint is removed by changing the
condition in the routine \texttt{backgrnd\_field}, to point to some feature
which will not be used, e.g. thermal conduction.

A logarithmically stretched radial grid, and uniform grid in the meridional
half-plane $\theta$=[0,$\pi/2$] in spherical coordinates are used to perform
2D-axisymmetric SDMI simulations. The resolution is set to
$R\times\theta$=(217$\times100$) grid cells, in the physical domain from
the stellar surface to 30 stellar radii, as shown in Fig.~\ref{fig:initialcond}.

The total energy density is
$E=P/(\gamma-1)+\rho(\vec{v}\cdot\vec{v})/2+\vec{B}\cdot\vec{B}/8\pi$,
where $\gamma=5/3$ is the plasma polytropic index. The electric current is
given by the Ampere's law $\vec{J}=\nabla\times\vec{B}/4\pi$.

To prevent the thermal thickening of the accretion disk in simulations,
following ZF09 description, from the {\sc pluto} energy equation the
underbraced Ohmic and viscous heating terms are removed. This equals to
the assumption that all the heating is radiated away from the disk.
To do this, lines in the code with the viscous and resistive part
of the fluxes computation in \verb|parabolic_flux| subroutine
are commented out. The viscous and resistive fluxes are added in
the \texttt{rhs} subroutine, to ensure the inclusion of the correct
dissipative fluxes terms. Such simulations remain in the
non-ideal MHD regime because of the viscous term in the momentum
equation, and the resistive term in the induction equation\footnote{Without
modification of the energy equation, but with the power law cooling
introduced to account for the disk dissipative heating, a similar
outcome is obtained. It is advisable to use this, simpler method, in
the first attempt of the set-up.}.

The initial disk is set with the initial density set by KK00 with a
self-similar profile with an aspect ratio $\epsilon$:
\begin{flalign}
& \rho_{\rm d}=\rho_{\rm d0}\bigg\{\frac{\gamma-1}{\gamma\epsilon^2}\bigg[\frac{R_*}{R}-\left(1-\frac{\gamma\epsilon^2}{\gamma-1}\right)\frac{R_\star}{R\sin\theta}\bigg]\bigg\}^{1/(\gamma-1)}= & \\ \nonumber 
& =\rho_{\rm d0}\bigg\{\frac{2}{5\epsilon^2}\bigg[\frac{R_\star}{R}-\left(1-\frac{5}{2}\epsilon^2\right)\frac{R_\star}{R\sin\theta}\bigg]\bigg\}^{3/2}. &  \nonumber
\end{flalign}
The pressure is
\begin{flalign}
& P_{\rm d}=\epsilon^2\rho_{\rm d0}v_{\rm K\star}^2\left(\frac{\rho_{\rm d}}{\rho_{\rm d0}}\right)^\gamma= & \\ \nonumber
& =\epsilon^2\bigg[\frac{R_\star}{R}-\left(1-\frac{\gamma\epsilon^2}{\gamma-1}\right)\frac{R_\star}{R\sin{\theta}}\bigg]^{5/2}. &  \nonumber
\end{flalign}

The capital $R$ indicates the spherical radius, and $r=R\sin\theta$ is the
cylindrical radius. The disk unit density $\rho_{\rm d0}$ and Keplerian speed
$V_{\rm K\star}$ are both calculated in the disk midplane at R$_\star$.
The initial disk is truncated about the corotation radius.

The obtained disk is a polytropic hydrodynamical solution of the viscous
accretion disk in the full 3D, obtained by approximate expansion up to the
second order in the terms of $\epsilon=c_{\rm s}/v_{\rm K}$, the disk aspect
ratio measured on the midplane of the disk, where
$c_{\rm s}=\sqrt{P_{\rm d}/\rho_{\rm d}}$ and $v_{\rm K}$ are the isothermal
sound speed and the Keplerian speed in the disk.

The viscosity and resistivity are defined explicitly in separate subroutines,
by a second-order finite difference approximation for the dissipative fluxes,
with checking of the time step. Both are parameterized by the Shakura-Sunyaev
prescription as $\alpha c^2/\Omega$. To avoid the issues related to the
backflow in the disk, here is addressed only the case with
$\alpha_{\rm v}=1$. The viscosity is defined by
$\eta_{\rm v}=\frac{2}{3}\rho\alpha_{\rm v}P_0r^{3/2}$, where $P_0$ is
the initial pressure in the disk. The resistivity is
$\eta_{\rm m}=\frac{2}{5}\alpha_{\rm m}P_0r^{3/2}$. The diffusive parameters
$\alpha_{\rm v}$ and $\alpha_{\rm m}$ are defined in separate subroutines
\texttt{visc\_nu} and \texttt{res\_eta}. A condition for inclusion of the
diffusive term is in both routines defined by the 
$\beta=P_{\rm mag}/P_{\rm hyd}>0.5$, meaning that the magnetic pressure is
prevailing. In both subroutines, the diffusive term is taken into account
only when tracer value is unity, otherwise it is set to zero.

The initial disk velocity profile is, by KK00:
\begin{flalign} 
& v_{Rd}=-\alpha_{\rm v}\epsilon^2\bigg[10-\frac{32}{3}\Lambda
\alpha_{\rm v}^2-\Lambda\left(5-\frac{1}{\epsilon^2\tan^2\theta}\right)\bigg]
\sqrt{\frac{GM_*}{R\sin^3\theta}} & \\ \nonumber
& v_{R\varphi}=\bigg[\sqrt{1-\frac{5\epsilon^2}{2}}+\frac{2}{3}
\epsilon^2\alpha_{\rm v}^2\Lambda\left(1-\frac{6}{5\epsilon^2\tan^2\theta}\right)
\bigg]\sqrt{\frac{GM_*}{R\sin\theta}} &  \nonumber
\end{flalign}
where
\begin{equation}
\Lambda=\frac{11}{5}/\left(1+\frac{64}{25}\alpha_{\rm v}^2\right).
\end{equation}

The initial corona is a non-rotating, polytropic corona with $\gamma=5/3$,
in the hydrostatic balance. It is defined by the density and pressure
given by:
\begin{flalign}
& \rho_{\rm c}=\rho_{\rm c0}(R_\star/R)^{1/(\gamma-1)},& \\ \nonumber
& P_{\rm c}=\rho_{\rm c0}\frac{\gamma-1}{\gamma}
\frac{GM_\star}{R_\star}\left(\frac{R_\star}{R}\right)^{\gamma/(\gamma-1)}, {\rm where}\ 
\rho_{\rm c0}\ll\rho_{\rm d0} & \nonumber
\end{flalign}
is the ratio between the initial coronal and disk density, set to 0.01.

The initial stellar magnetic dipole field is set with the field axis
aligned with the stellar rotation axis. There is no resistivity in the
magnetosphere outside of the disk. It means that the reconnection of the
magnetic field is a numerical, not physical dissipation.

In the {\bf internal boundary} part, which enables change in the quantities
inside the computational domain in {\sc pluto}, the density in the
grid cell just above the star is corrected when it falls below some limit
value (I set it to $5\times10^{-8}$), to avoid too small density near the star.
The pressure is corrected in such a way to conserve the same sound speed
in the corona. To maintain the self-consistency, velocities are also changed,
to conserve the momentum. Here is also ensured that the scalar tracer value
is always set to zero in the corona. Around the reconnection sheet and
outflows, the tracer scalar can obtain spurious values, here it is prevented.

The numerical heating in the corona is prevented by enforcing the
conservation of the entropy S, to keep the values close to the initial
conditions. The maximum of the entropy is atop the star, so it is kept
throughout the computational box below the value
$S_{max}=P_c/\rho_c^{\gamma}=2\rho_{c0}^{-2/3}/(5R_*)=8.61774$. For the
minimum, the small number $S_{min}=0.01$ is set. From this are obtained
corrected values for the pressure interval in the computational box,
by $P=\max[\min(P,S_{max}\rho^{\gamma}),S_{min}\rho^{\gamma}]$.

In the {\bf inner boundary conditions}, the density, pressure and toroidal components of
the velocity and magnetic field are prescribed from the active zones into the
boundary. The linear extrapolation is used, with Van Leer limiter in the density
and the magnetic field, and minmod limiter in the pressure and velocity. For the
numerical stability in the corona, in the cases with $v_R>0$, introduced is a
correction of the pressure by a free parameter $T_f$ in the inner radial boundary
condition. It should be set to the number of order a few hundred. It adjusts the
ram pressure $\rho v^2$ atop the star (stellar wind) from the default
$T_f=2v_R^2/5$ to $T_f=(2/5-T_f)*v_R^2$ so that the pressure in the corona is
given by $P=(2/5-T_f*v_R^2)\rho_cR_i^{-5/2}$. The first term here is the coronal
initial pressure $P_c=2\rho_cR_i^{-5/2}/5$. Since $P\sim\rho T$ in the ideal gas
law, we can think of 2/5 as an effective temperature, where $T_f$ is an
ad-hoc correction.

In the axisymmetric 2D set-up, only the toroidal component of the magnetic field
is set in the staggered grid, because in the Constrained Transport
method, a staggered grid is used\footnote{When the staggered grid is used,
in \texttt{boundary} subroutine one has to correct the misplaced call to the
user-defined boundary at the grid cell centers, to come after the
assignment of the normal component of staggered $\vec{\rm B}$, instead before it.}.
I use a specially tailored boundary condition for $B_\varphi$, in which 
$\Omega_{\rm eff}=\Omega-\rm{v}_pB_\varphi/rB_p$ is derived from the
condition for the stellar surface as a rotating perfect conductor. The magnetic
torque to drive the plasma rotation atop the star is set, with the matching
measured by the comparison of the stellar angular velocity and the effective
rotation rate of the field lines by the ratio $\Omega_{\rm eff}/\Omega_*$.
This prescribes rotation of the matter atop the star and the effective rotation
rate of the field lines to $\Omega_{\rm eff}$, with $v_p$ and $B_p$
standing for the poloidal velocity and magnetic field, respectively. In addition,
in the Constrained Transport method subroutine \texttt{ct} toroidal component of
the EMF is set to zero. Then in the stellar reference frame, the electric field
is zero, with the flow speed parallel to the magnetic field.

I do not follow the suggestion in ZF09 to impose a continuity of the
speed along the field lines on the outflowing matter, nor the continuity
of the axisymmetric MHD invariant $k=4\pi\rho v_{\rm p}/B_{\rm p}$ on the
infalling material, as it did not show to improve the result.

With those corrections, the stellar rotation rate is set. I start the
simulation with $\Omega_\star$, not with the slower rotation of the
star as described in ZF09.

In the {\bf outer boundary conditions}, in the coronal part of the domain,
the logarithmic extrapolation in the density and pressure is set. In the
radial and meridional components of the velocity an outflow is set, and a
linear extrapolation with the minmod limiter for the azimuthal velocity
component and Van Leer limiter in the toroidal magnetic field component.
At the disk outer radial boundary, I introduce the initial hydrodynamical
values, anticipating the thickening of the disk for 25\%. Since the velocity
could roll back above the thickened disk, inflow to twice the initial disk
height in the corona is prevented by setting the velocities to zero. The
toroidal magnetic field is linearly extrapolated with a Van Leer limiter.

The simulations are stopped after 100 stellar rotations. In some cases,
the simulation stops earlier, because of a too small timestep. It is caused
by the numerical instability in the rarefied corona, not because of the
instability in the disk.

\subsection{Normalization and physical units}
Normalized equations are solved in the code. The unit length, velocity
and mass are chosen with the stellar radius $R_\star$, the Keplerian speed at
the stellar surface $v_{\rm K\star}$ and mass $M_\star$. The time unit is then
$t_0=R_\star/v_{\rm K\star}$.  Time t in the results is measured in
the number of stellar rotation periods P$_\star$. The mass flux rate is measured in
$\dot{M}_0=\rho_{\rm d0}R_\star^2v_{\rm K\star}^3$, which is the free
parameter in simulation, from which is determined $\rho_{\rm d0}$,
the disk density. The initial coronal density is defined as a free parameter
in the code, $\rho_{\rm c0}=0.01\rho_{\rm d0}$.
The magnetic field unit is defined by $B_0=v_{\rm K\star}
\sqrt{\rho_{\rm d0}}$. Torque in the simulations is measured in
the units of $\dot{J}_0=\rho_{\rm d0}R_\star^3v_{\rm K\star}^2$.

\begin{table}
\caption{Typical values and scaling for different central objects. The mass
M$\star$,
radius R$\star$, period P$\star$ and equatorial stellar magnetic field
B$_\star$ are chosen, to derive the rest of the quantities. The code units
should be multiplied by the factors given in the table, to apply it to
different cases.}
\label{rescal} 
\centering                          
\begin{tabular}{ c c c c }        
\hline    
    & YSOs & WDs & NSs \\
\hline\hline
M$_\star$(M$_\sun$) & 0.5 & 1 & 1.4 \\
R$_\star$ & 2R$_\sun$ & 5000km & 10km \\
P$_\star$ & 4.6d & 6.1s & 0.46ms \\
B$_\star$ (G) & 500 & 5$\times10^5$ & 10$^8$ \\
\hline
$\rho_{d0}$(g/cm$^3$) & 1.2$\times10^{-10}$ & 9.4$\times10^{-9}$ &
4.6$\times10^{-6}$\\
v$_0$(km/s)  & 218 & 5150 & 136000 \\
$\dot{M}_0(M_\sun/yr)$ & 5.7$\times10^{-7}$ & 1.9$\times10^{-9}$  &
$10^{-9}$\\
B$_0$(G) & 200 & 5$\times10^{4}$ & 2.93$\times10^{7}$\\
\hline\hline     
\end{tabular}
\end{table}
Simulations can be rescaled to different objects by using the scaling
coefficients from the Table~\ref{rescal}. In the case of compact objects, one
should keep in mind that the radial extension of the domain, measured from
the axis of rotation, should not reach the light cylinder,
$R_{\rm \ell c}\Omega_\star=c$, where the azimuthal velocity equals the
speed of light. This limitation was not mentioned in the previous
publications with SDMI, so we write distances of the light cylinder in the
different cases are listed in the Table~\ref{lcyls}.
\begin{table}
\caption{Position of the light cylinder as a function of stellar rotation
rate in some typical cases. Stellar rotation rate is expressed in the units
of stellar breakup rotation rate, and position of the light cylinder in the
stellar radii, R$_\star$.}
\label{lcyls} 
\centering                          
\begin{tabular}{ c c c c }        
\hline    
 $\Omega_\star/\Omega_{\rm br}$ & $R_{\rm \ell c}({\rm YSO})$ & $R_{\rm
\ell c}({\rm WD})$ & $R_{\rm \ell c}({\rm NS})$ \\
\hline\hline
0.05 & 27454 & 1164 & 44 \\
0.1 & 13727 & 582 & 22 \\
0.2 & 6864 & 291 & 11 \\
0.25 & 9151 & 233 & 8.8 \\
0.5 & 2745 & 116  & 4.4 \\
0.75 & 1830 & 78 & 2.9 \\
1.0 & 1373 & 58 & 2.2 \\
\hline\hline     
\end{tabular}
\end{table}
\section{``Atlas'' of the results}
For comparison of matter density and poloidal magnetic field distribution in
the solutions, I present the results in an ``Atlas''.

To clearly show the accretion column, a zoom is done into 2/3 of the radial
domain in the simulations. The extent to which disk is magnetically
connected with the star is shown with a sample of the poloidal
magnetic field lines, assigned with the corresponding values of
the flux function.

Snapshots are shown in the quasi-stationary state in all 64 simulations.
They are grouped by the increasing stellar magnetic field and rotation
rates, with four panels showing the solutions with increasing resistivity.


\begin{figure*}
\includegraphics[width=\columnwidth]{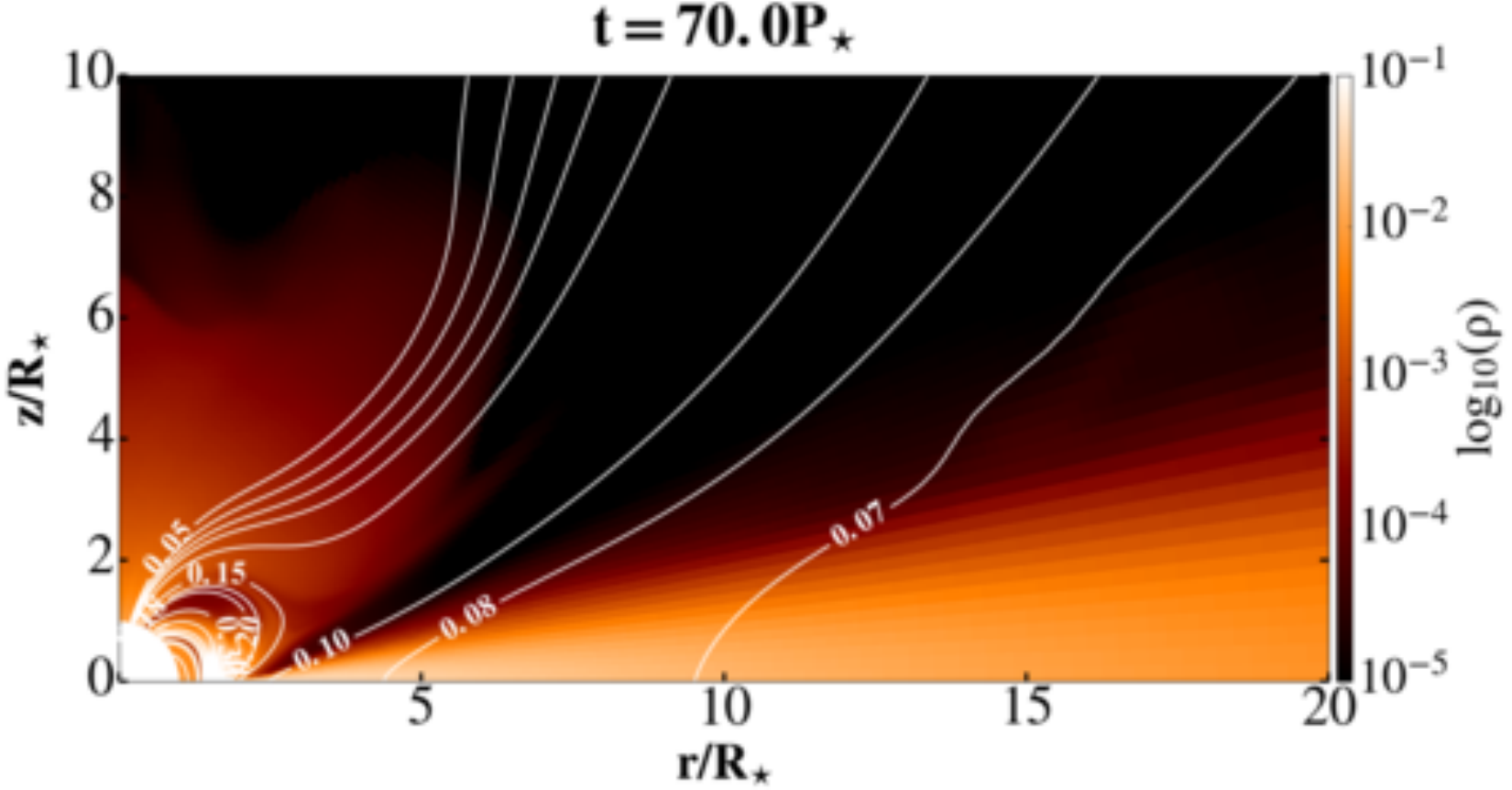}
\includegraphics[width=\columnwidth]{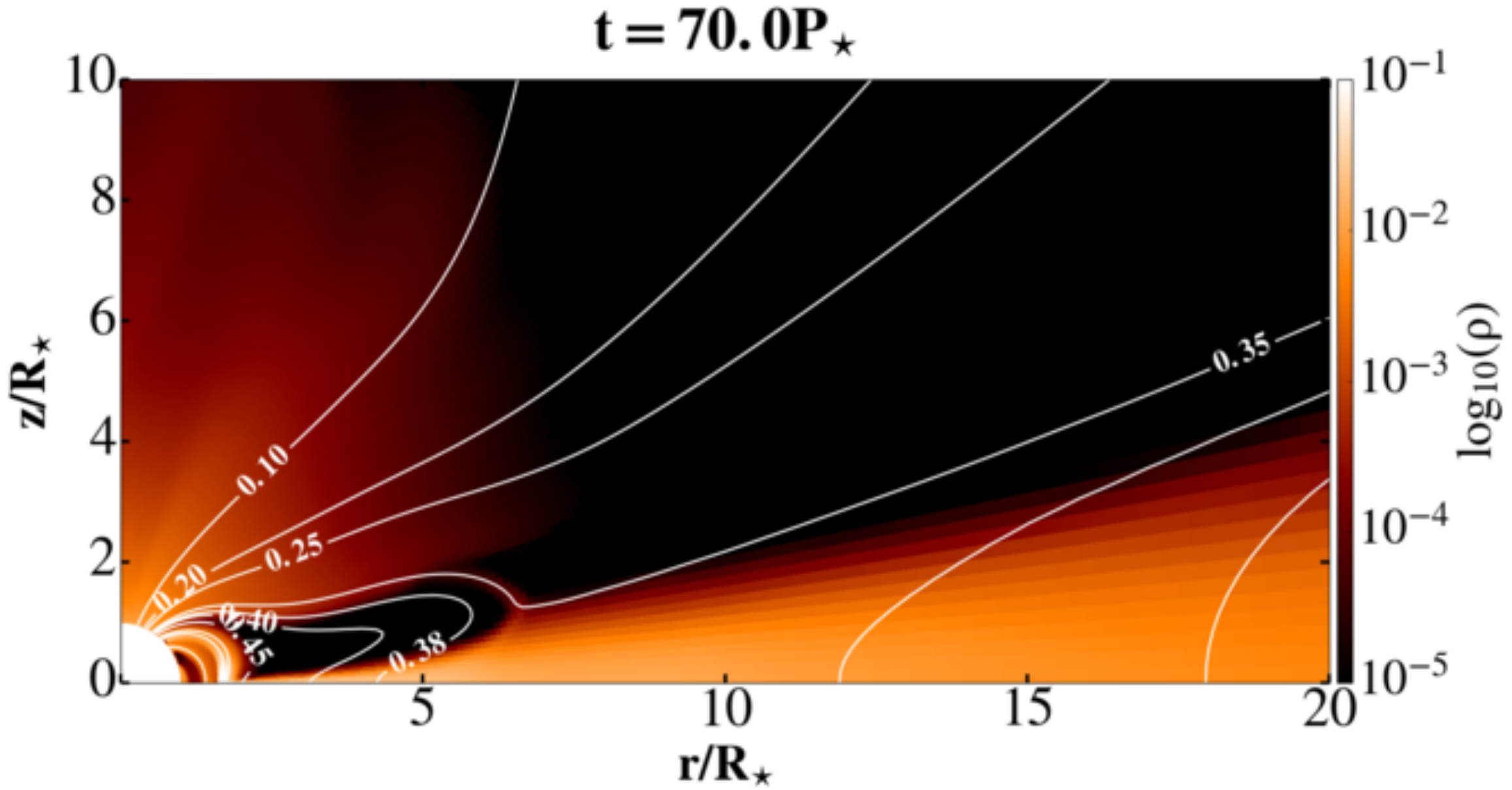}
\includegraphics[width=\columnwidth]{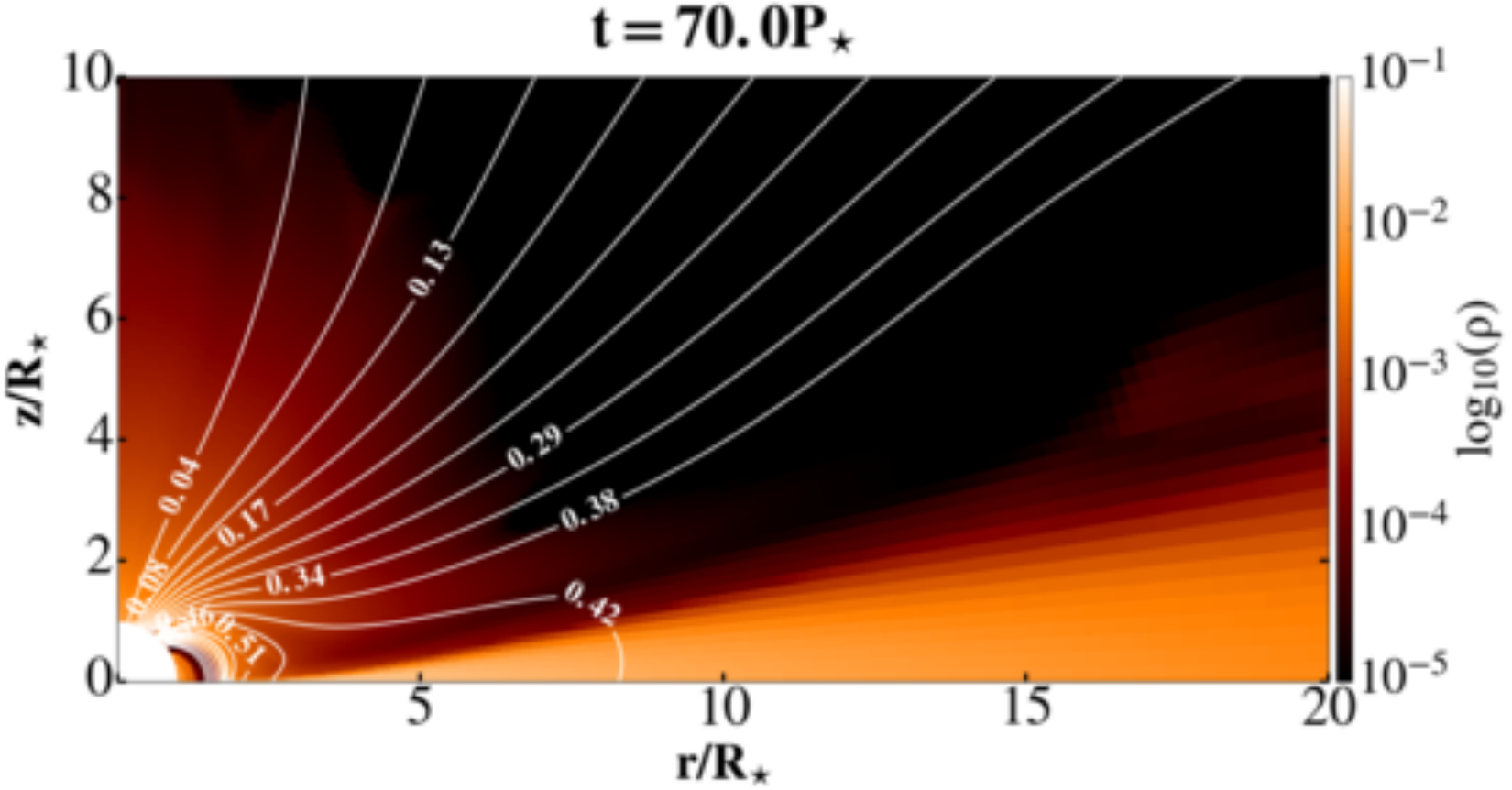}
\includegraphics[width=\columnwidth]{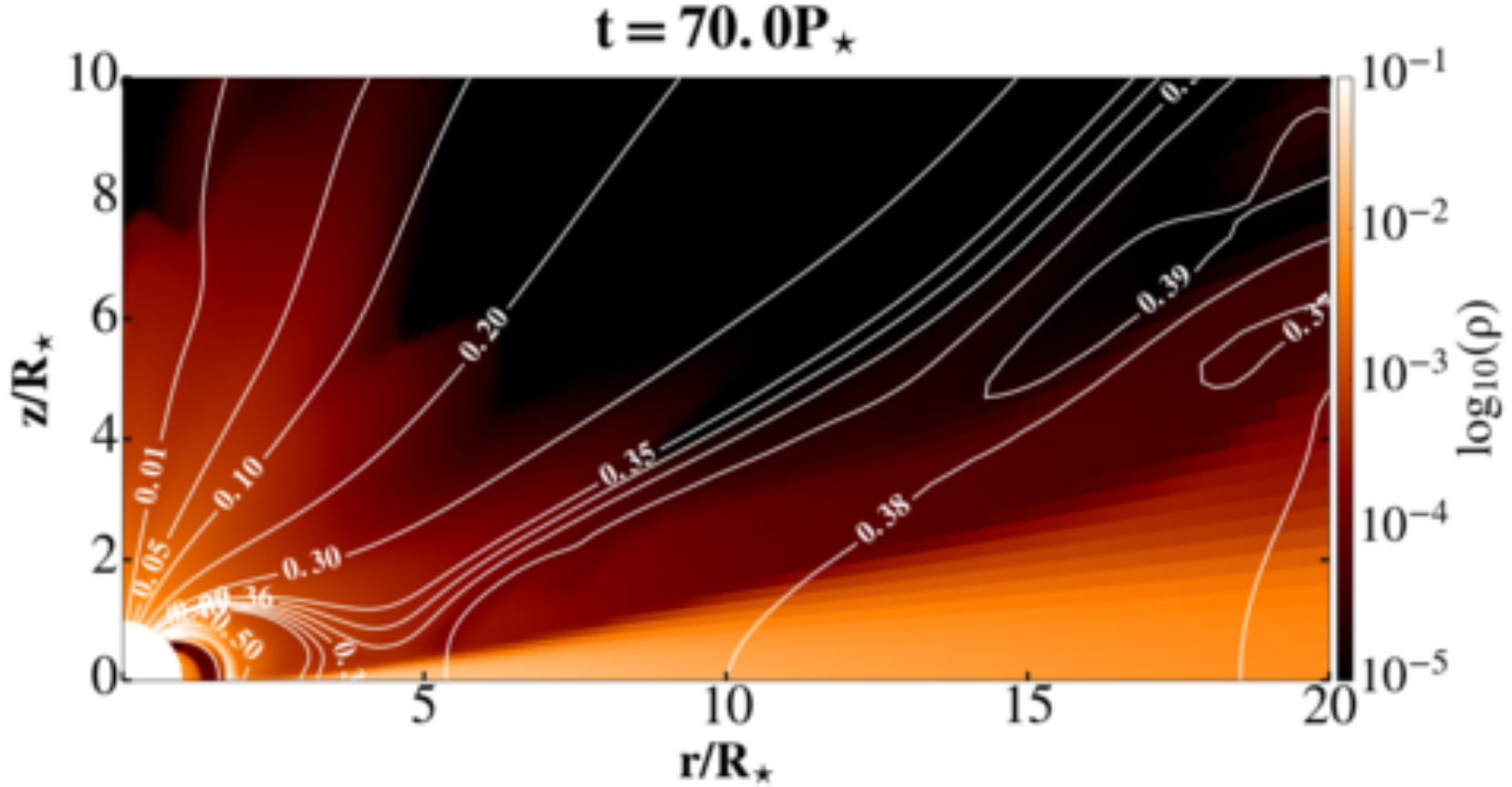} 
\caption{Matter density and poloidal magnetic field distribution in
the quasi-stationary interval in $\mu=0.35$ (0.25 kG) case with
$\Omega_\star$=0.05, with $\alpha_m=0.1$, 0.4, 0.7 and 1.0.}
\end{figure*}
\begin{figure*}
\includegraphics[width=\columnwidth]{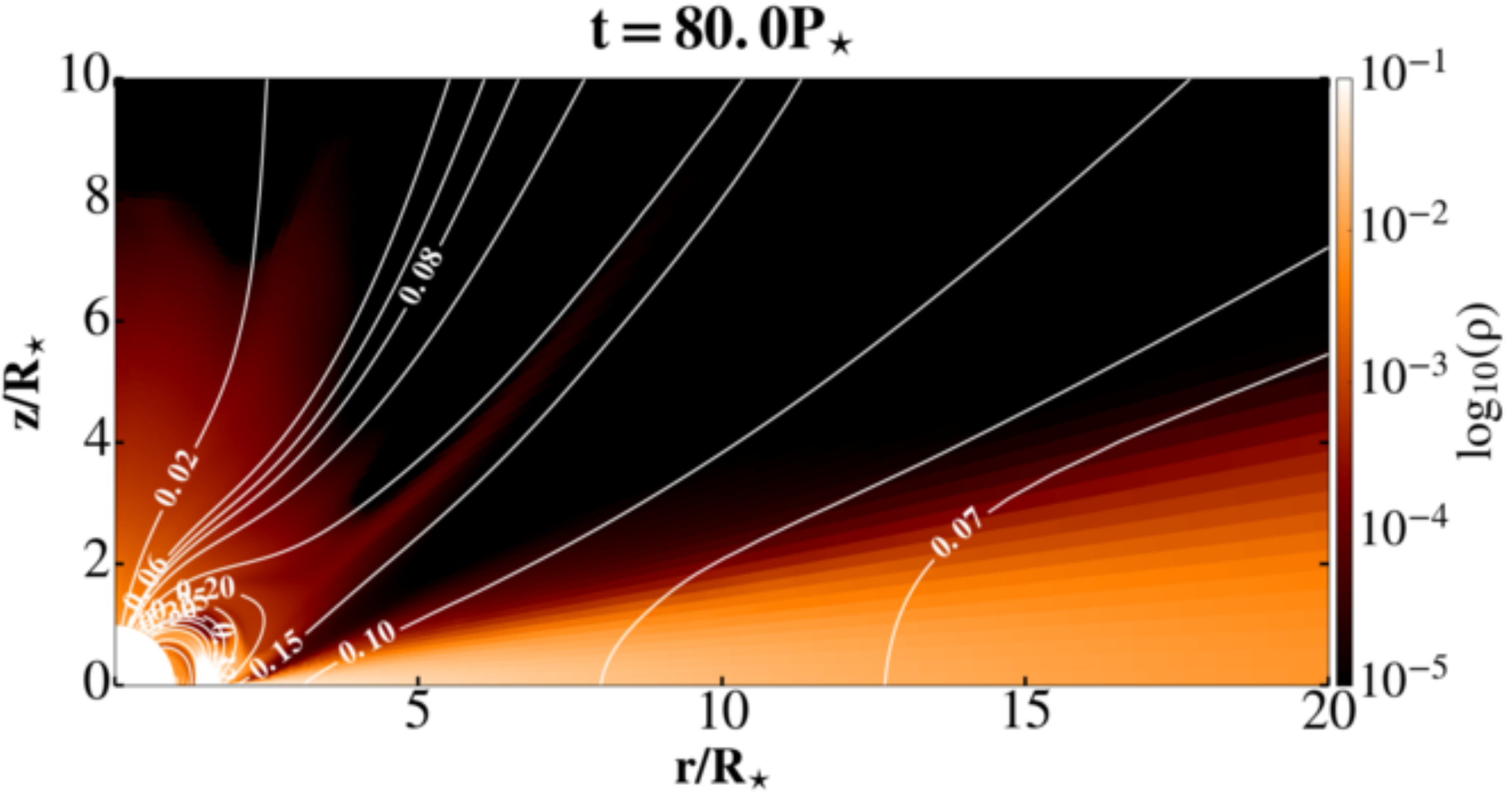}
\includegraphics[width=\columnwidth]{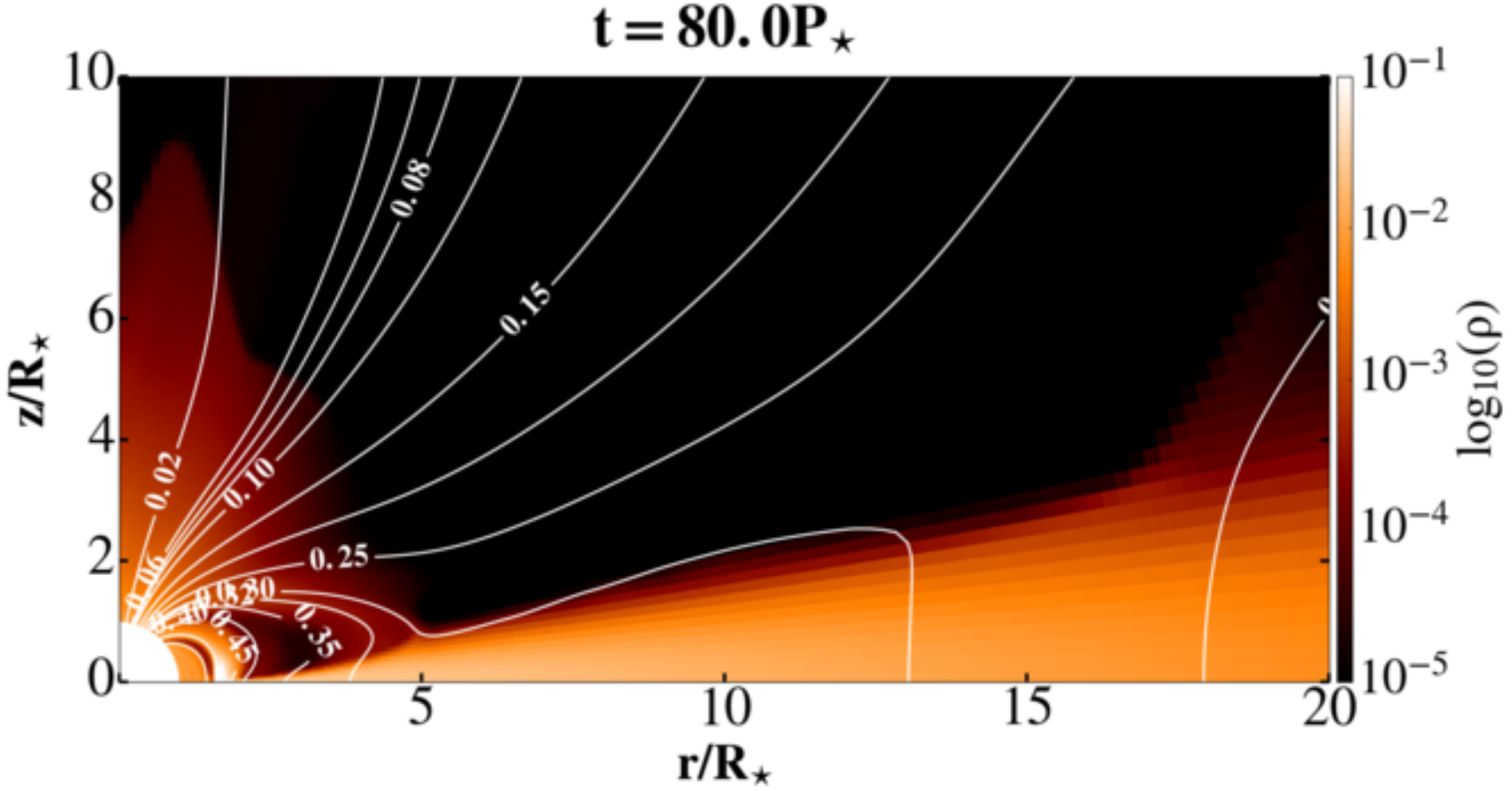}
\includegraphics[width=\columnwidth]{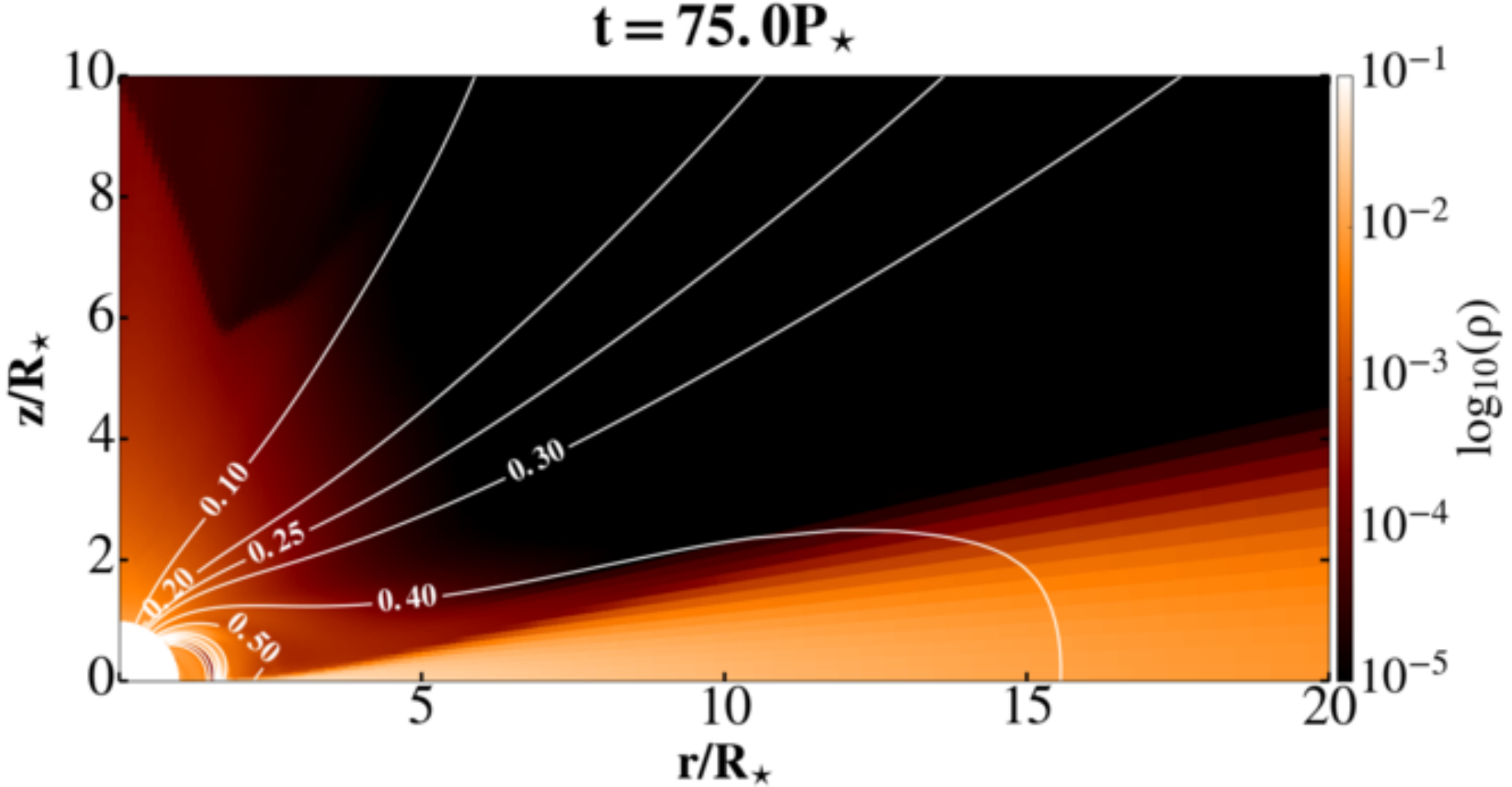}
\includegraphics[width=\columnwidth]{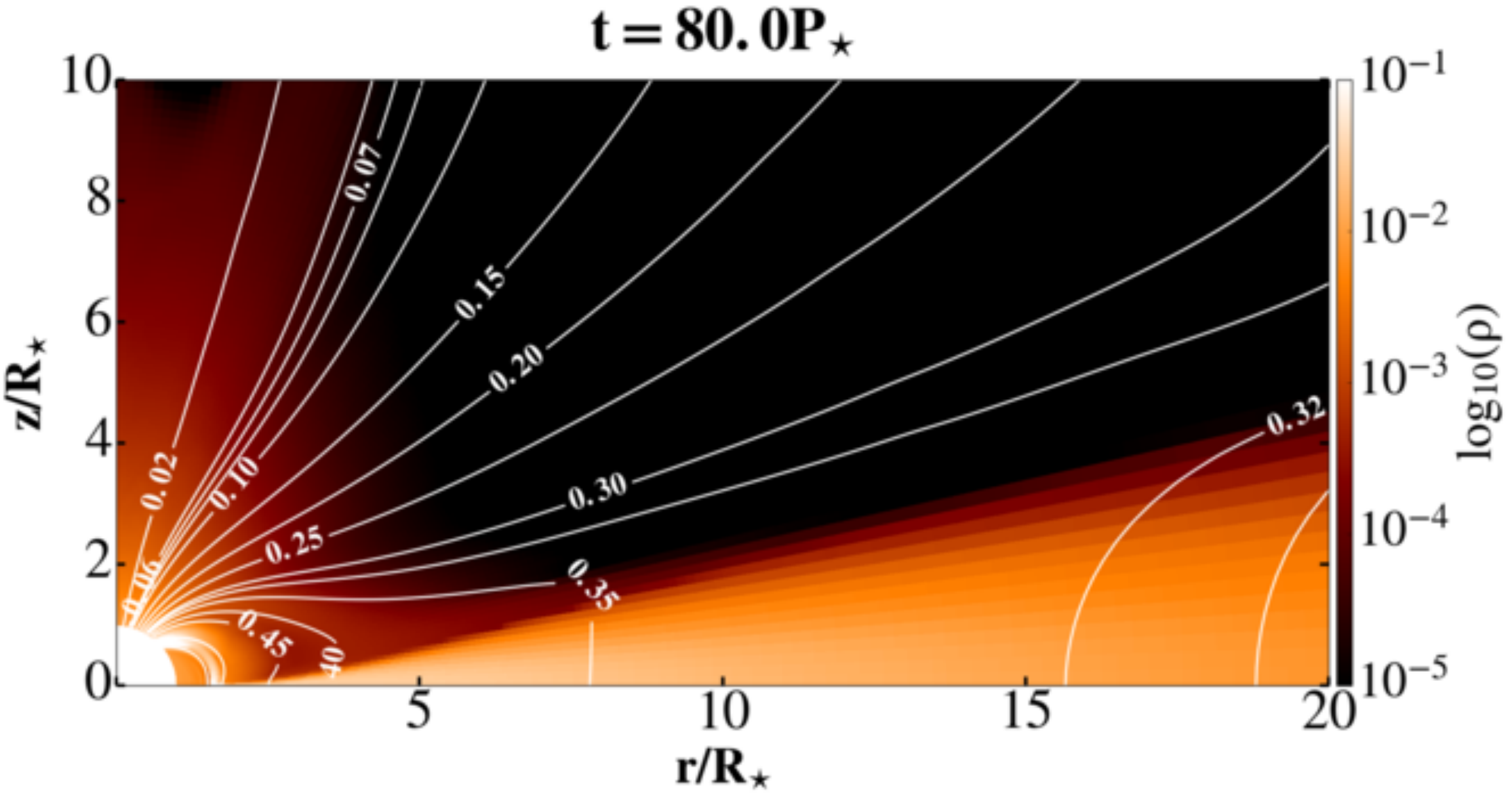} 
\caption{Matter density and poloidal magnetic field distribution in
the quasi-stationary interval in $\mu=0.35$ (0.25 kG) case with
$\Omega_\star$=0.1, with $\alpha_m=0.1$, 0.4, 0.7 and 1.0.}
\end{figure*}
\begin{figure*}
\includegraphics[width=\columnwidth]{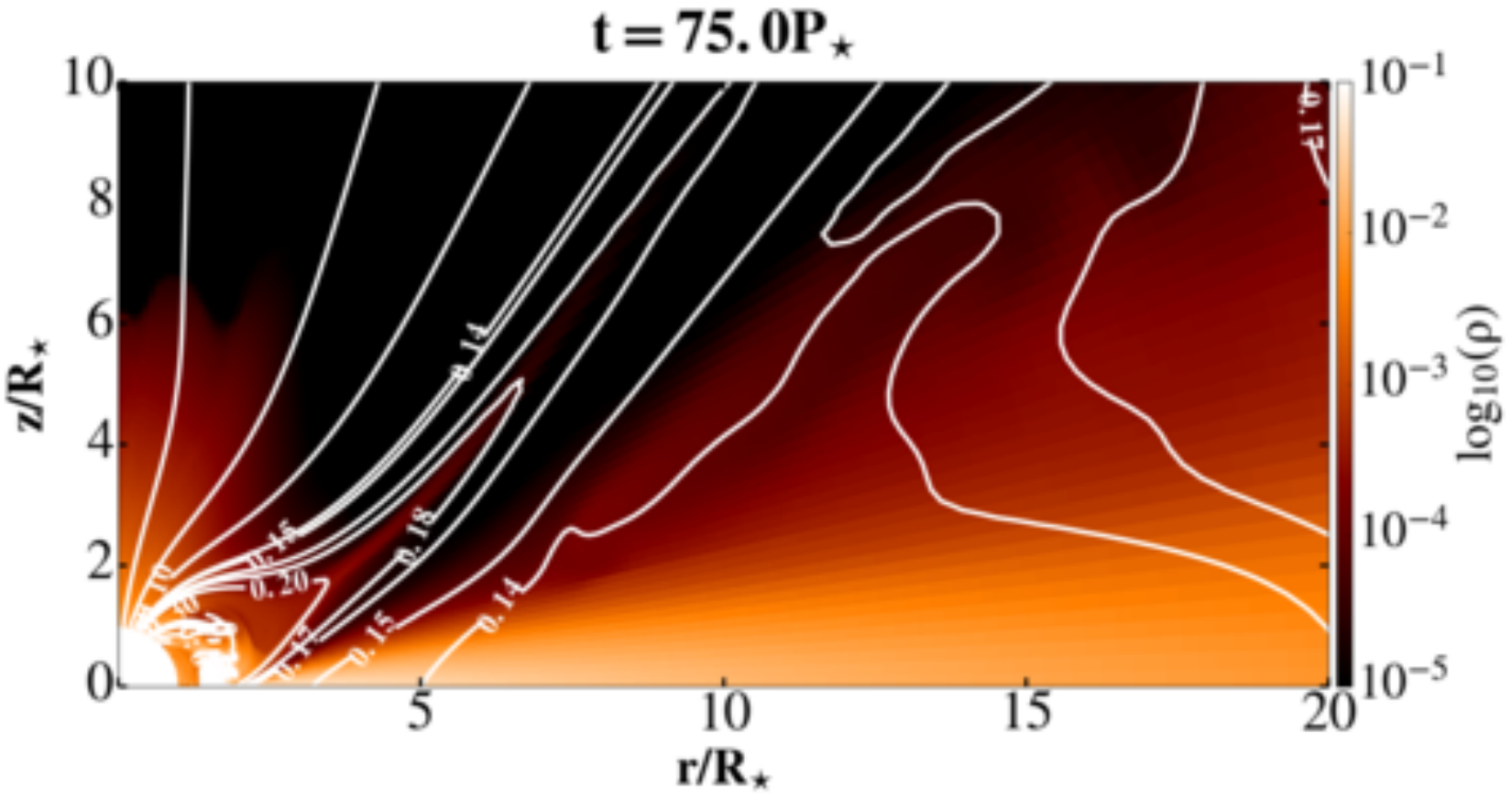}
\includegraphics[width=\columnwidth]{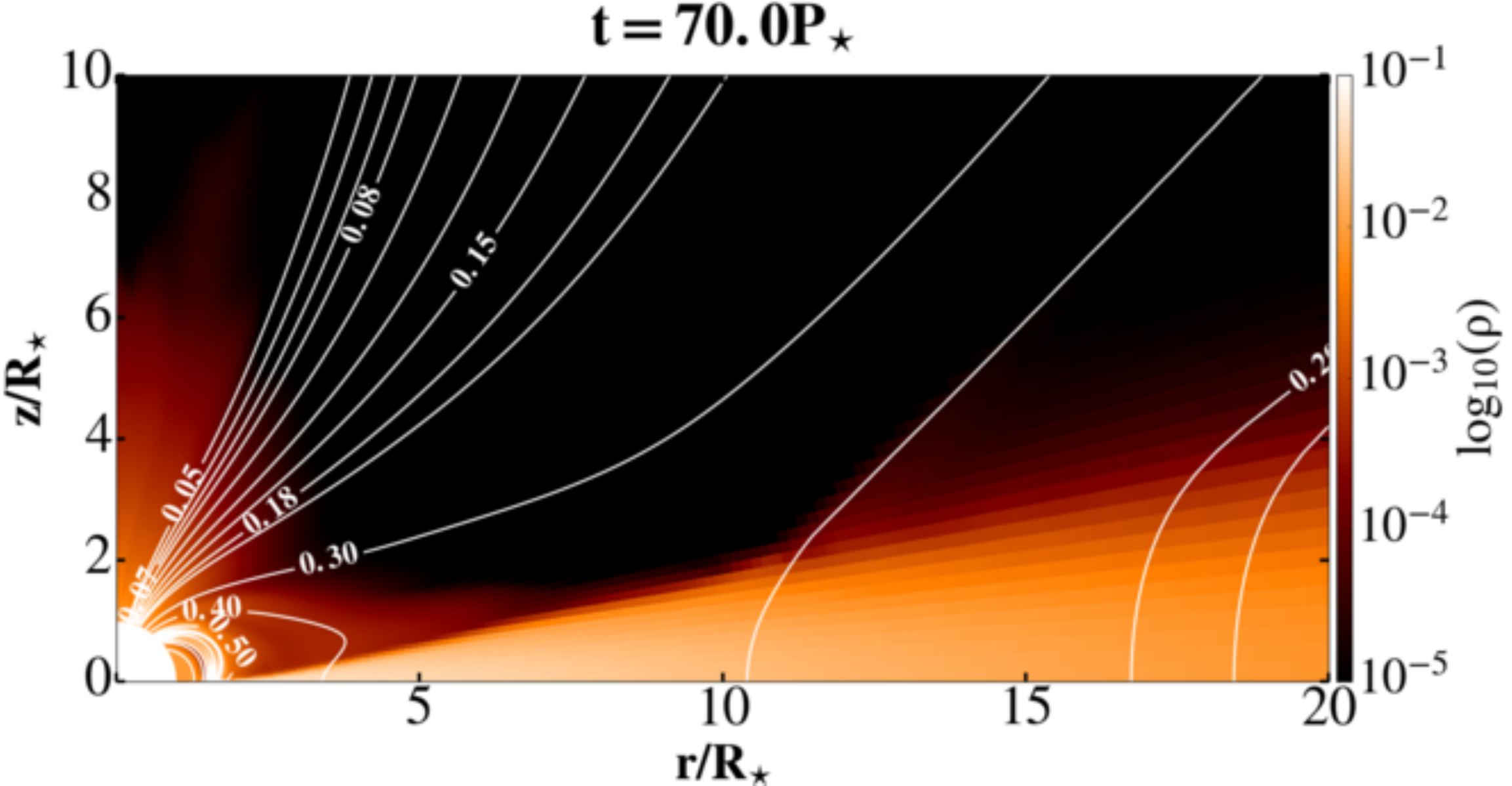}
\includegraphics[width=\columnwidth]{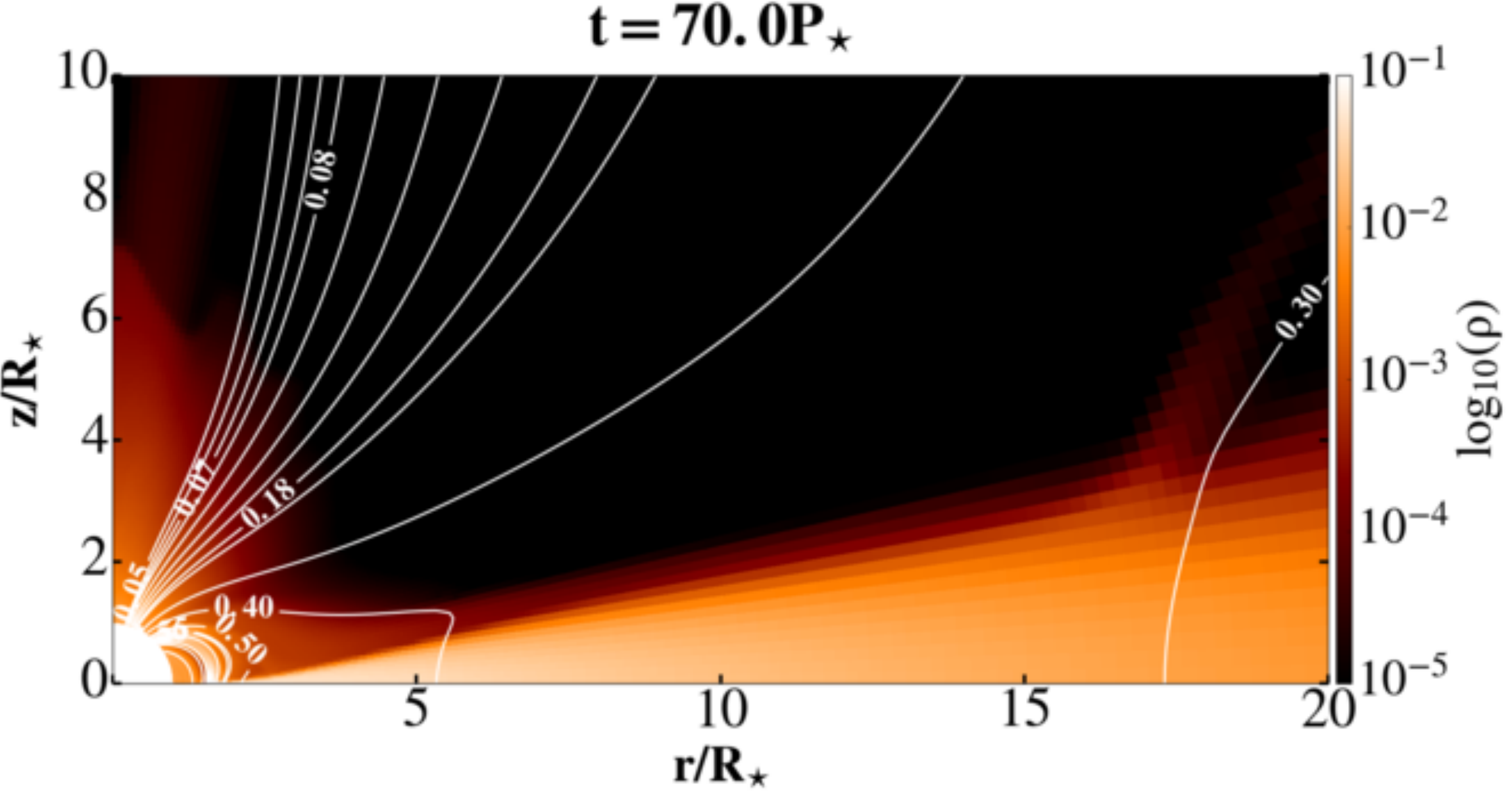}
\includegraphics[width=\columnwidth]{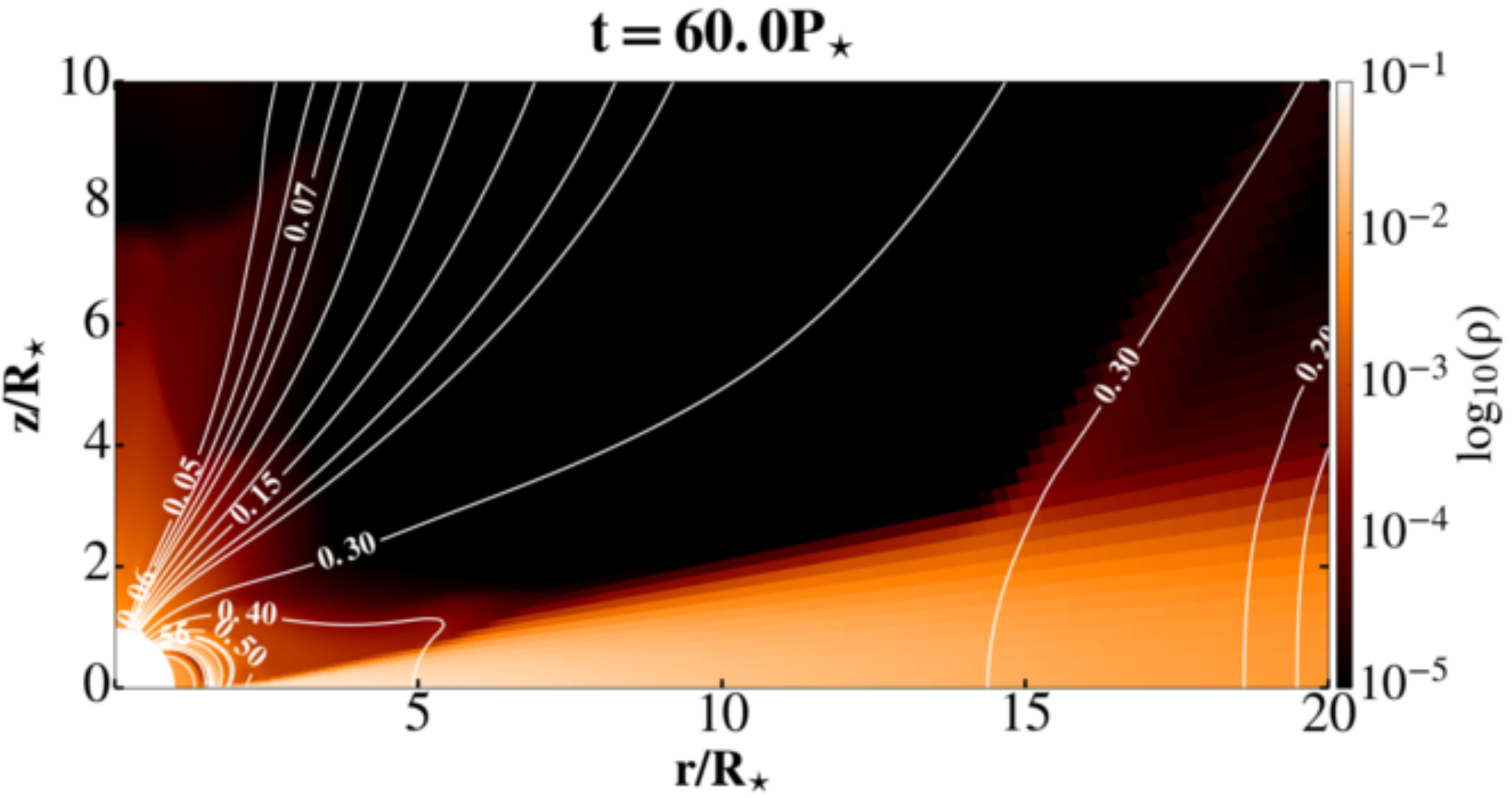} 
\caption{Matter density and poloidal magnetic field distribution in
the quasi-stationary interval in $\mu=0.35$ (0.25 kG) case with
$\Omega_\star$=0.15, with $\alpha_m=0.1$, 0.4, 0.7 and 1.0. }
\end{figure*}
\begin{figure*}
\includegraphics[width=\columnwidth]{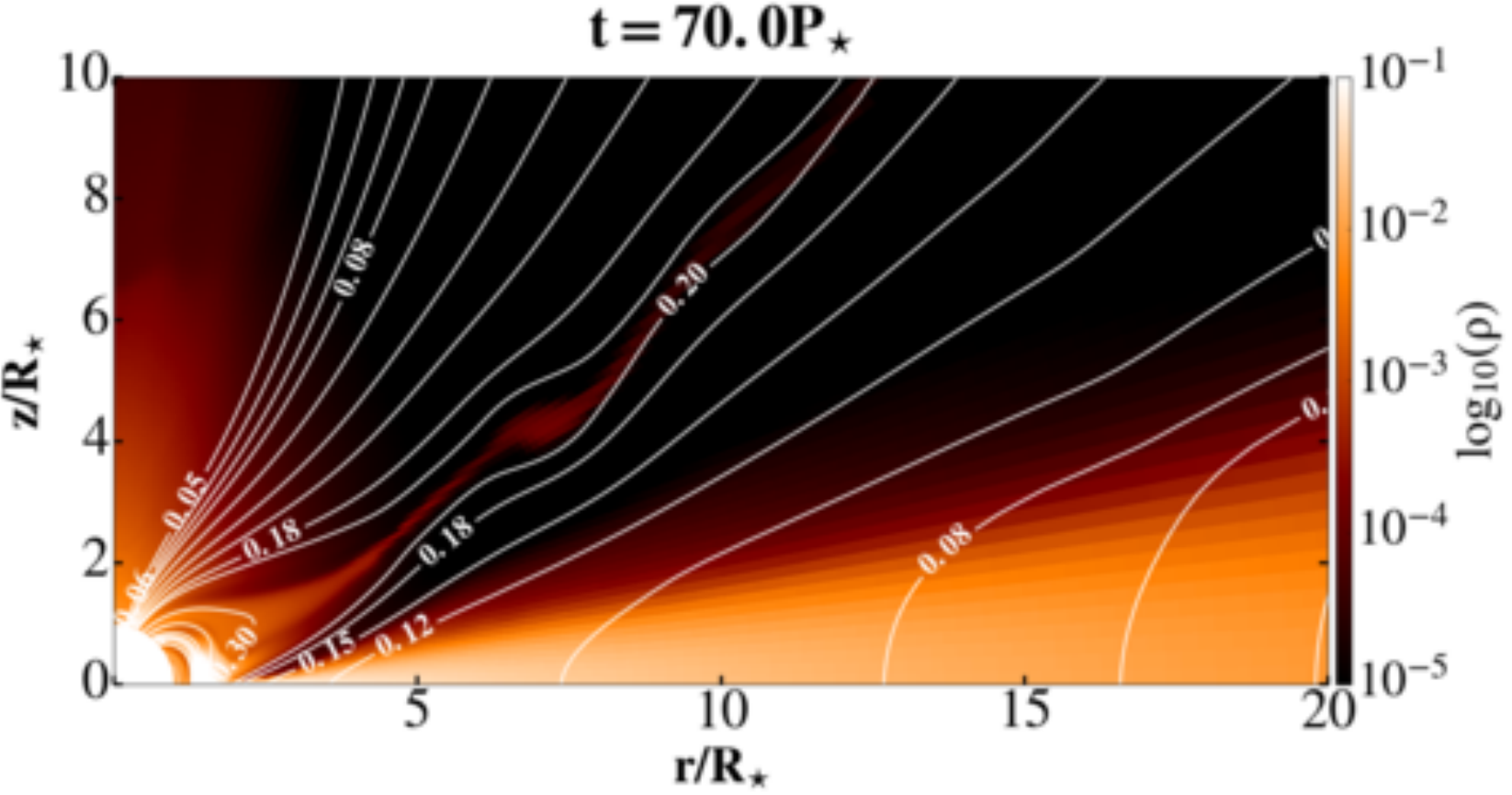}
\includegraphics[width=\columnwidth]{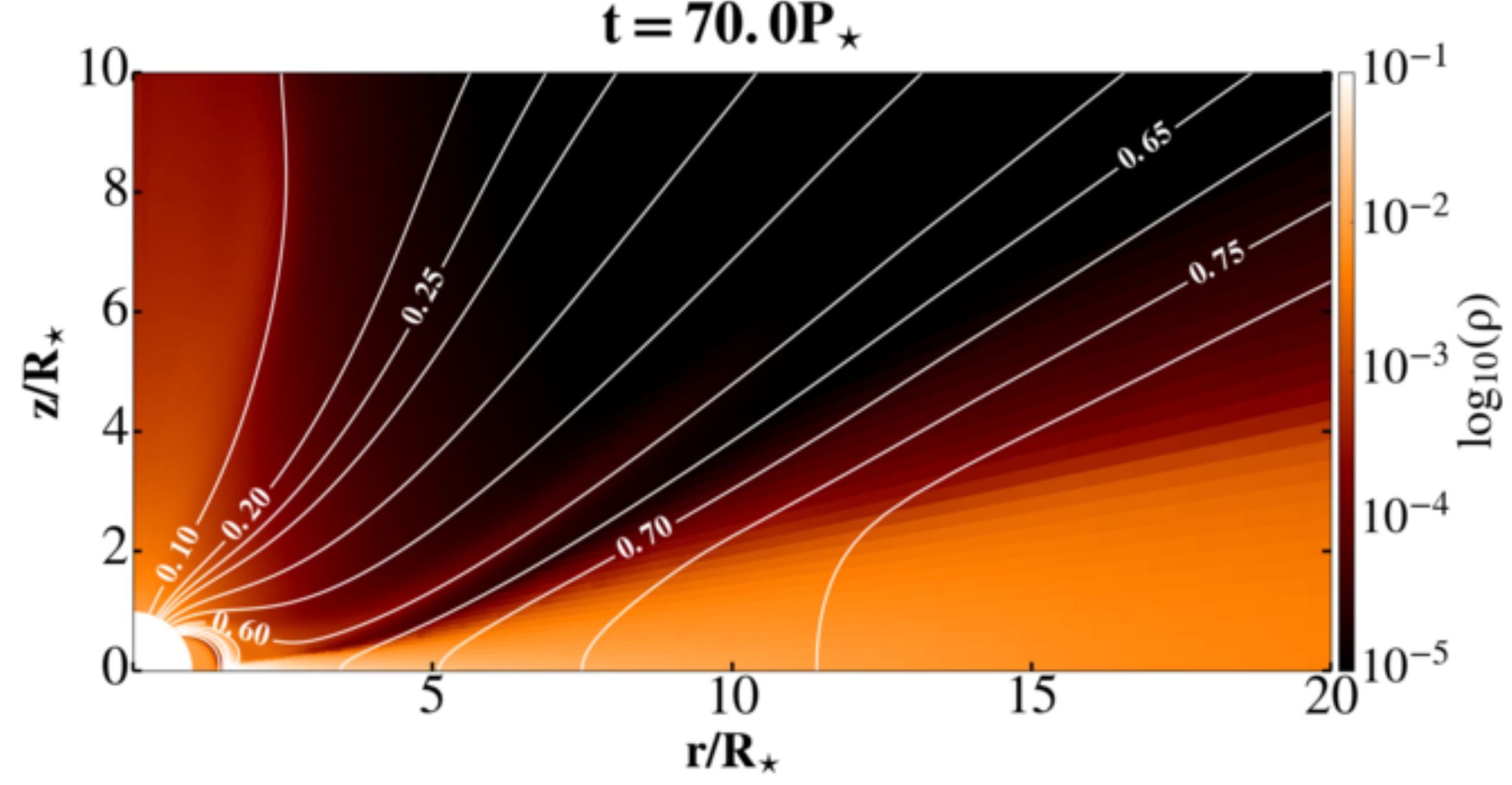}
\includegraphics[width=\columnwidth]{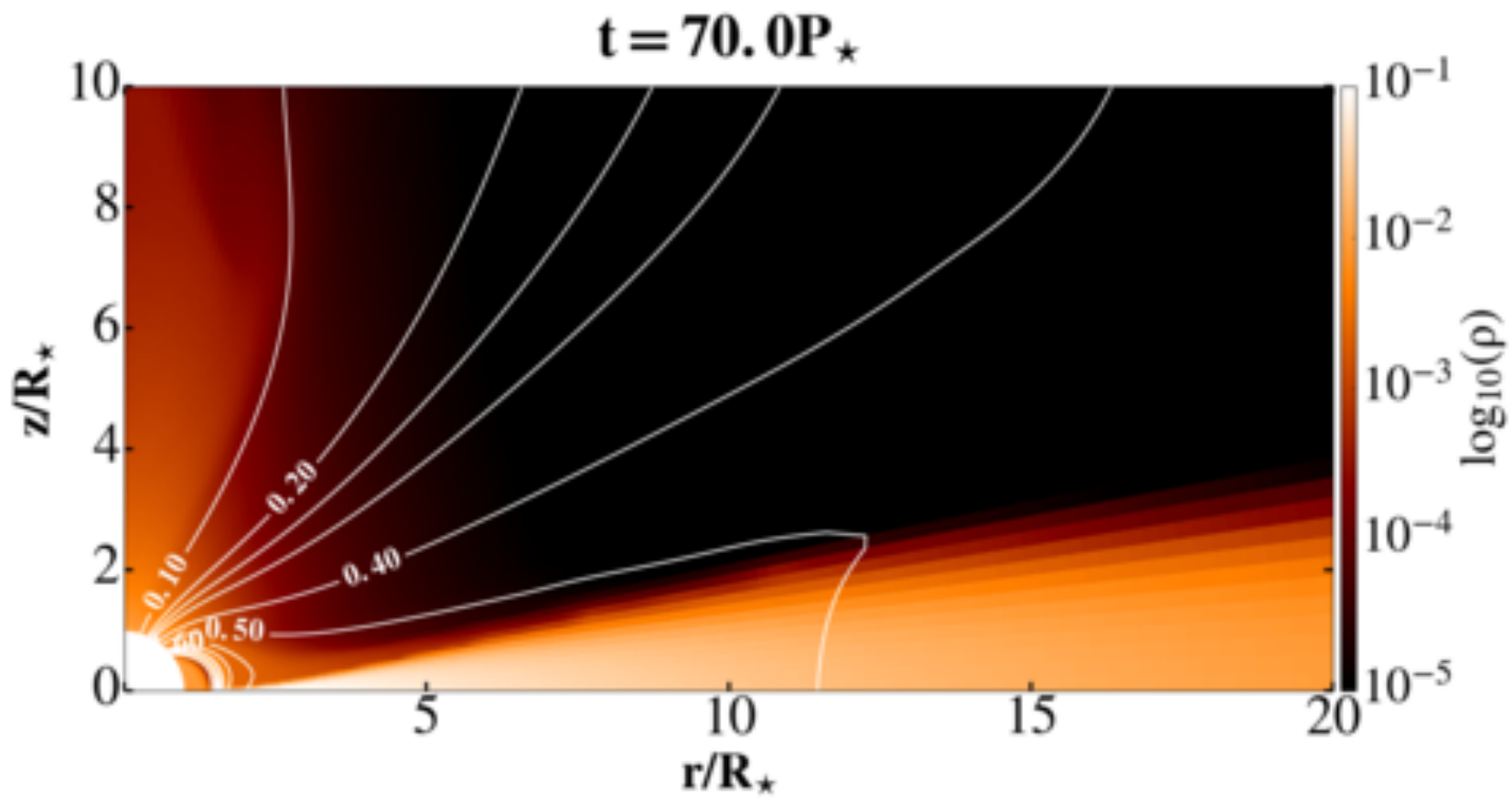}
\includegraphics[width=\columnwidth]{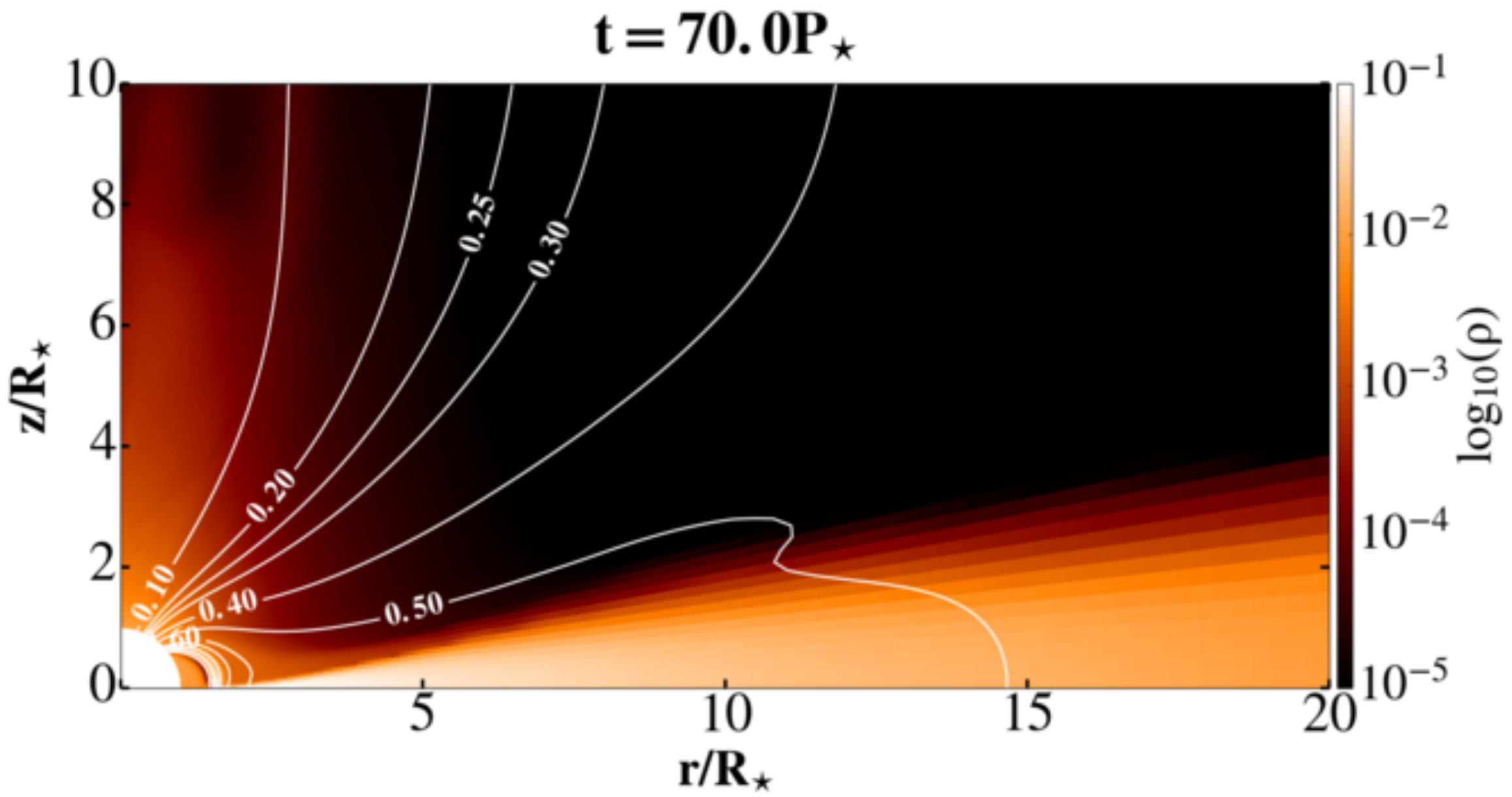} 
\caption{Matter density and poloidal magnetic field distribution in
the quasi-stationary interval in $\mu=0.35$ (0.25 kG) case with
$\Omega_\star$=0.2, with $\alpha_m=0.1$, 0.4, 0.7 and 1.0. }
\end{figure*}
\begin{figure*}
\includegraphics[width=\columnwidth]{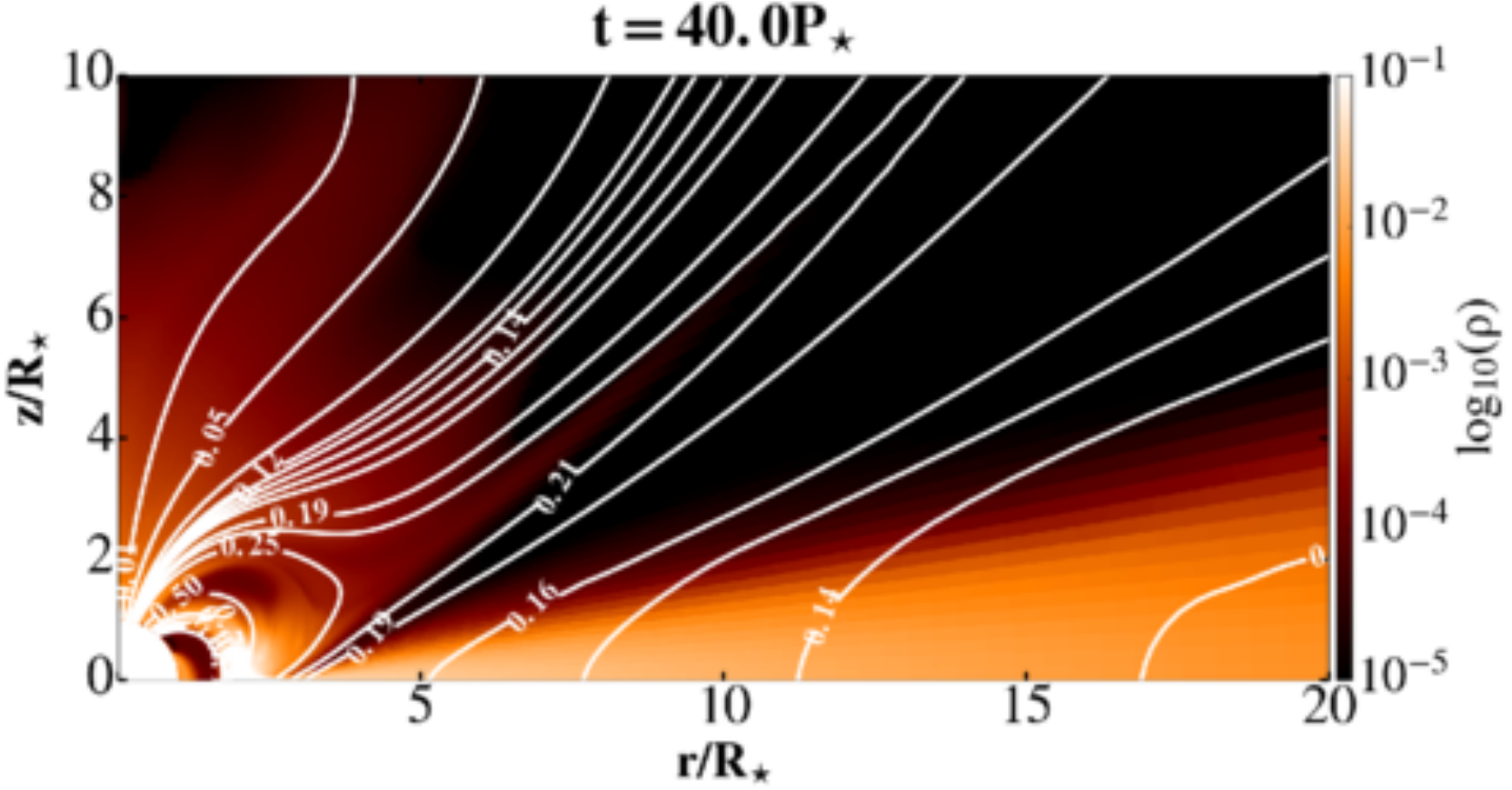}
\includegraphics[width=\columnwidth]{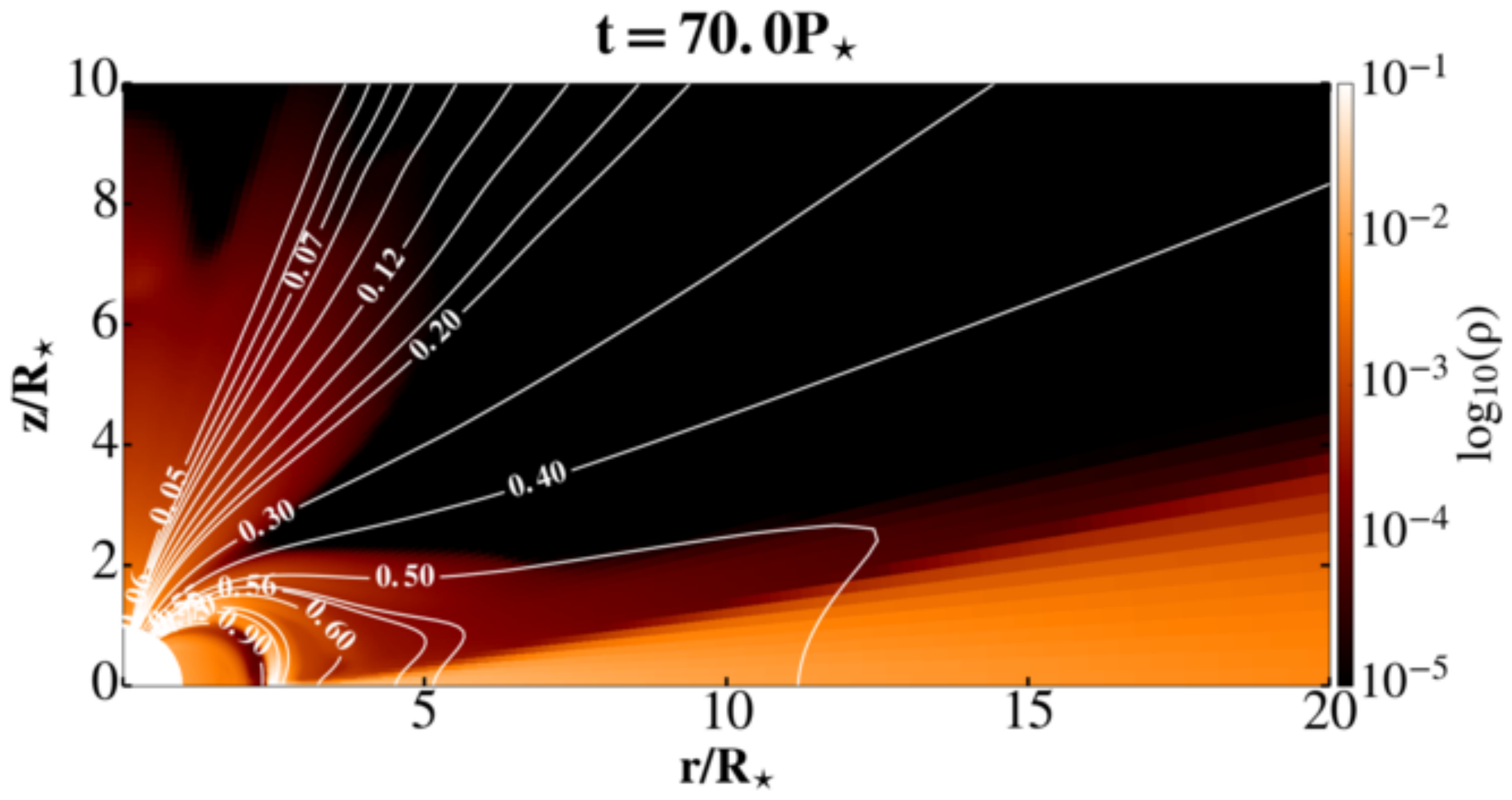}
\includegraphics[width=\columnwidth]{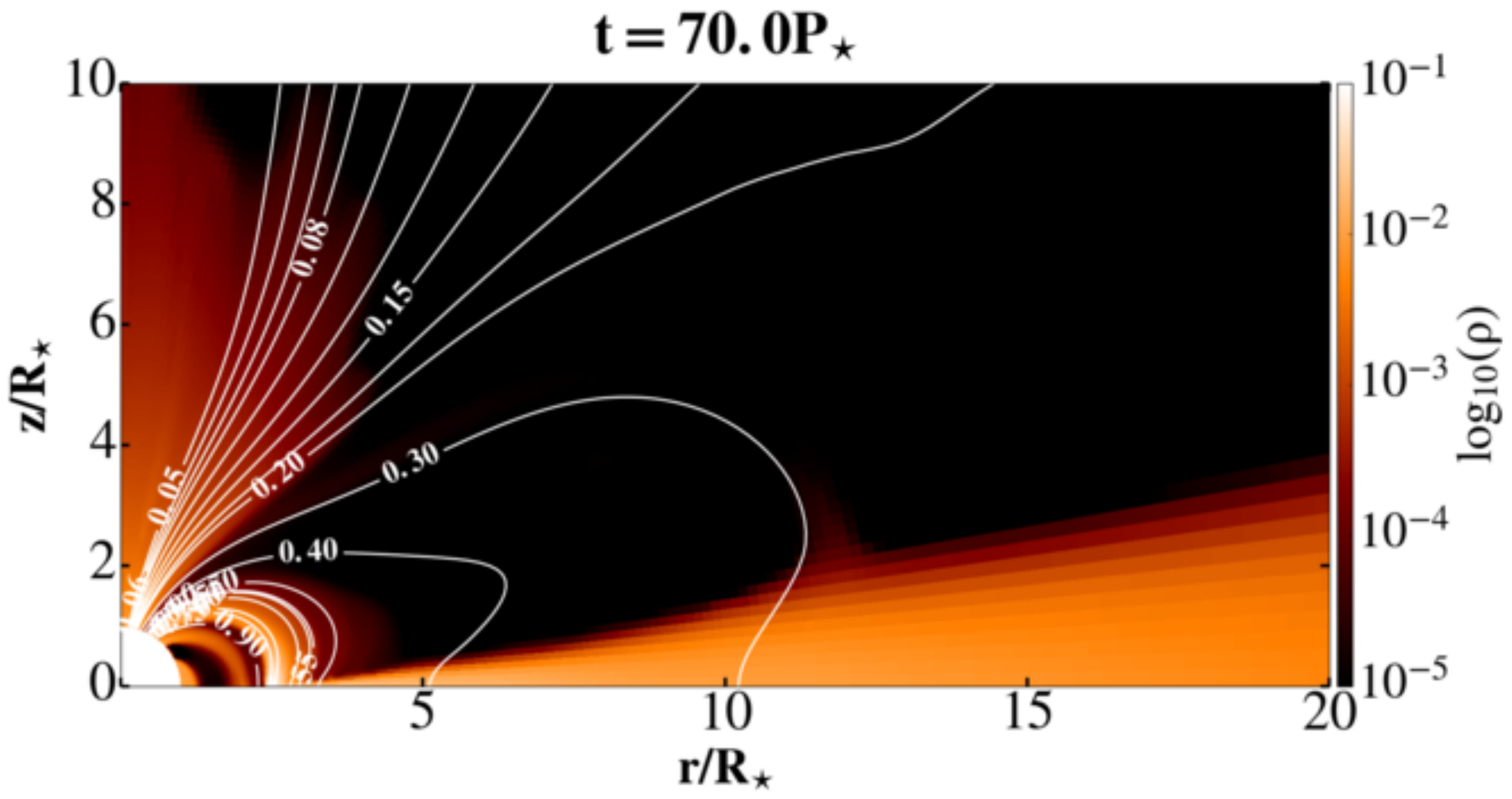}
\includegraphics[width=\columnwidth]{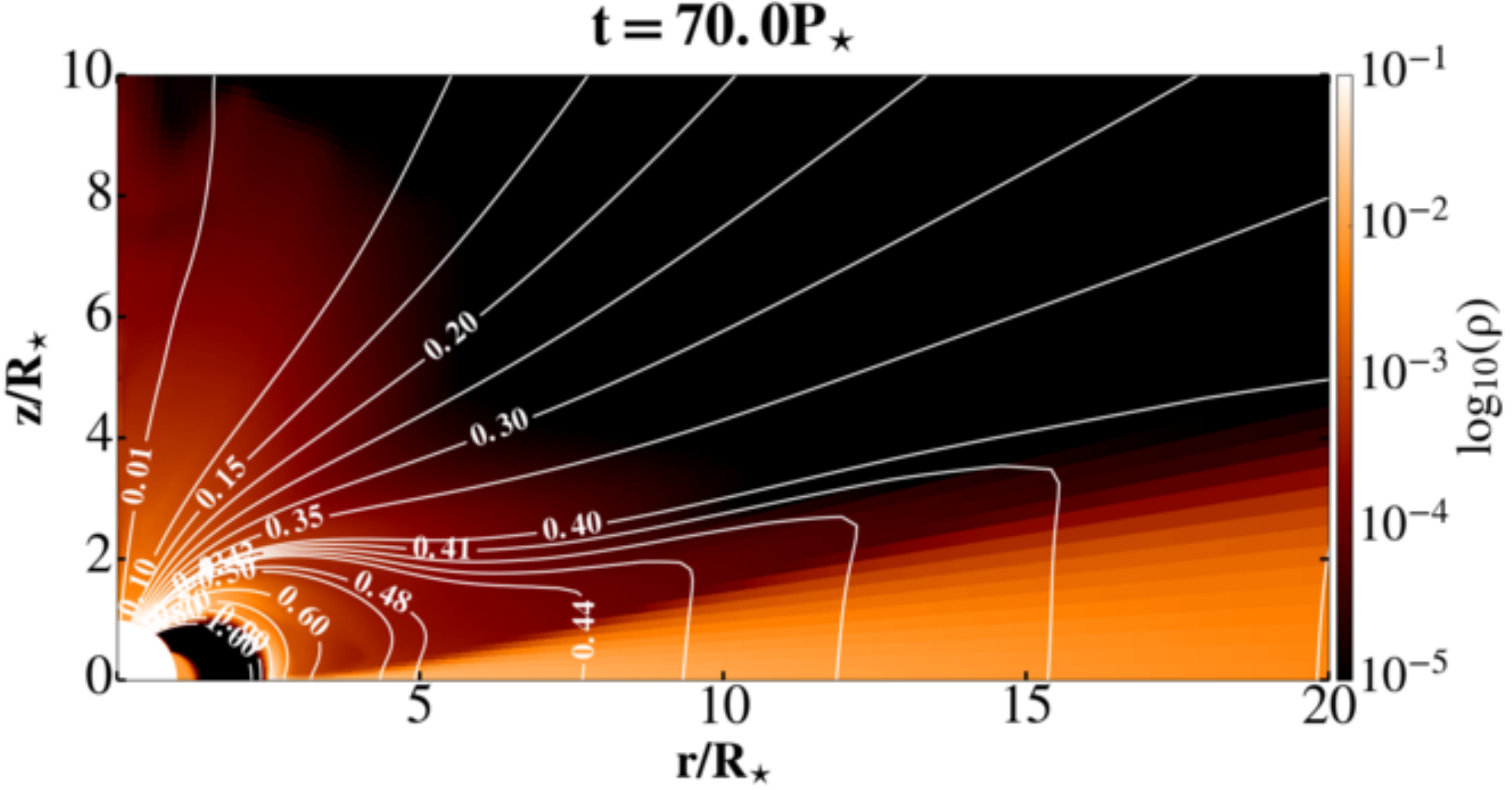} 
\caption{Matter density and poloidal magnetic field distribution in
the quasi-stationary interval in $\mu=0.7$ (0.5 kG) case with
$\Omega_\star$=0.05, with $\alpha_m=0.1$, 0.4, 0.7 and 1.0. }
\end{figure*}
\begin{figure*}
\includegraphics[width=\columnwidth]{{plotmu.7Av1Am.1om.1}.pdf}
\includegraphics[width=\columnwidth]{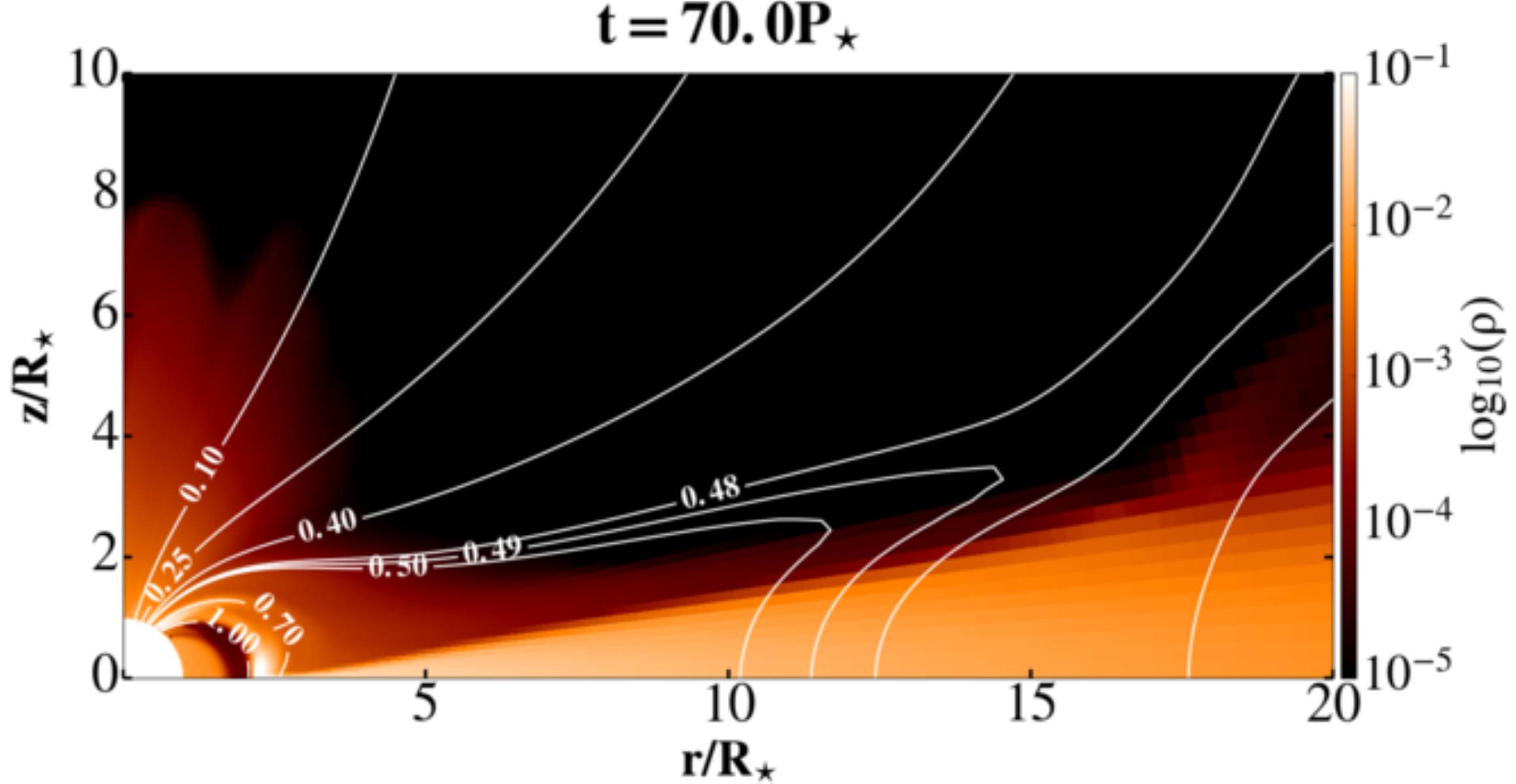}
\includegraphics[width=\columnwidth]{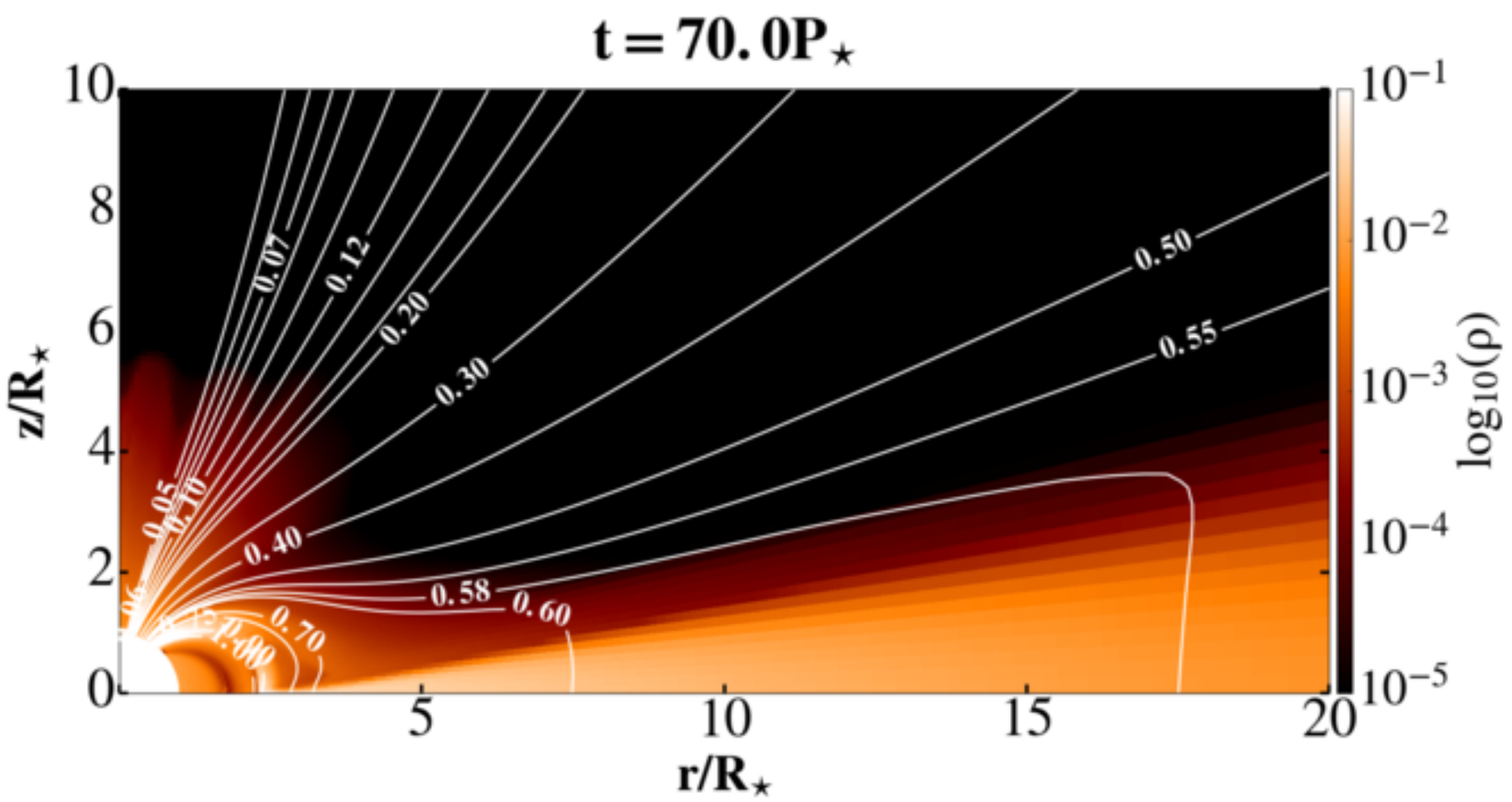}
\includegraphics[width=\columnwidth]{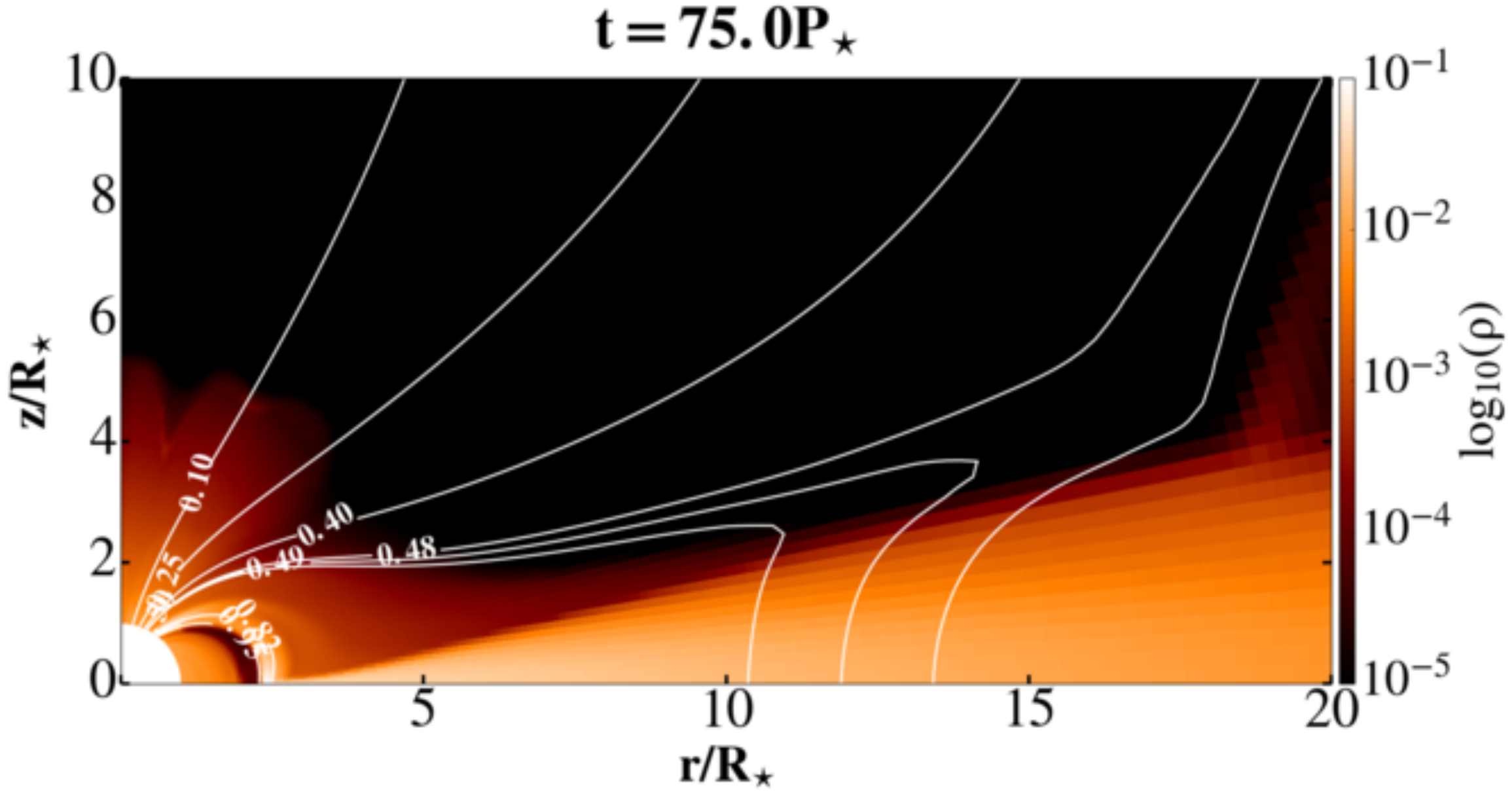} 
\caption{Matter density and poloidal magnetic field distribution in
the quasi-stationary interval in $\mu=0.7$ (0.5 kG) case with
$\Omega_\star$=0.1, with $\alpha_m=0.1$, 0.4, 0.7 and 1.0. }
\end{figure*}
\begin{figure*}
\includegraphics[width=\columnwidth]{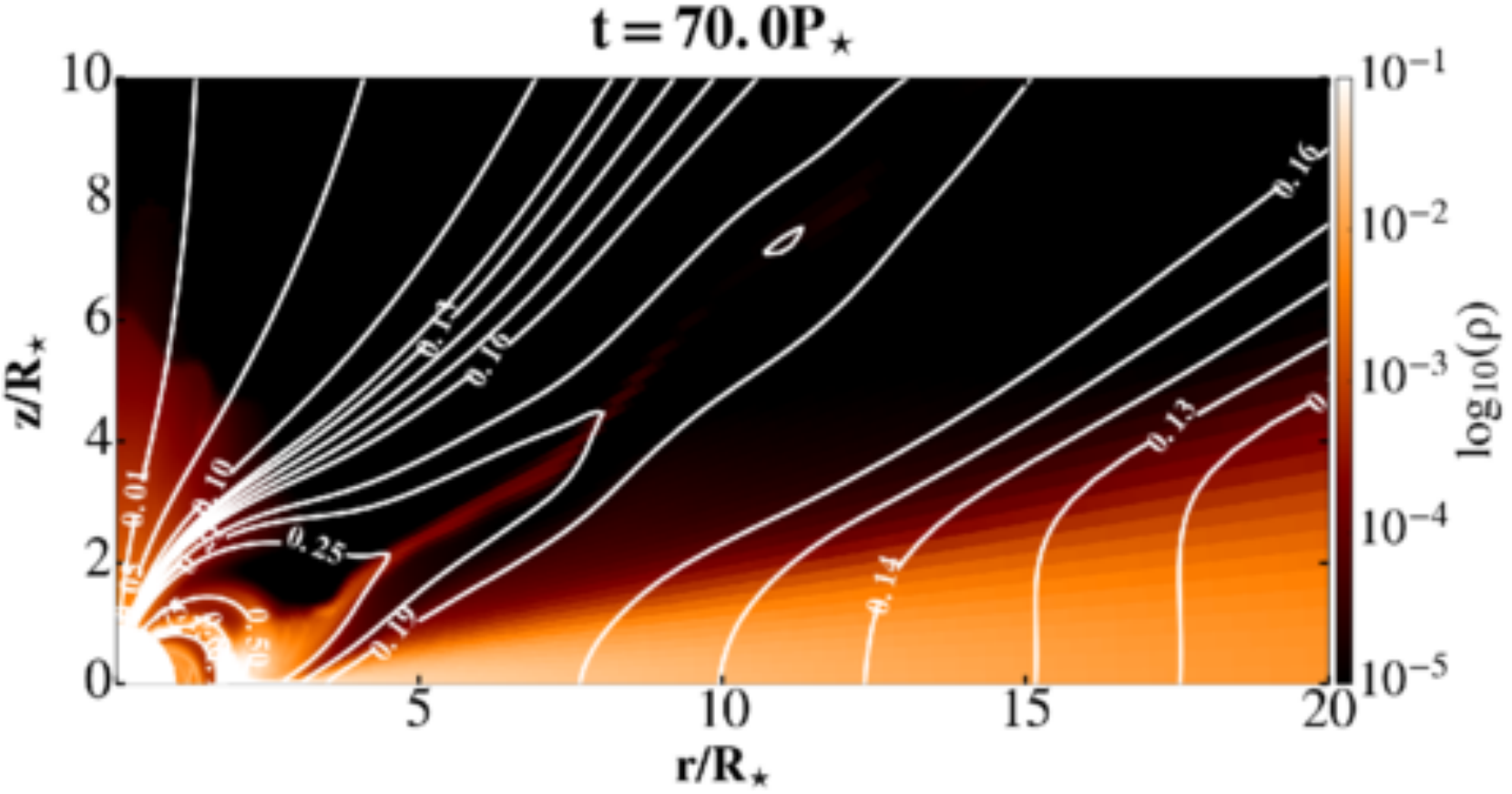}
\includegraphics[width=\columnwidth]{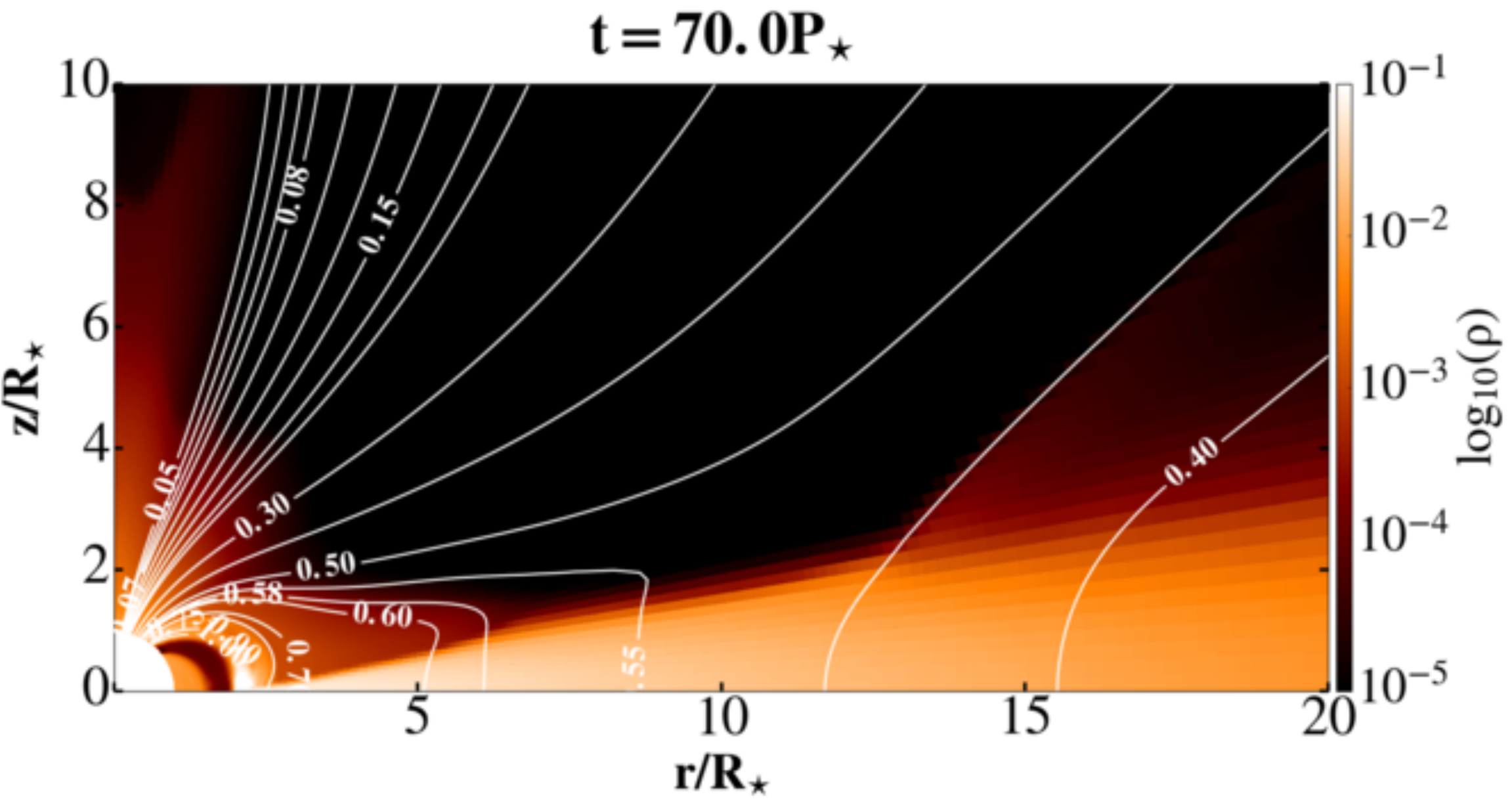}
\includegraphics[width=\columnwidth]{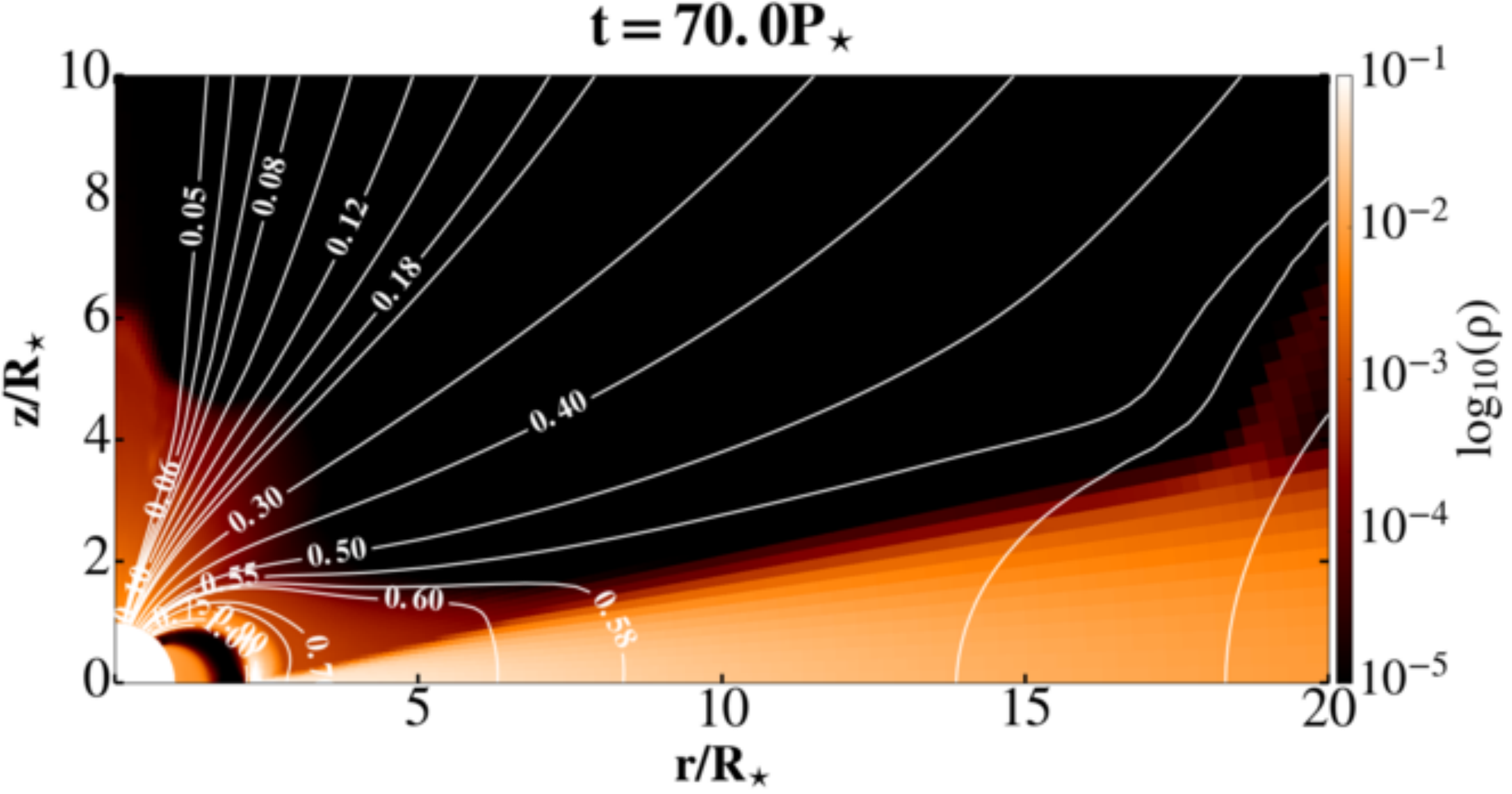}
\includegraphics[width=\columnwidth]{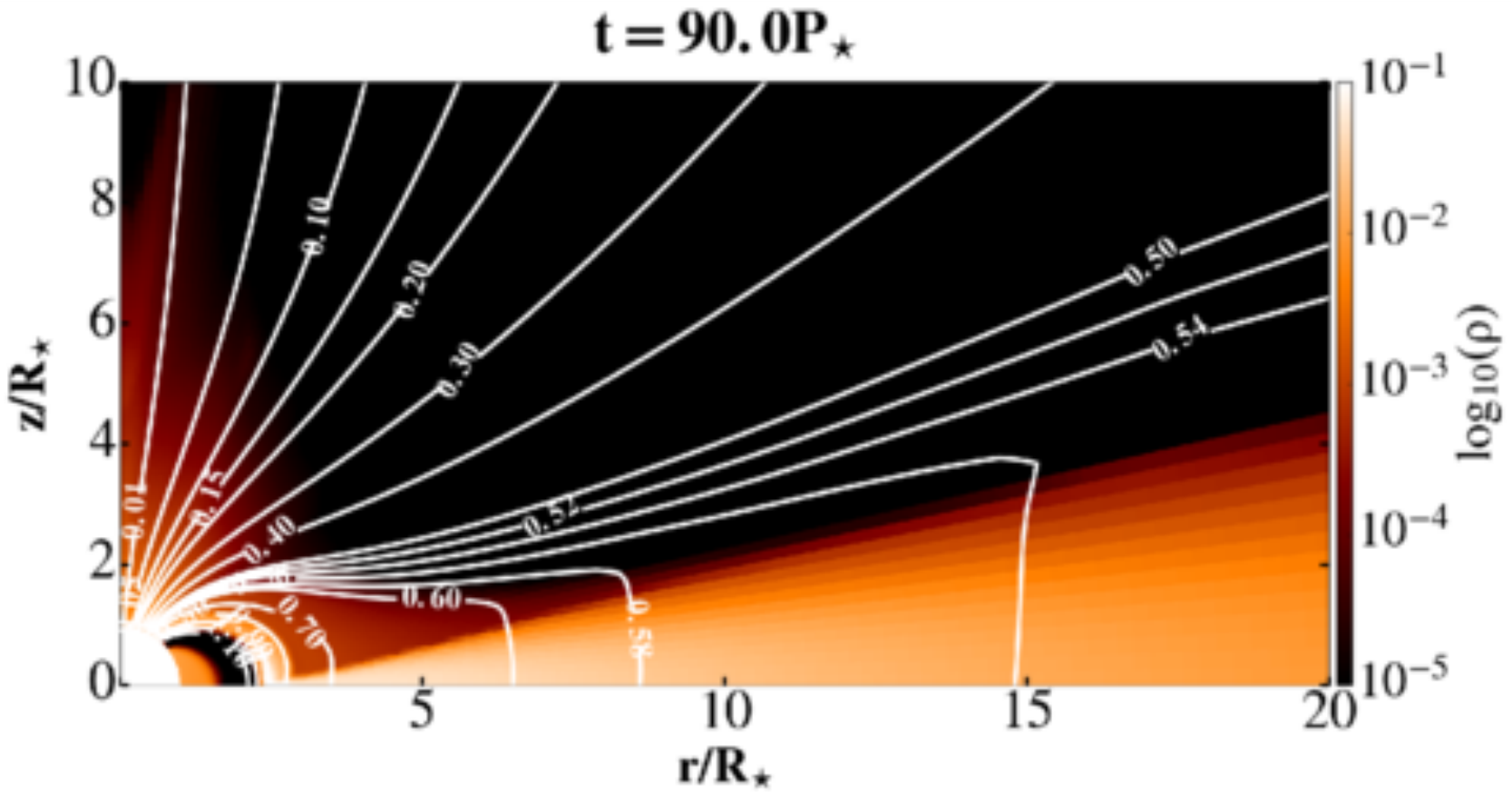} 
\caption{Matter density and poloidal magnetic field distribution in
the quasi-stationary interval in $\mu=0.7$ (0.5 kG) case with
$\Omega_\star$=0.15, with $\alpha_m=0.1$, 0.4, 0.7 and 1.0. }
\end{figure*}
\begin{figure*}
\includegraphics[width=\columnwidth]{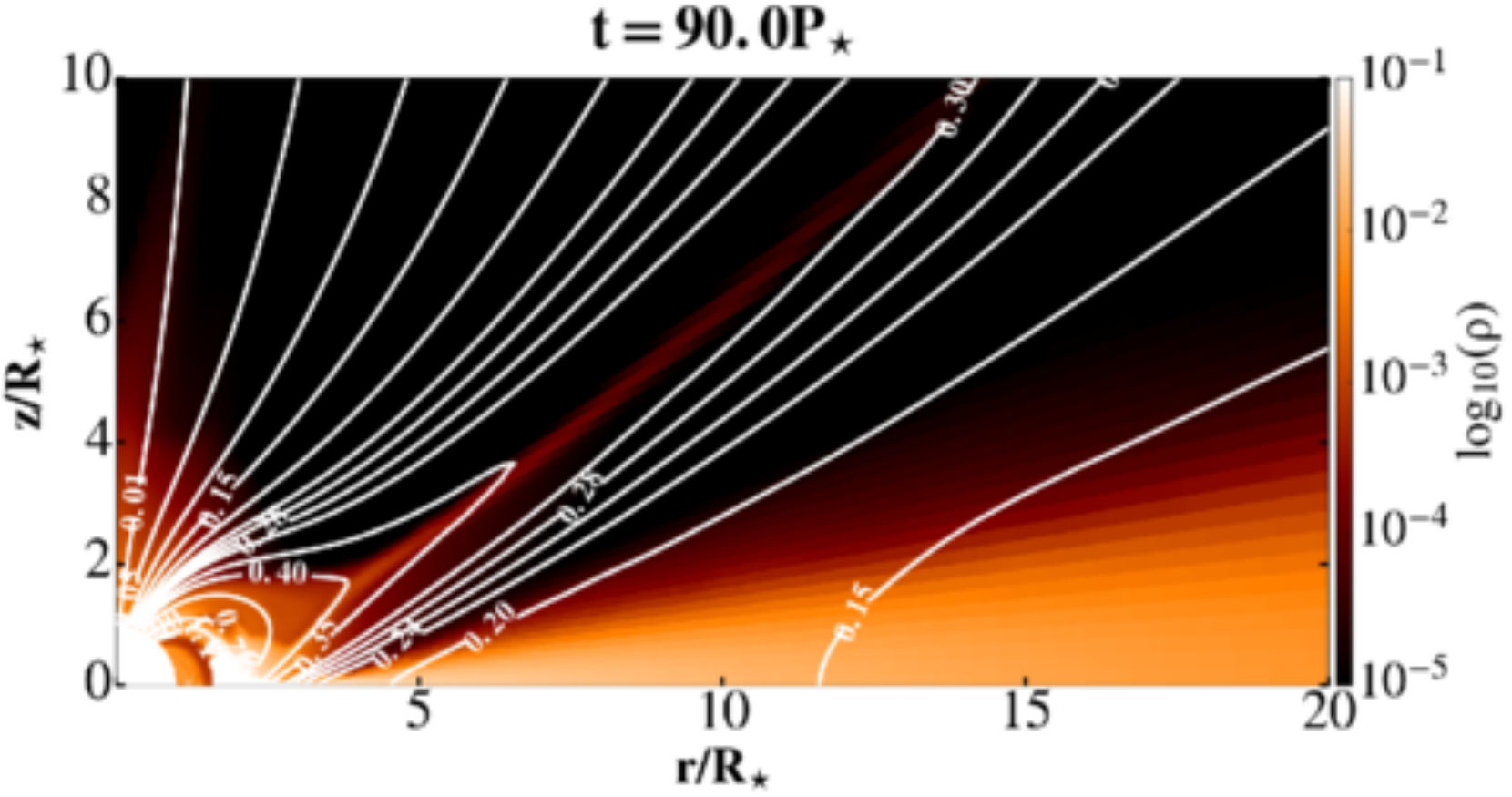}
\includegraphics[width=\columnwidth]{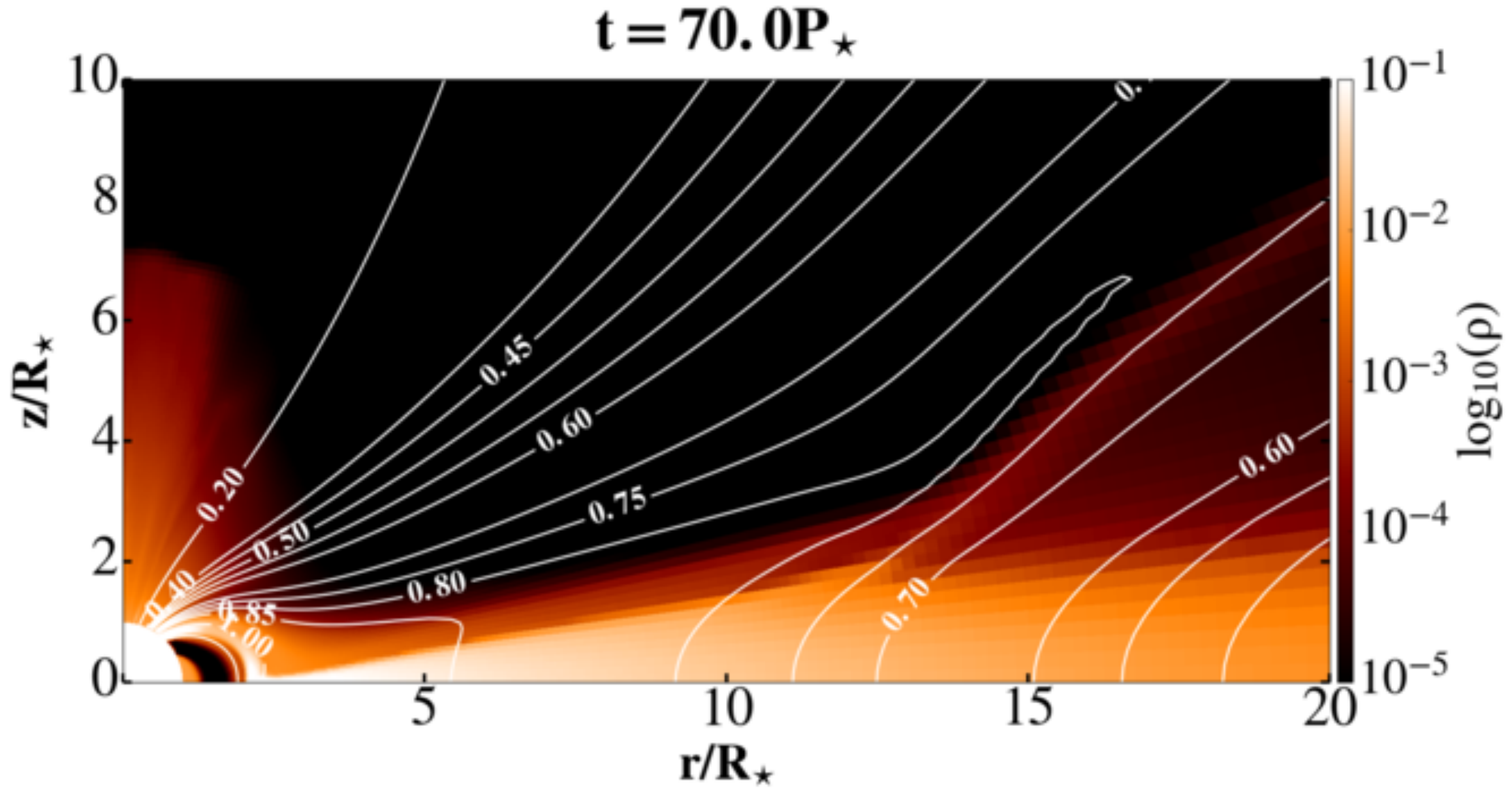}
\includegraphics[width=\columnwidth]{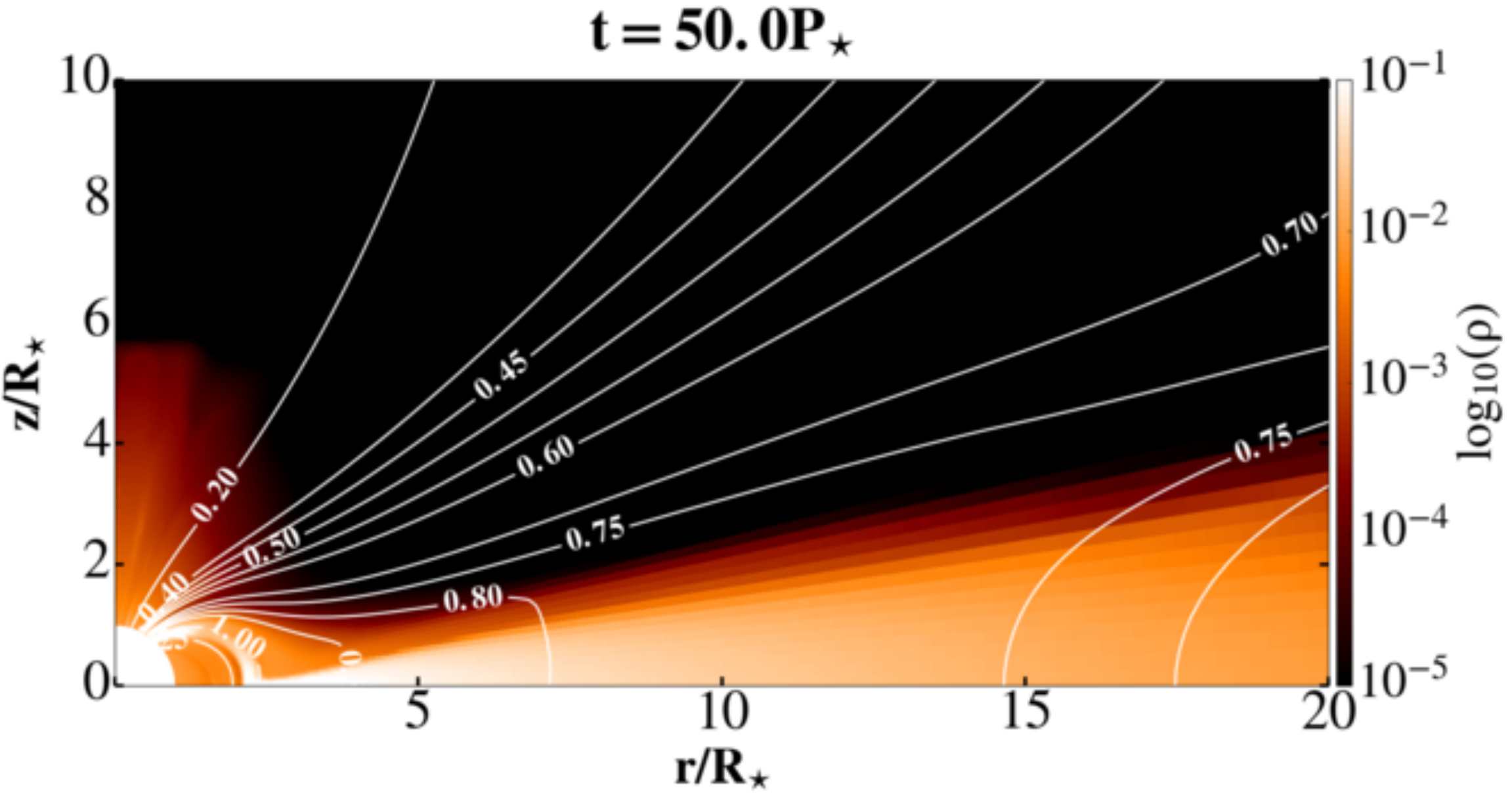}
\includegraphics[width=\columnwidth]{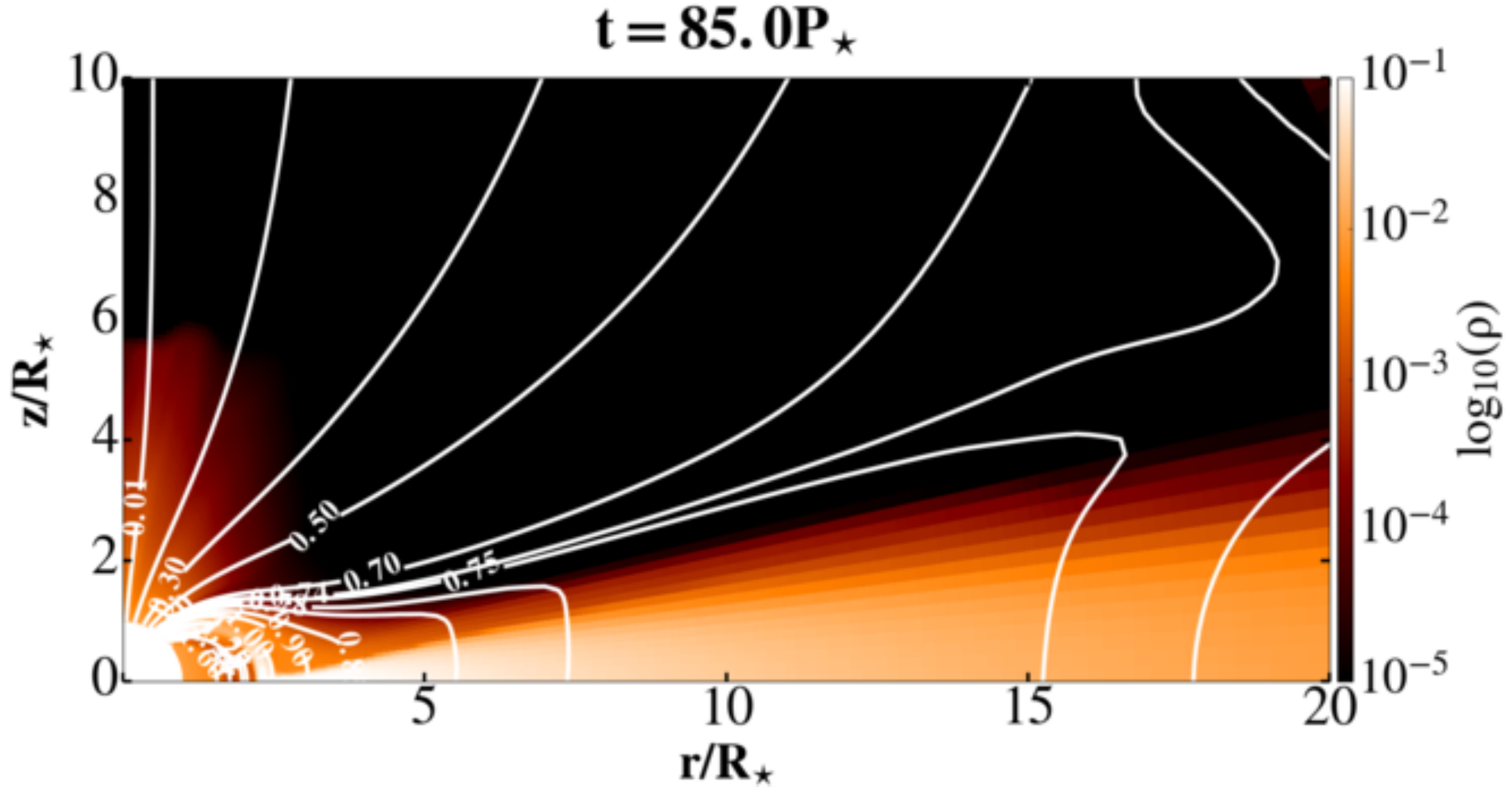} 
\caption{Matter density and poloidal magnetic field distribution in
the quasi-stationary interval in $\mu=0.7$ (0.5 kG) case with
$\Omega_\star$=0.2, with $\alpha_m=0.1$, 0.4, 0.7 and 1.0.}
\end{figure*}
\begin{figure*}
\includegraphics[width=\columnwidth]{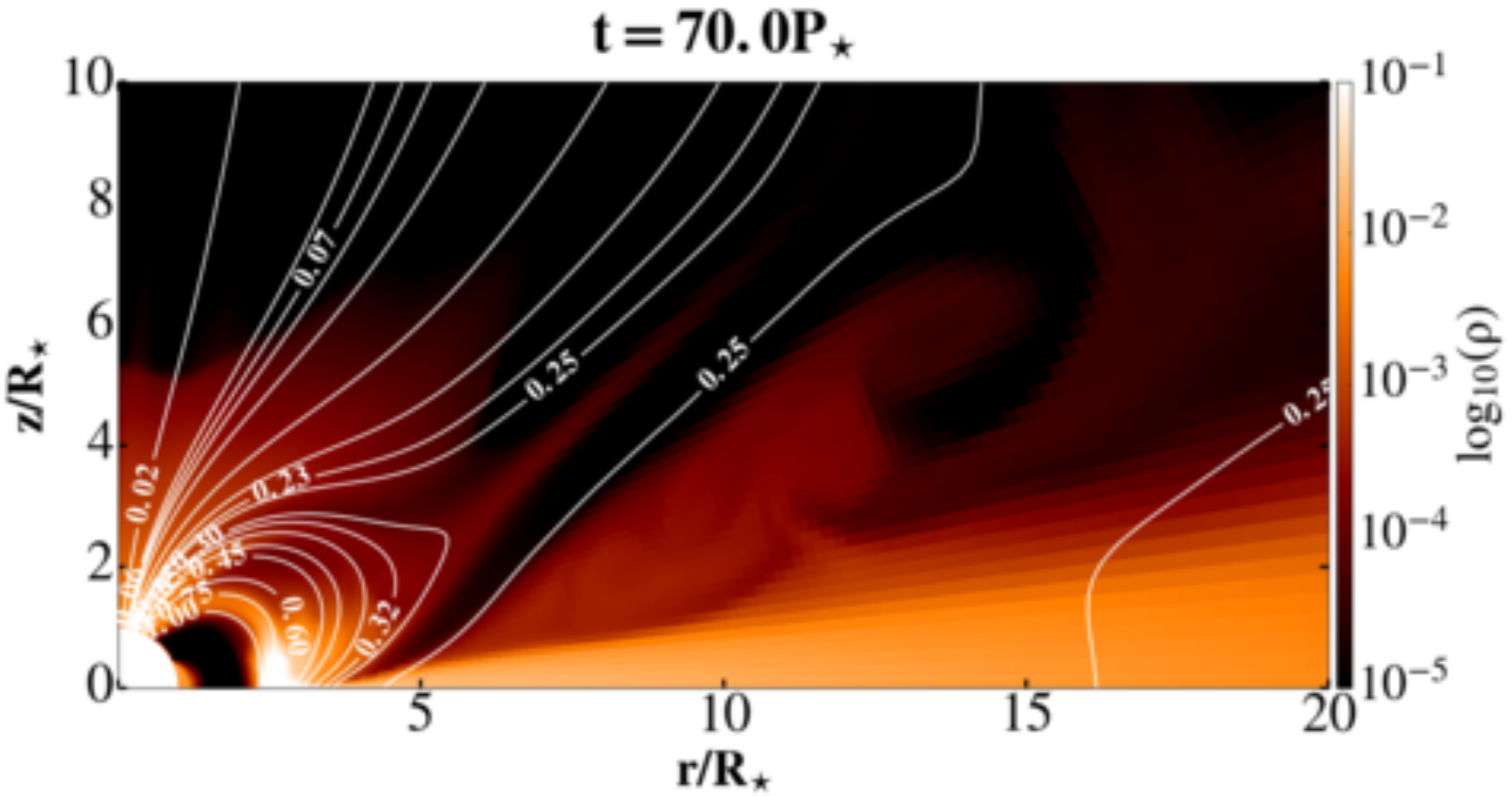}
\includegraphics[width=\columnwidth]{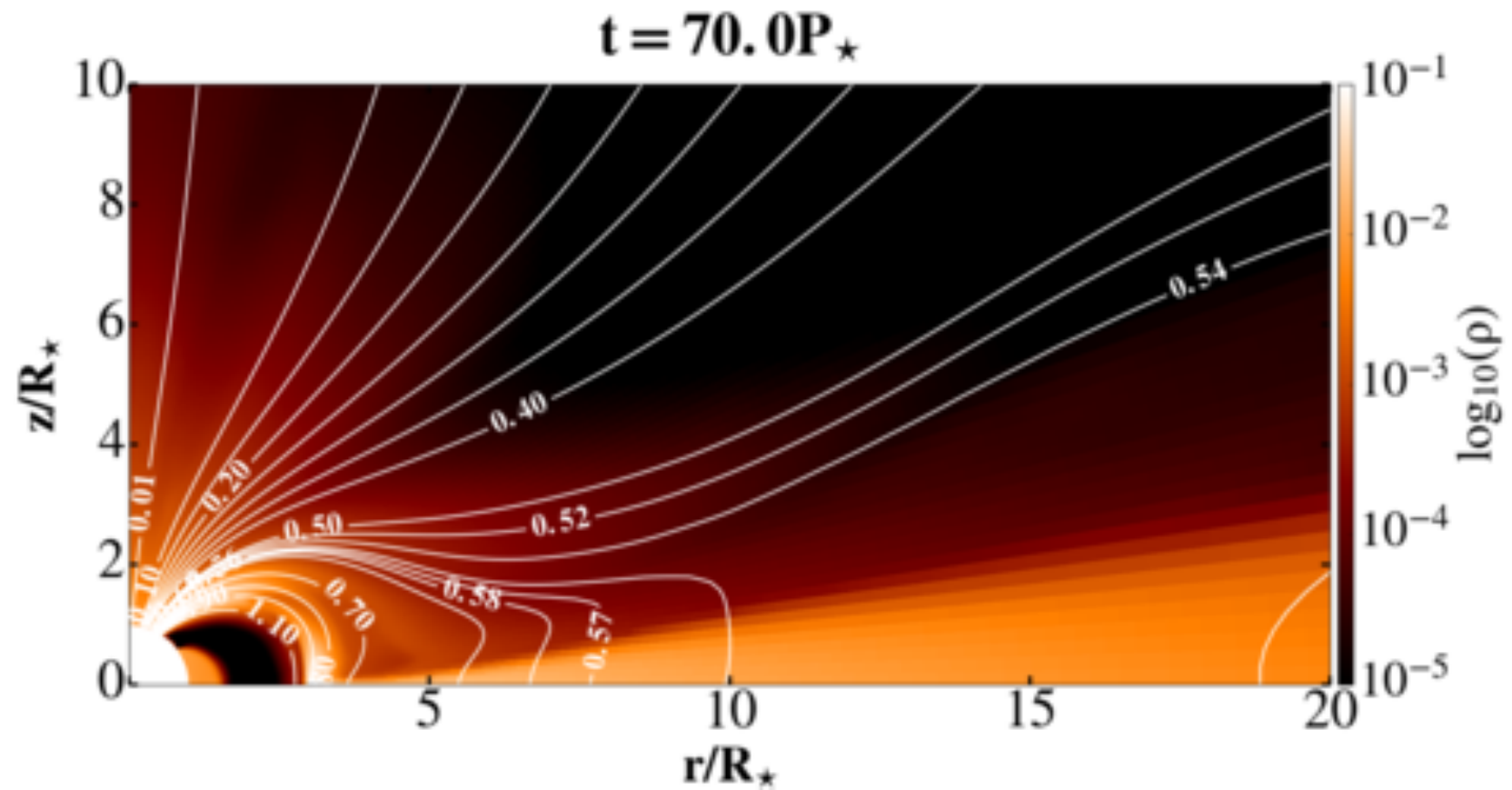}
\includegraphics[width=\columnwidth]{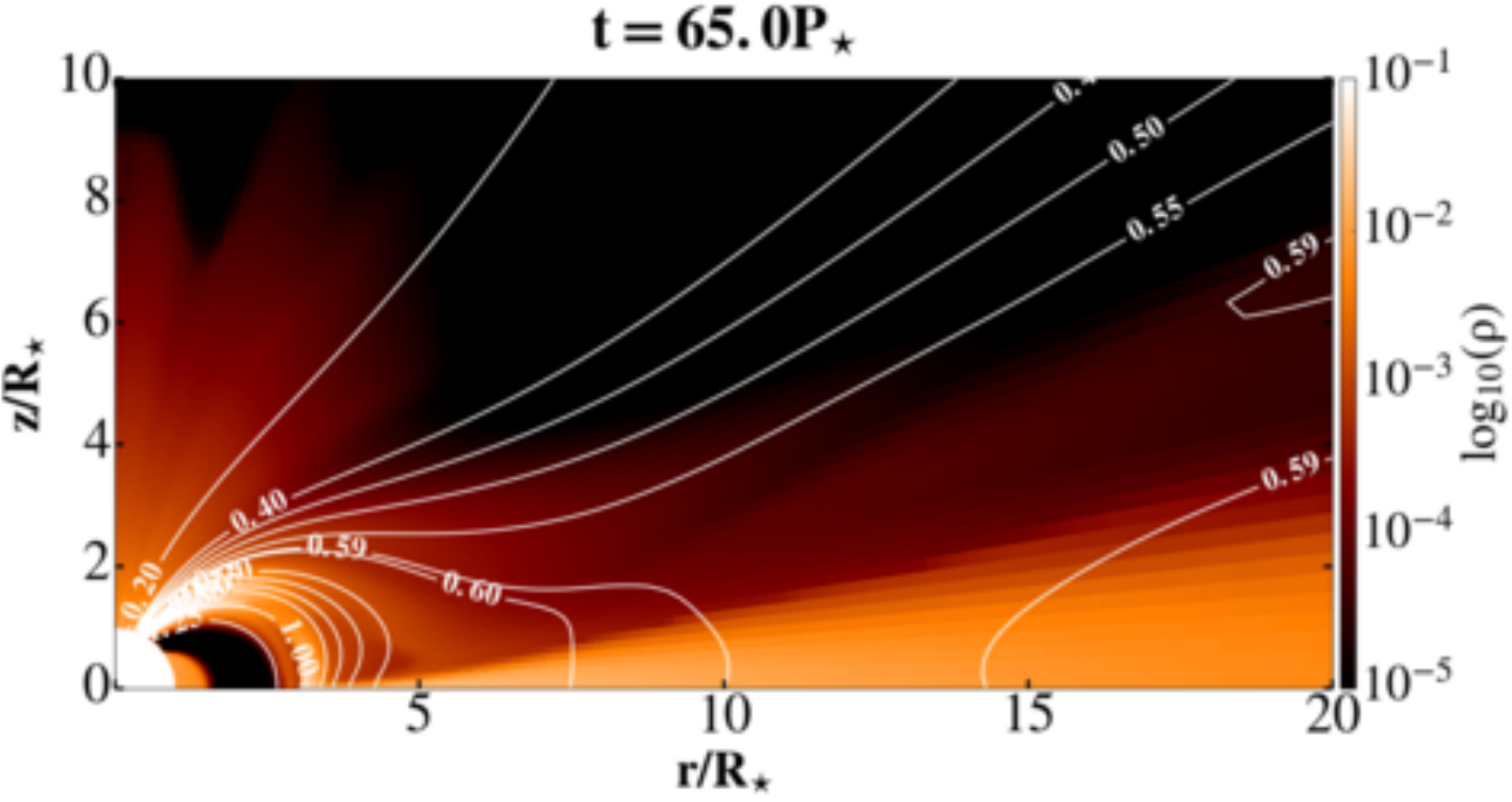}
\includegraphics[width=\columnwidth]{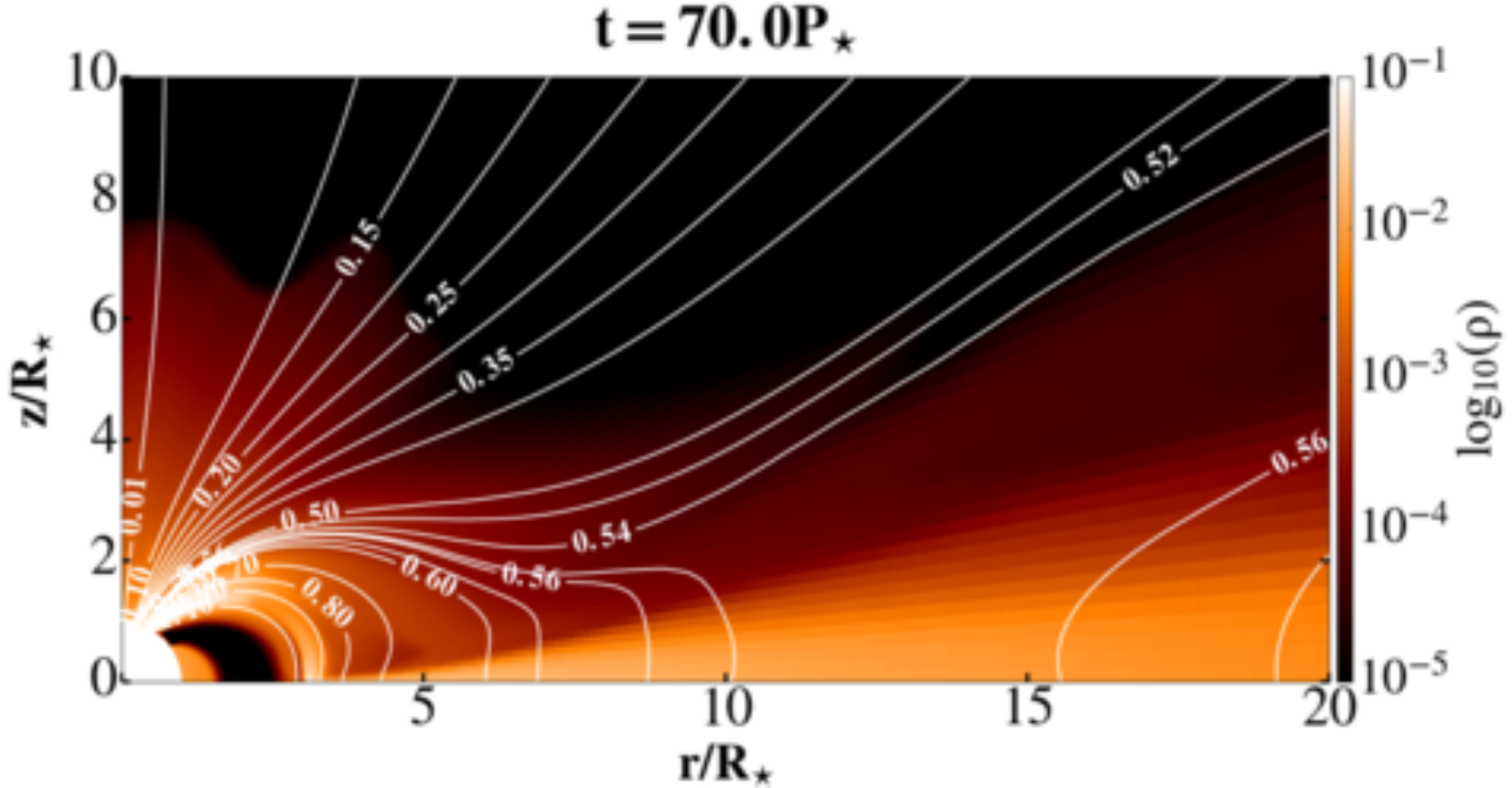} 
\caption{Matter density and poloidal magnetic field distribution in
the quasi-stationary interval in $\mu=1.05$ (0.75 kG) case with
$\Omega_\star$=0.05, with $\alpha_m=0.1$, 0.4, 0.7 and 1.0.}
\end{figure*}
\begin{figure*}
\includegraphics[width=\columnwidth]{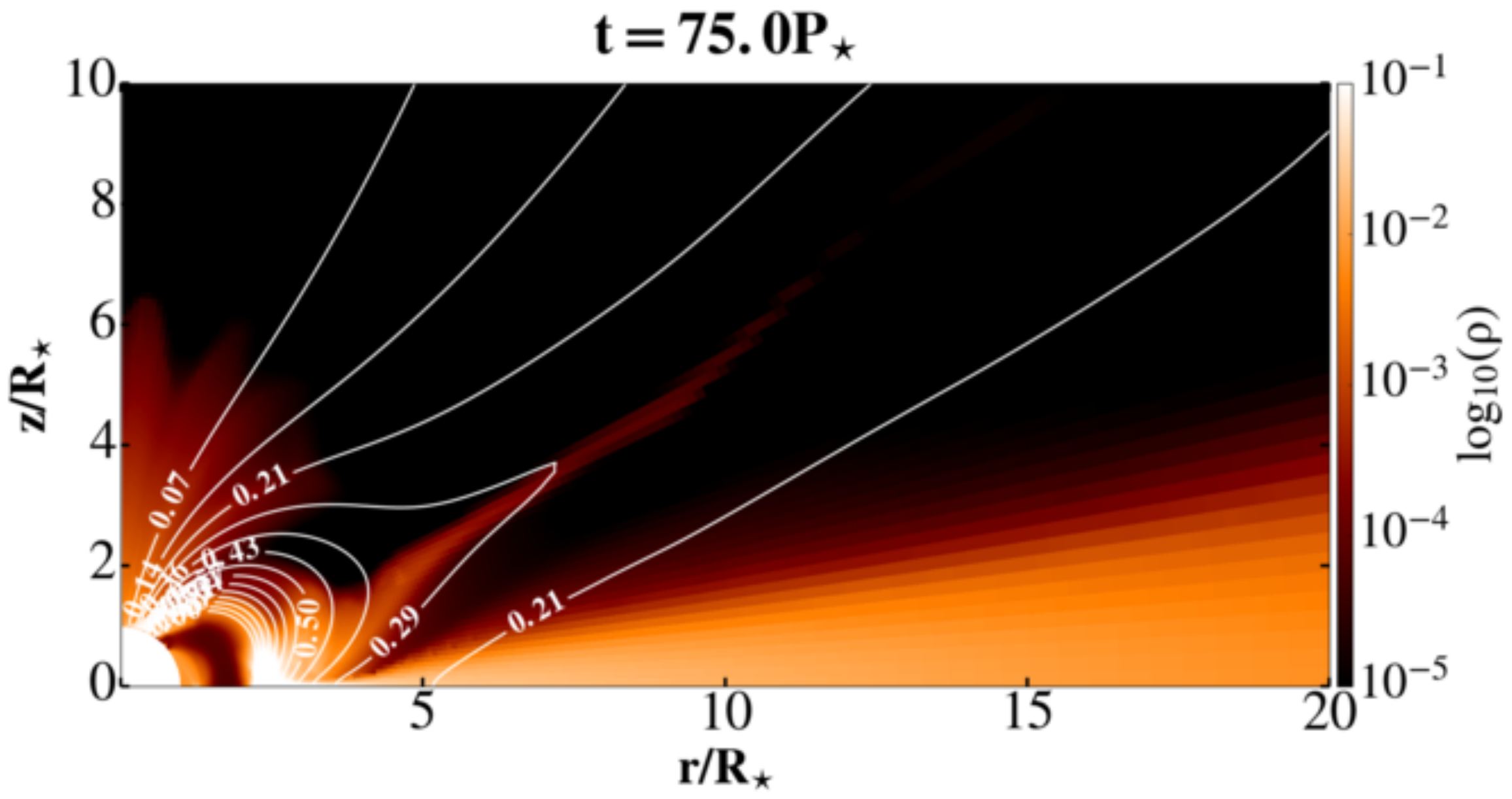}
\includegraphics[width=\columnwidth]{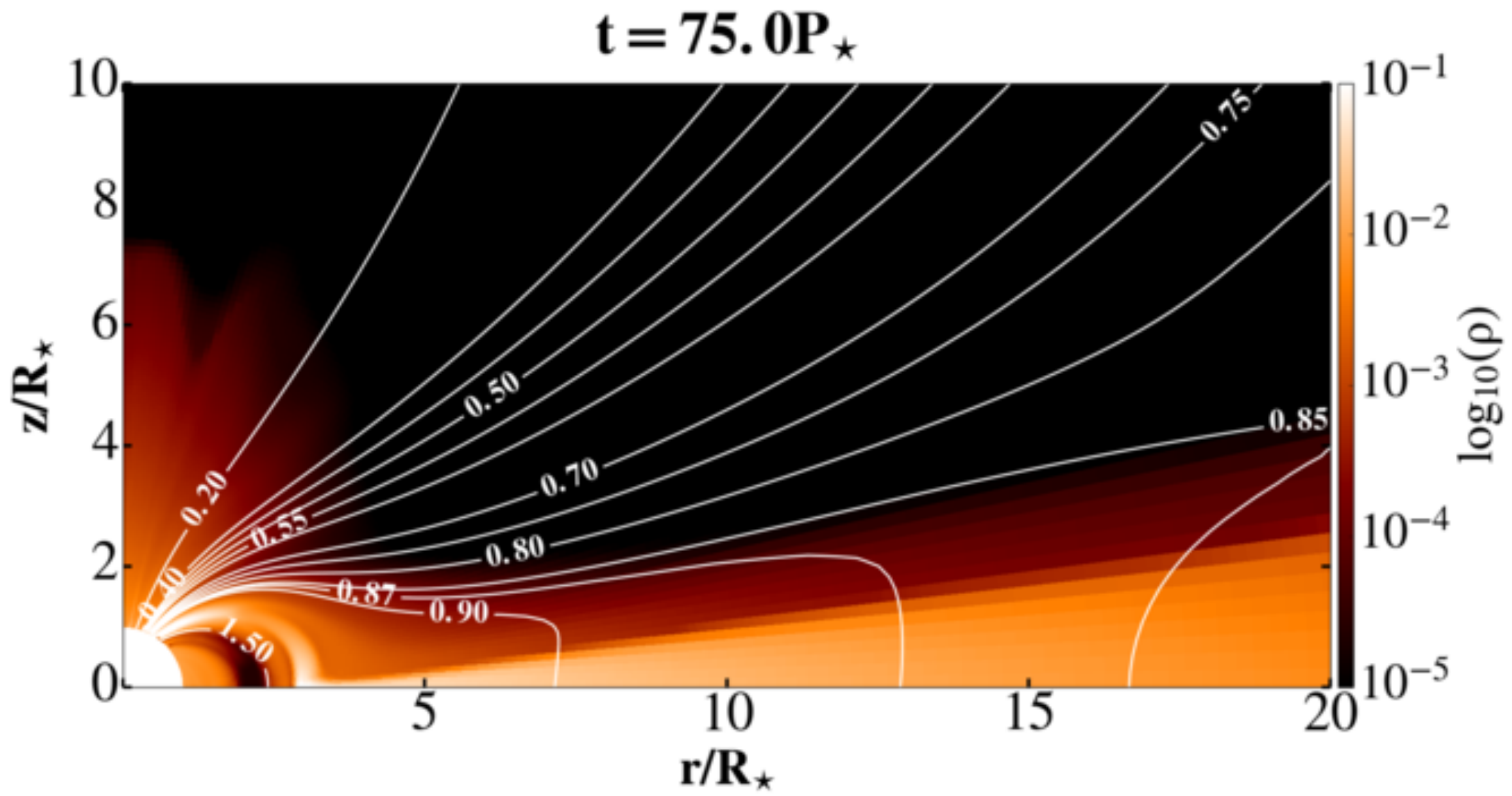}
\includegraphics[width=\columnwidth]{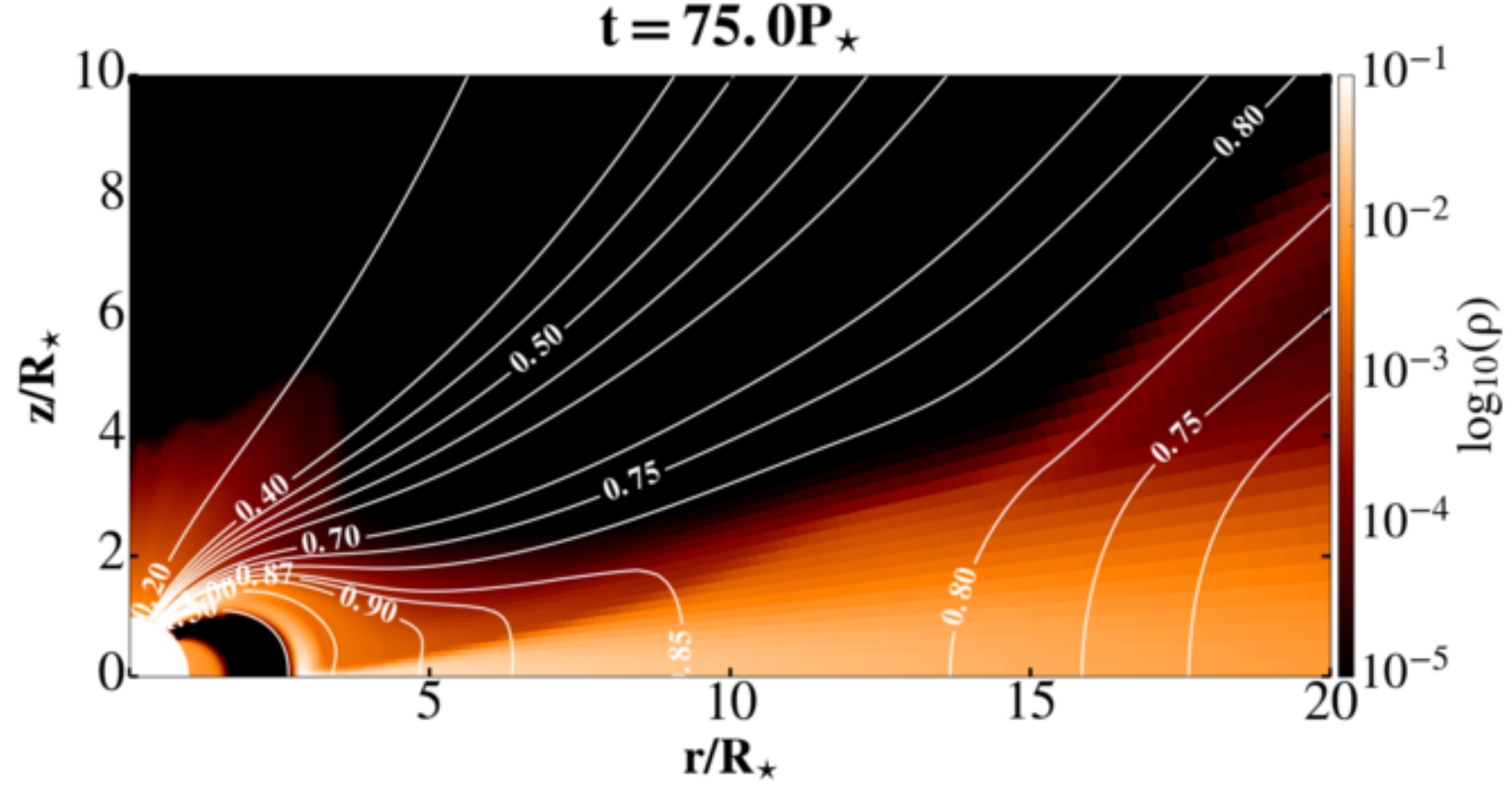}
\includegraphics[width=\columnwidth]{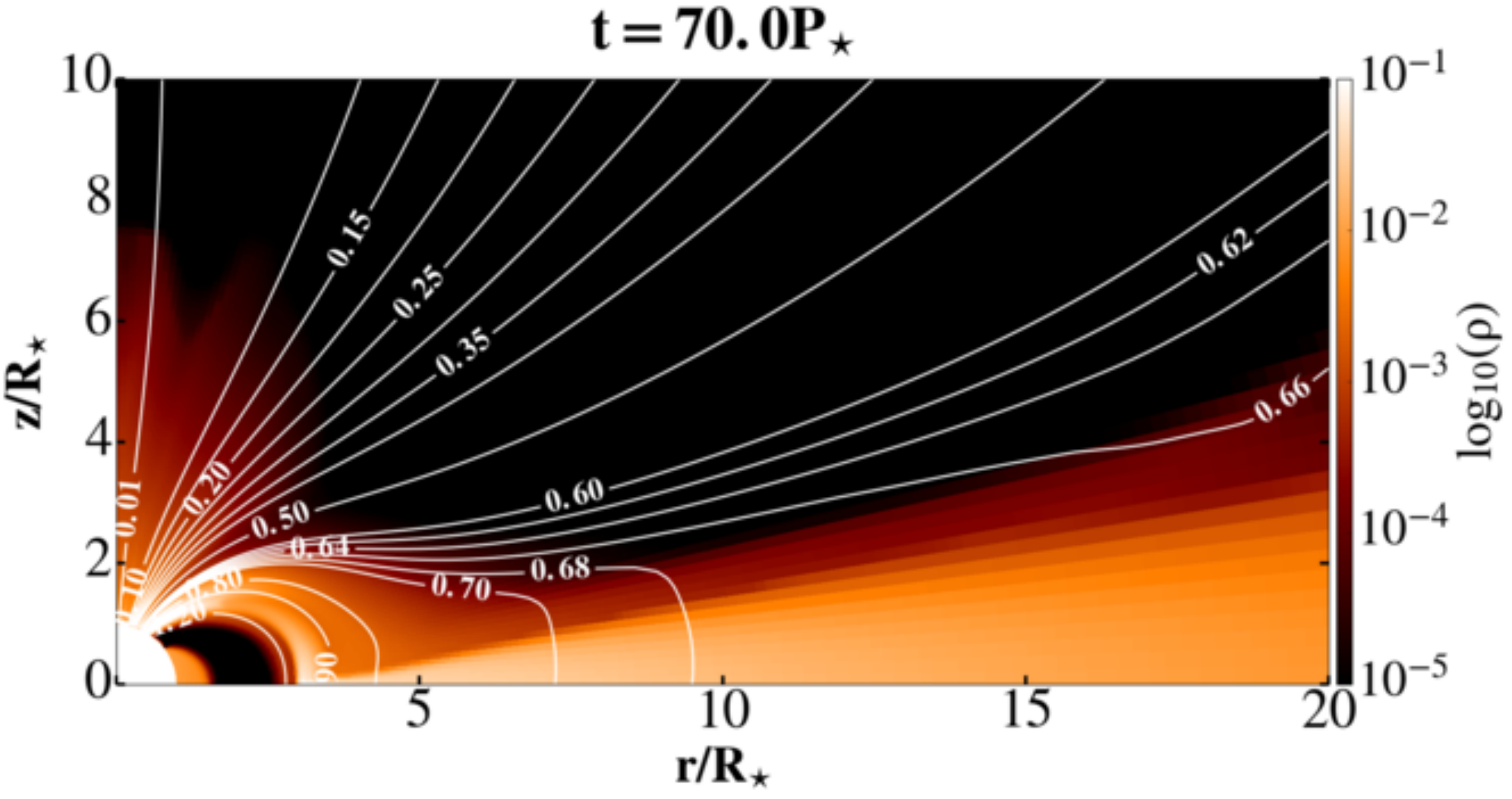} 
\caption{Matter density and poloidal magnetic field distribution in
the quasi-stationary interval in $\mu=1.05$ (0.75 kG) case with
$\Omega_\star$=0.1, with $\alpha_m=0.1$, 0.4, 0.7 and 1.0. }
\end{figure*}
\begin{figure*}
\includegraphics[width=\columnwidth]{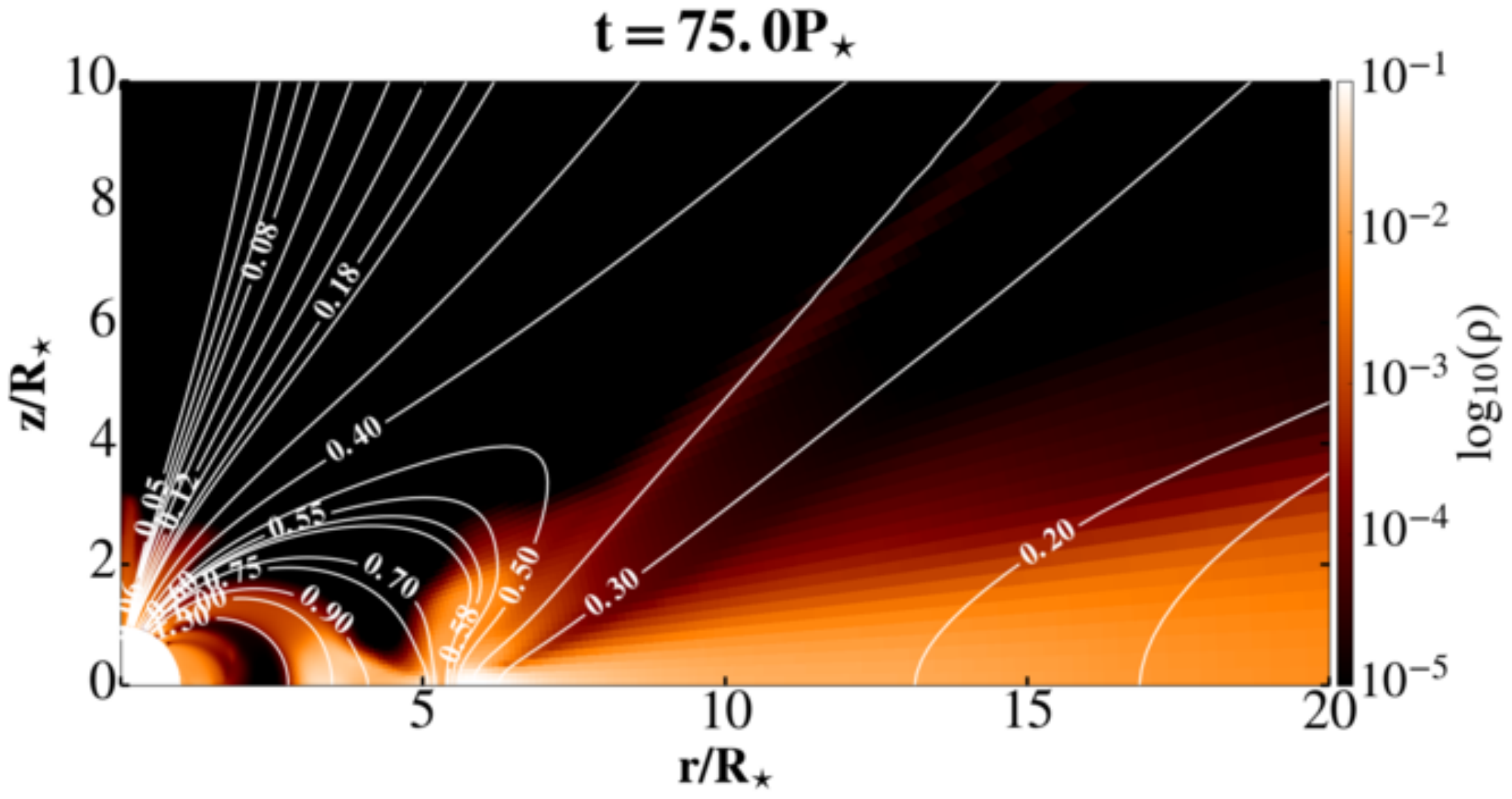}
\includegraphics[width=\columnwidth]{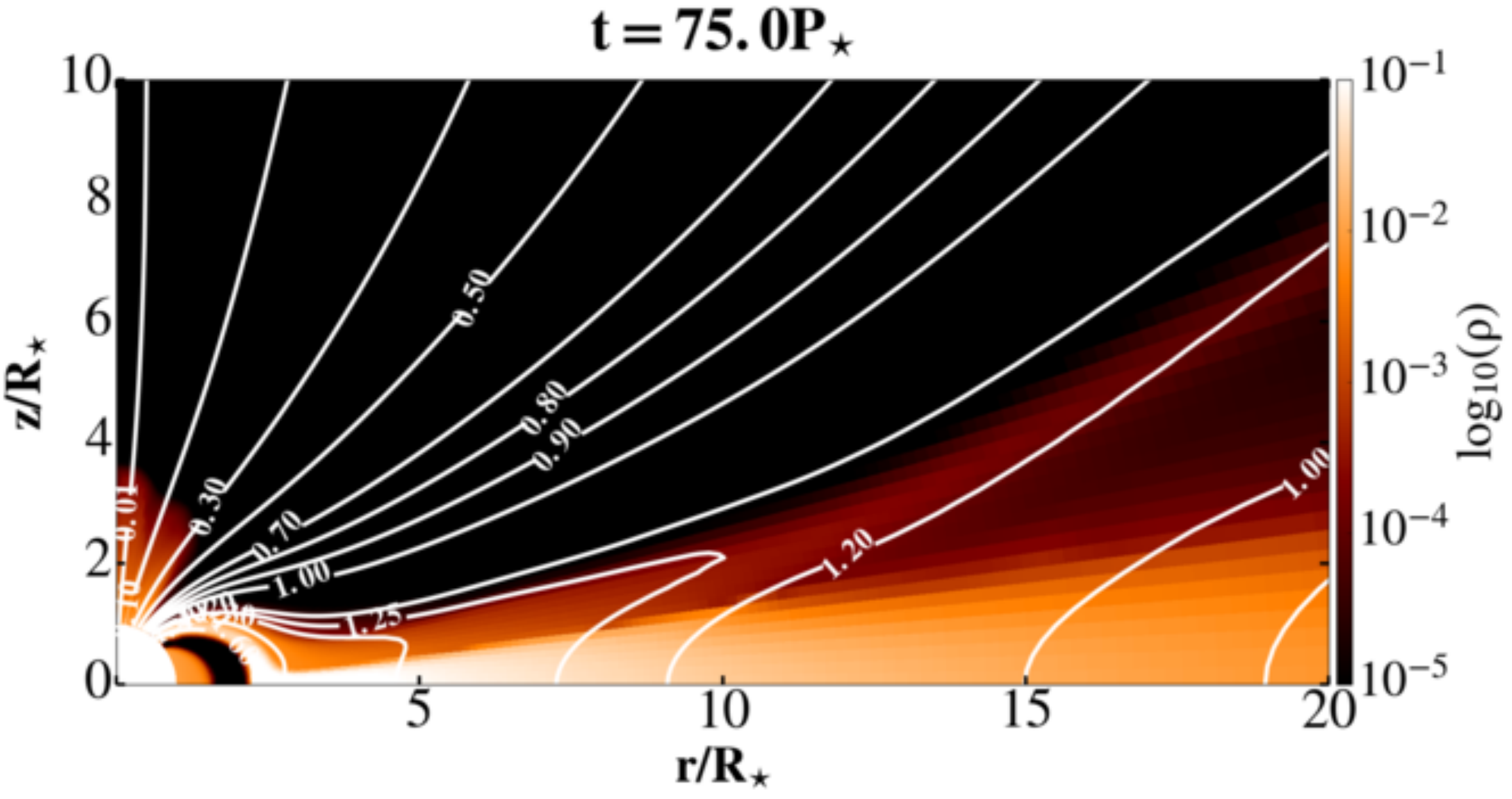}
\includegraphics[width=\columnwidth]{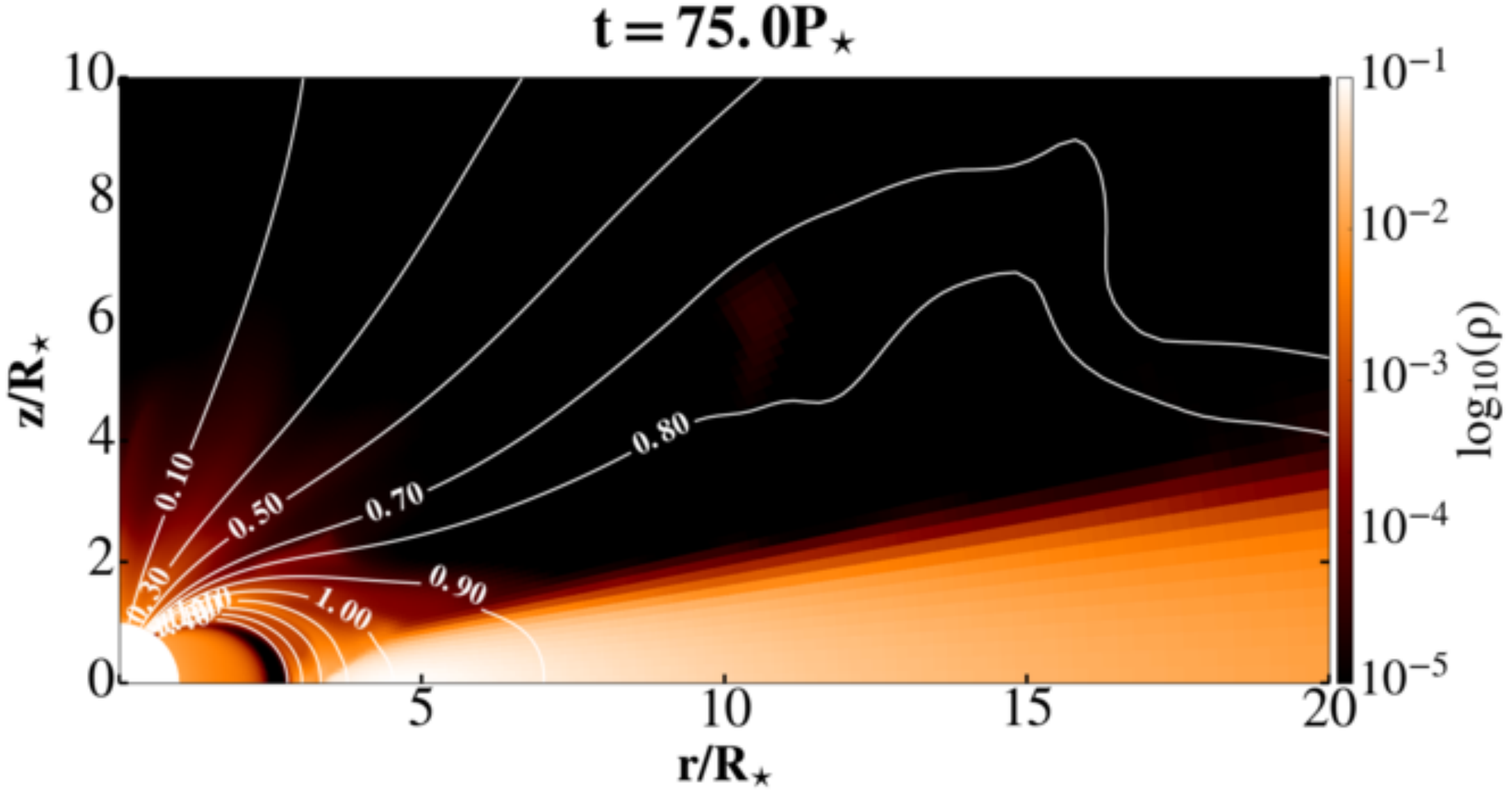}
\includegraphics[width=\columnwidth]{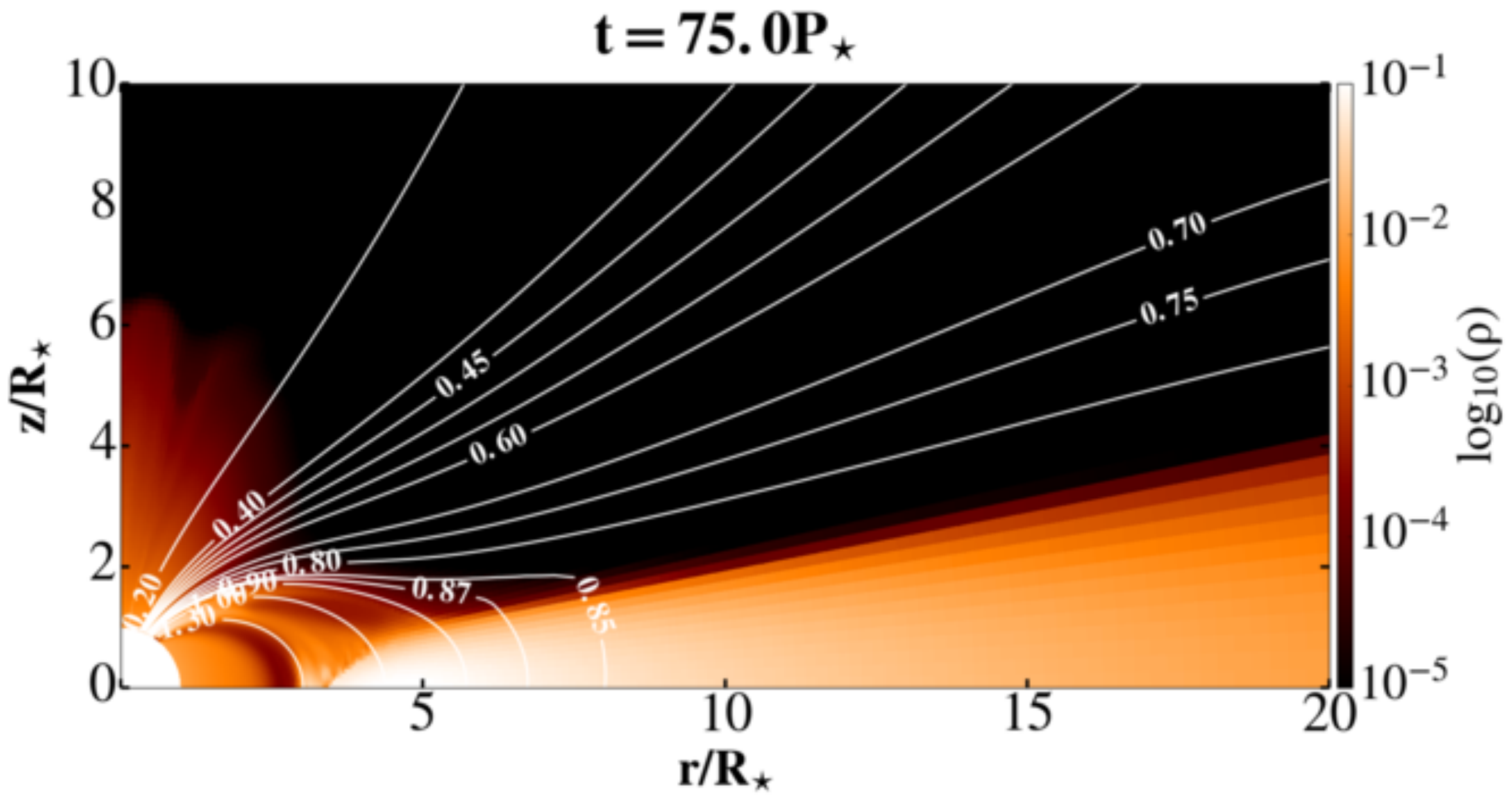} 
\caption{Matter density and poloidal magnetic field distribution in
the quasi-stationary interval in $\mu=1.05$ (0.75 kG) case with
$\Omega_\star$=0.15, with $\alpha_m=0.1$, 0.4, 0.7 and 1.0. }
\end{figure*}
\begin{figure*}
\includegraphics[width=\columnwidth]{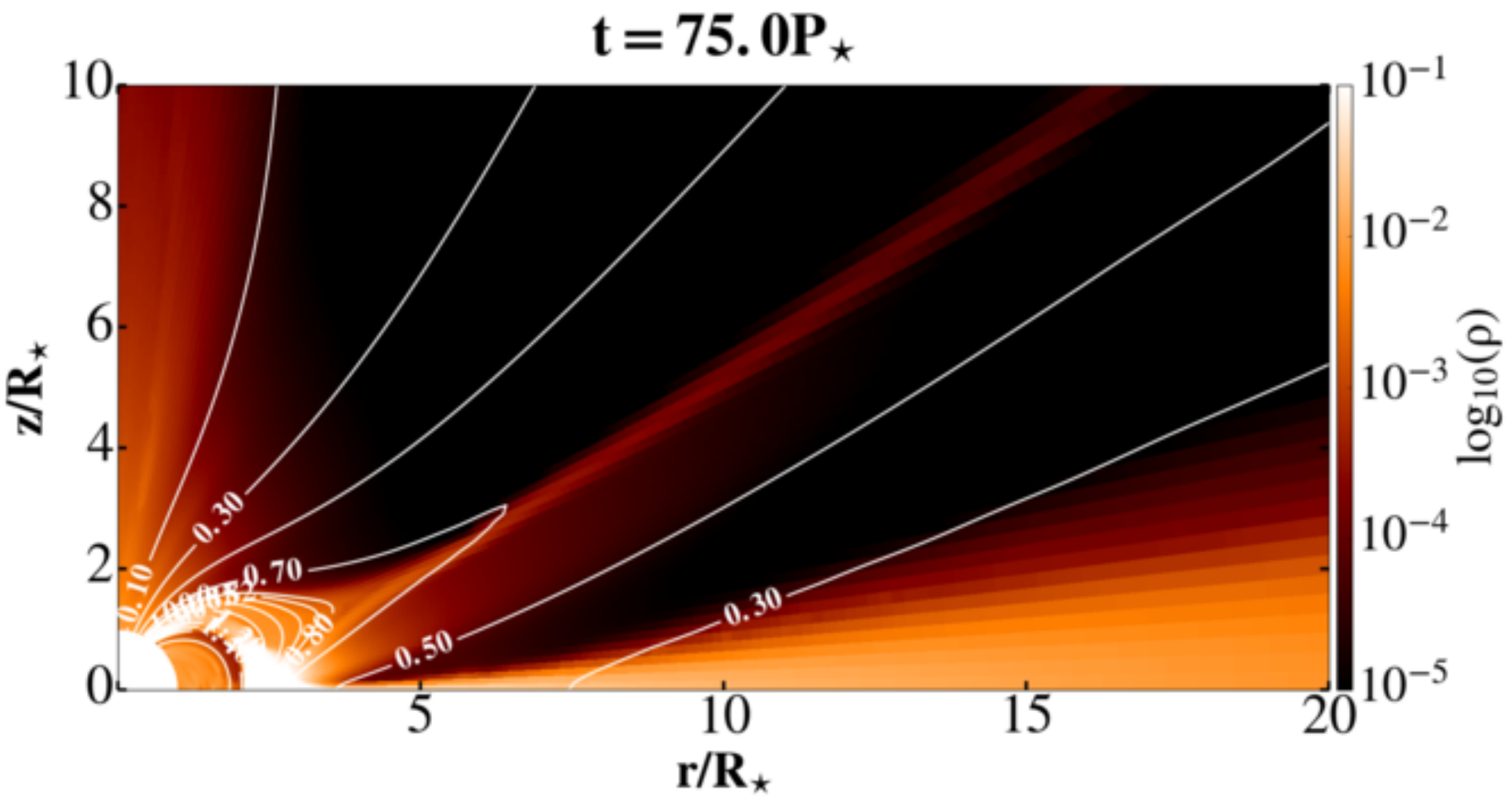}
\includegraphics[width=\columnwidth]{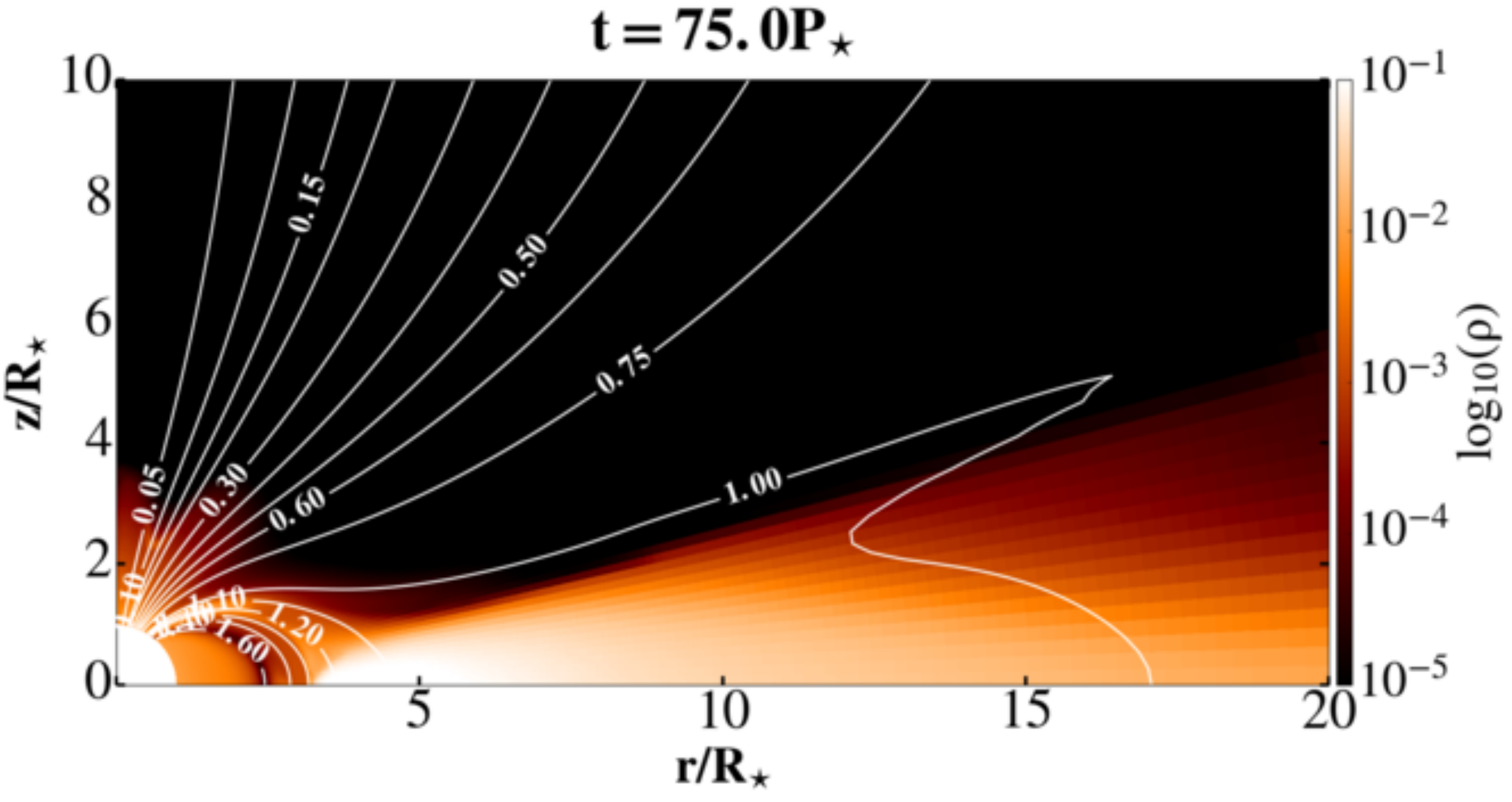}
\includegraphics[width=\columnwidth]{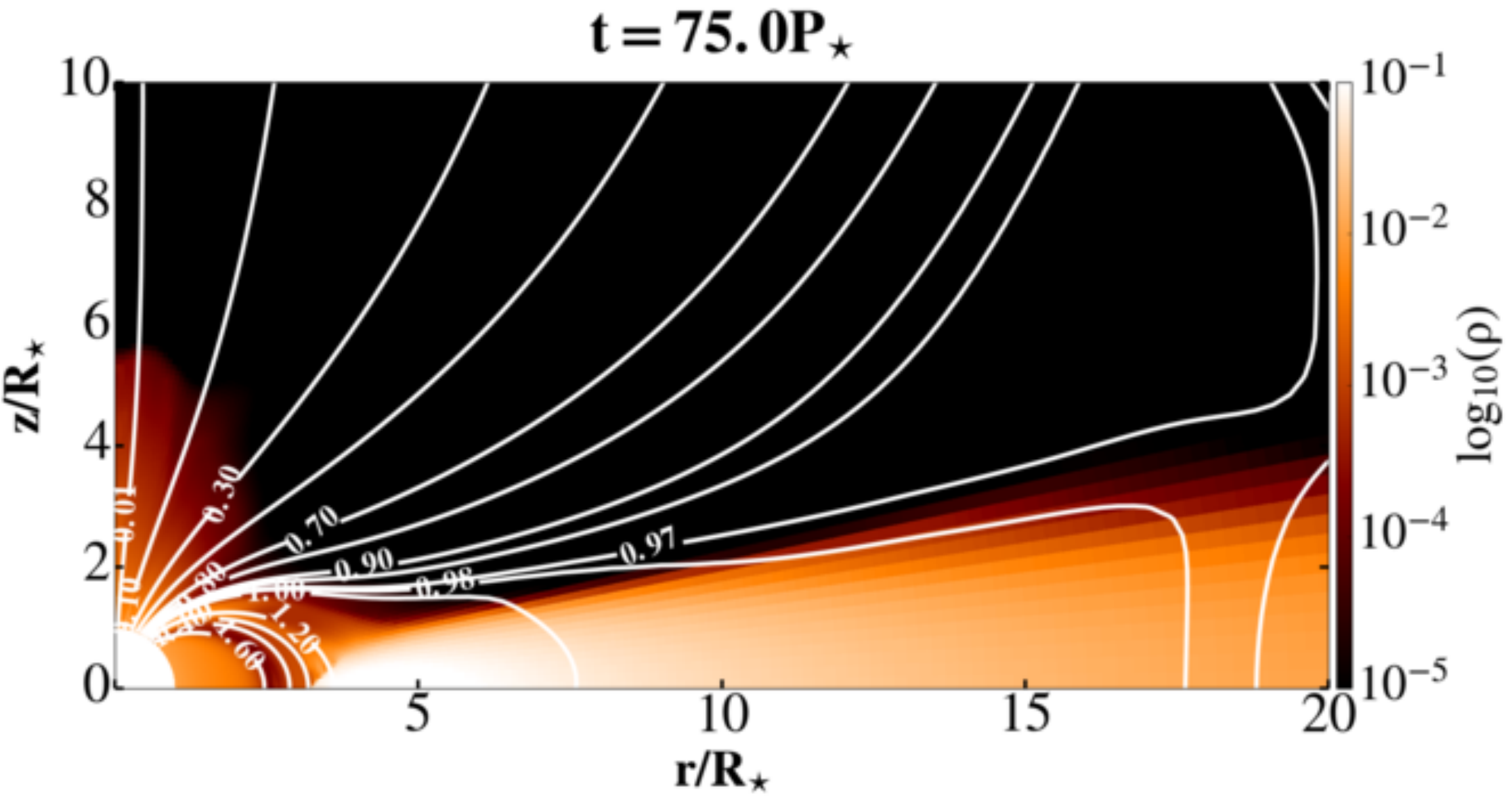}
\includegraphics[width=\columnwidth]{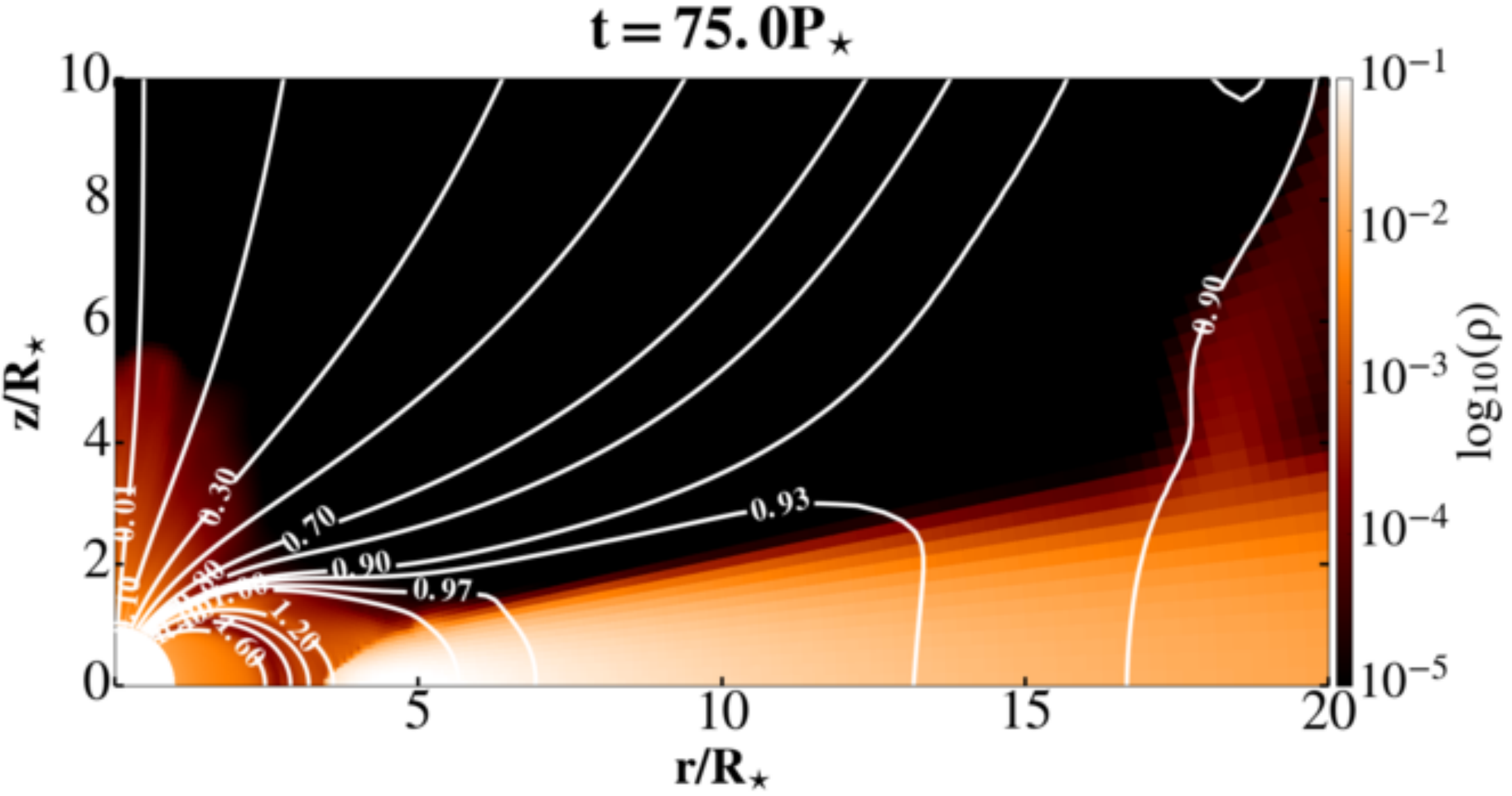} 
\caption{Matter density and poloidal magnetic field distribution in
the quasi-stationary interval in $\mu=1.05$ (0.75 kG) case with
$\Omega_\star$=0.2, with $\alpha_m=0.1$, 0.4, 0.7 and 1.0.}
\end{figure*}
\begin{figure*}
\includegraphics[width=\columnwidth]{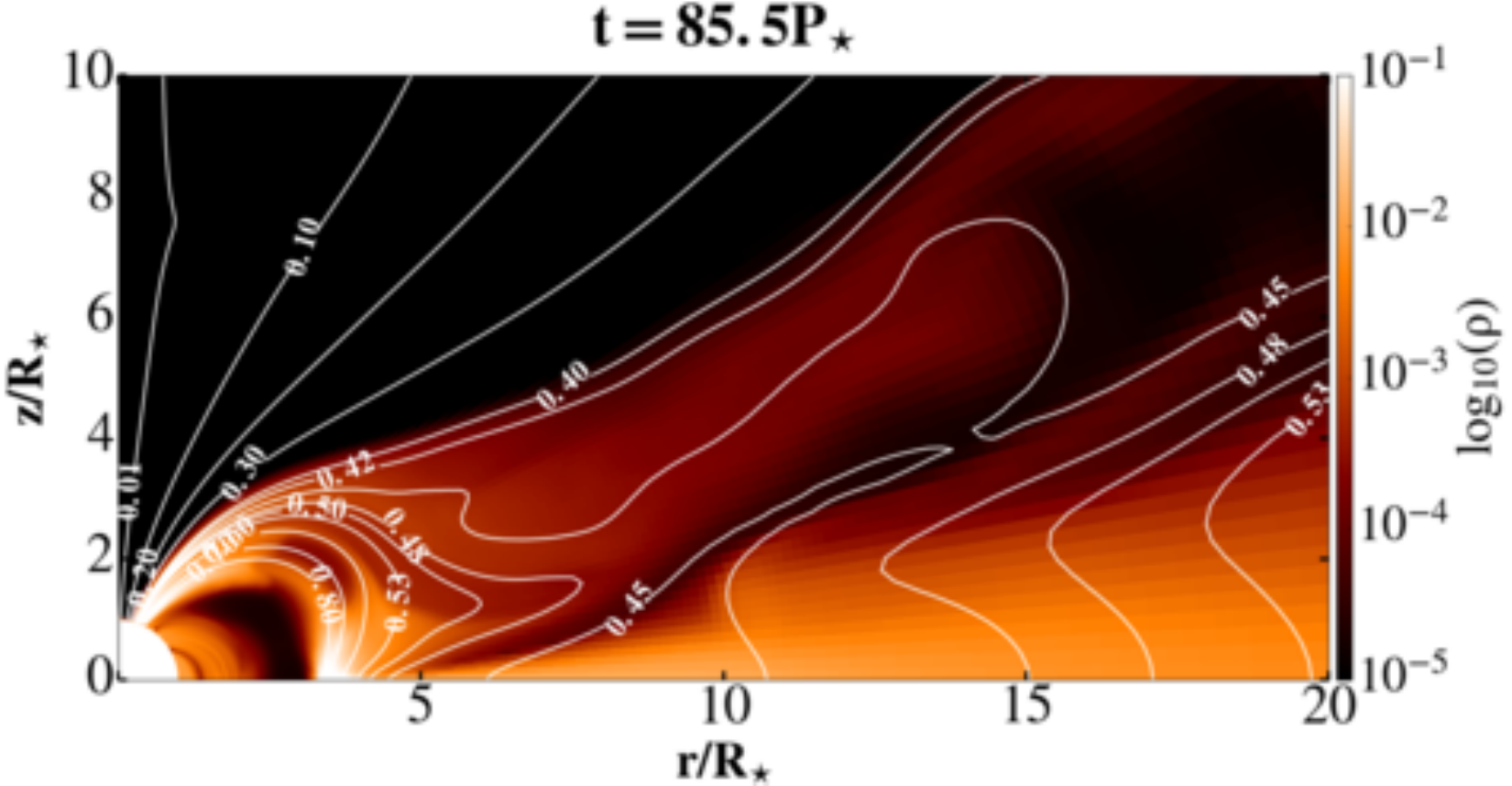}
\includegraphics[width=\columnwidth]{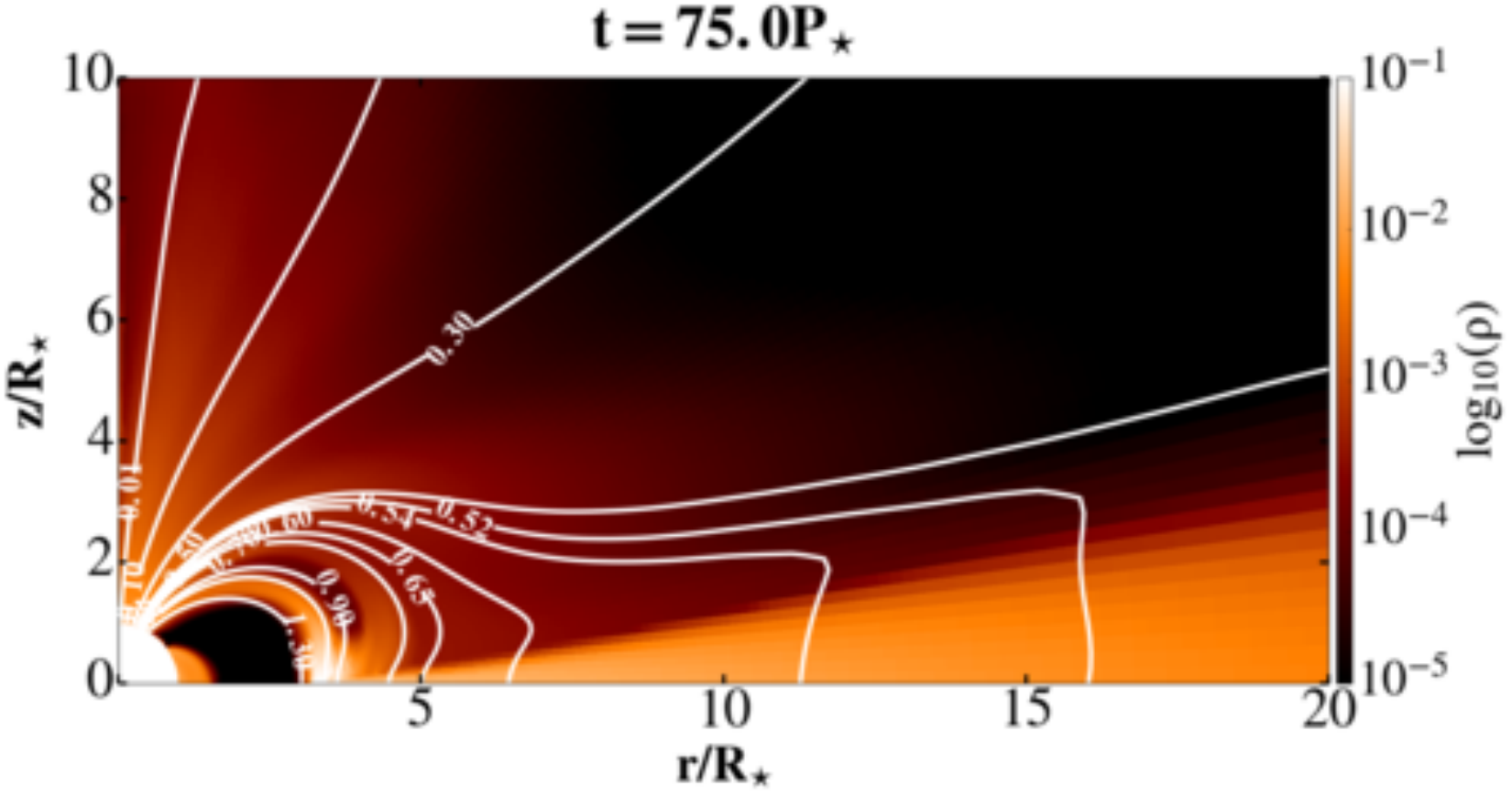}
\includegraphics[width=\columnwidth]{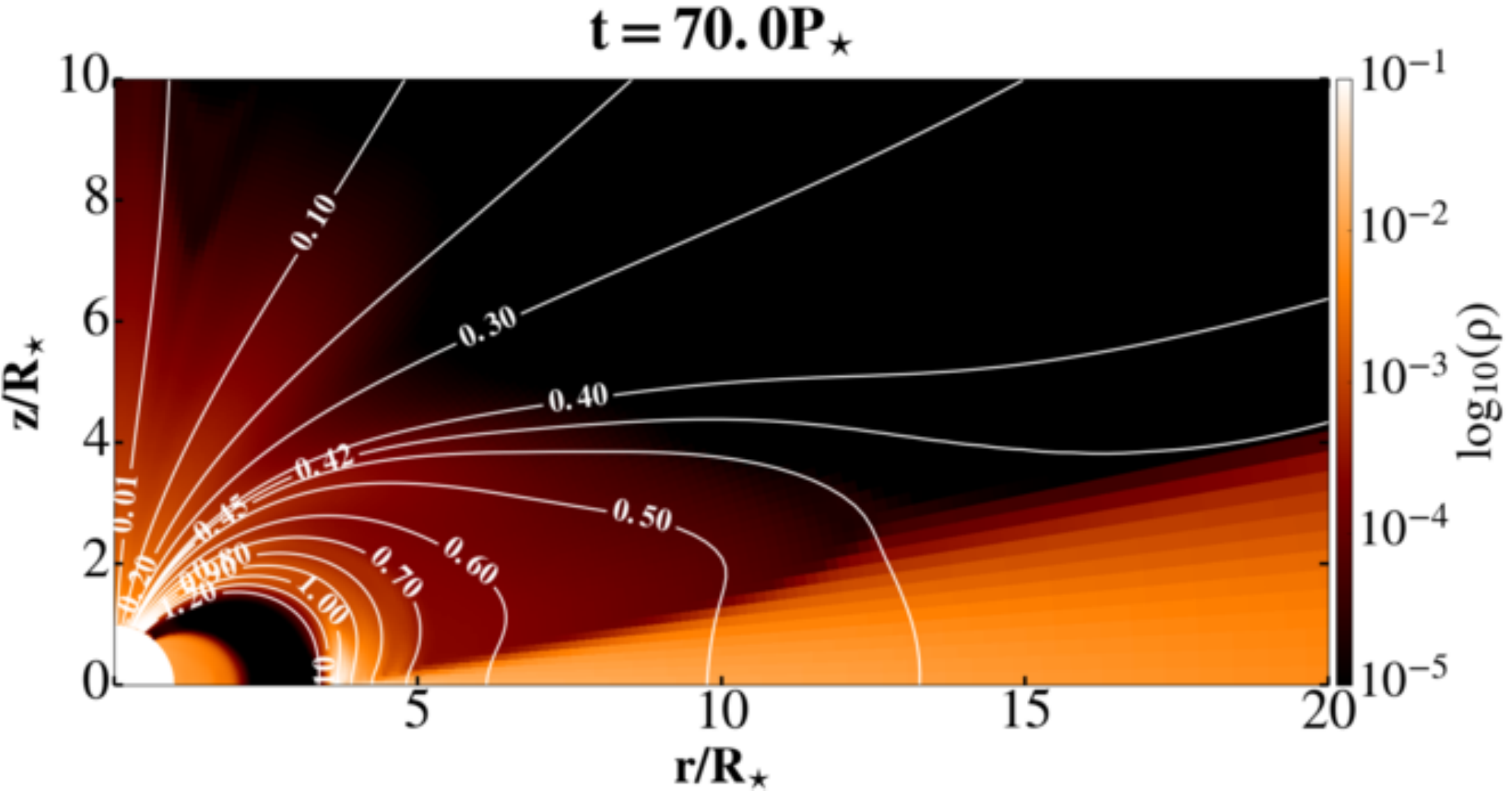}
\includegraphics[width=\columnwidth]{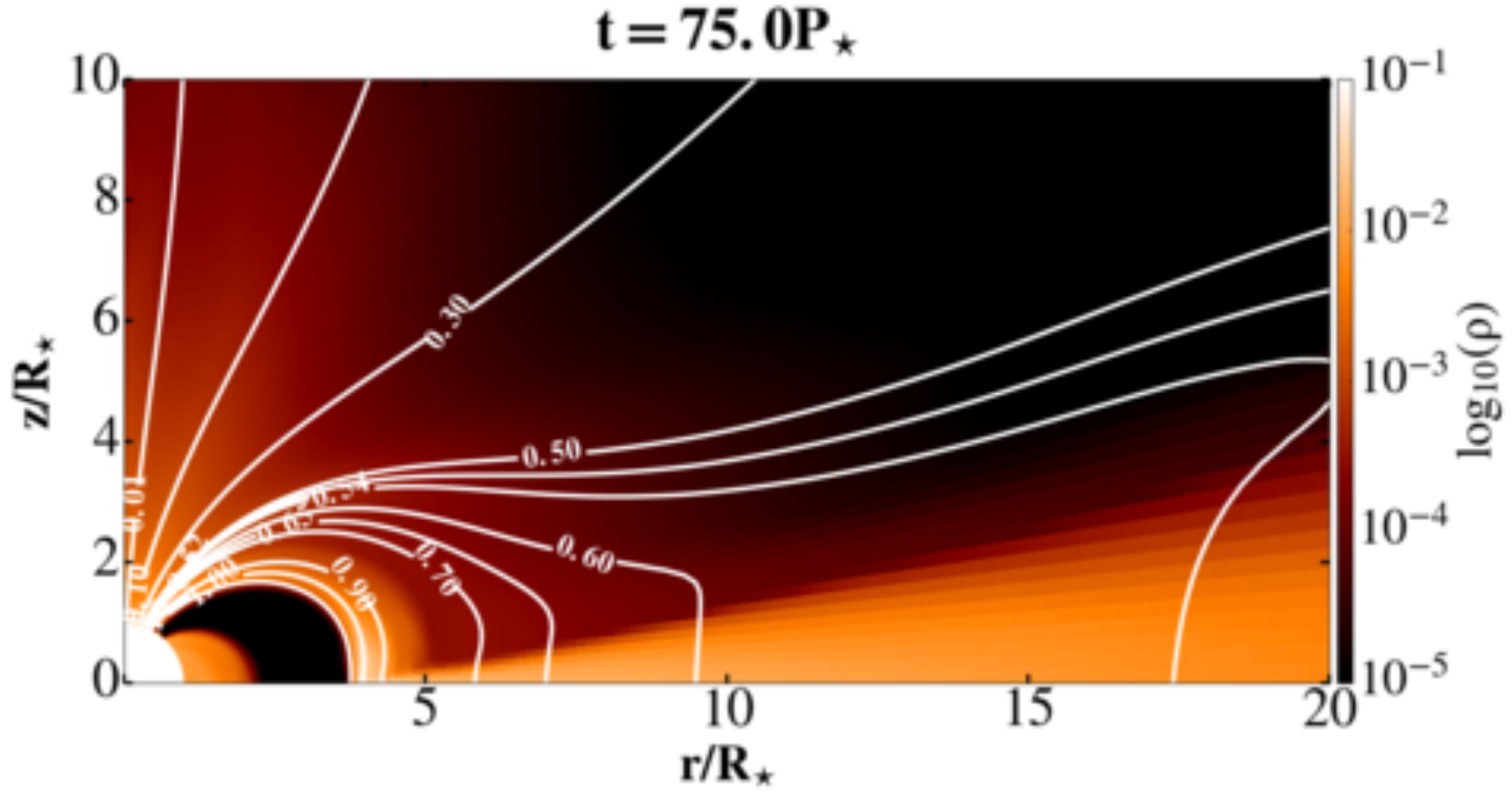} 
\caption{Matter density and poloidal magnetic field distribution in
the quasi-stationary interval in $\mu=1.4$ (1.0 kG) case with
$\Omega_\star$=0.05, with $\alpha_m=0.1$, 0.4, 0.7 and 1.0.}
\end{figure*}
\begin{figure*}
\includegraphics[width=\columnwidth]{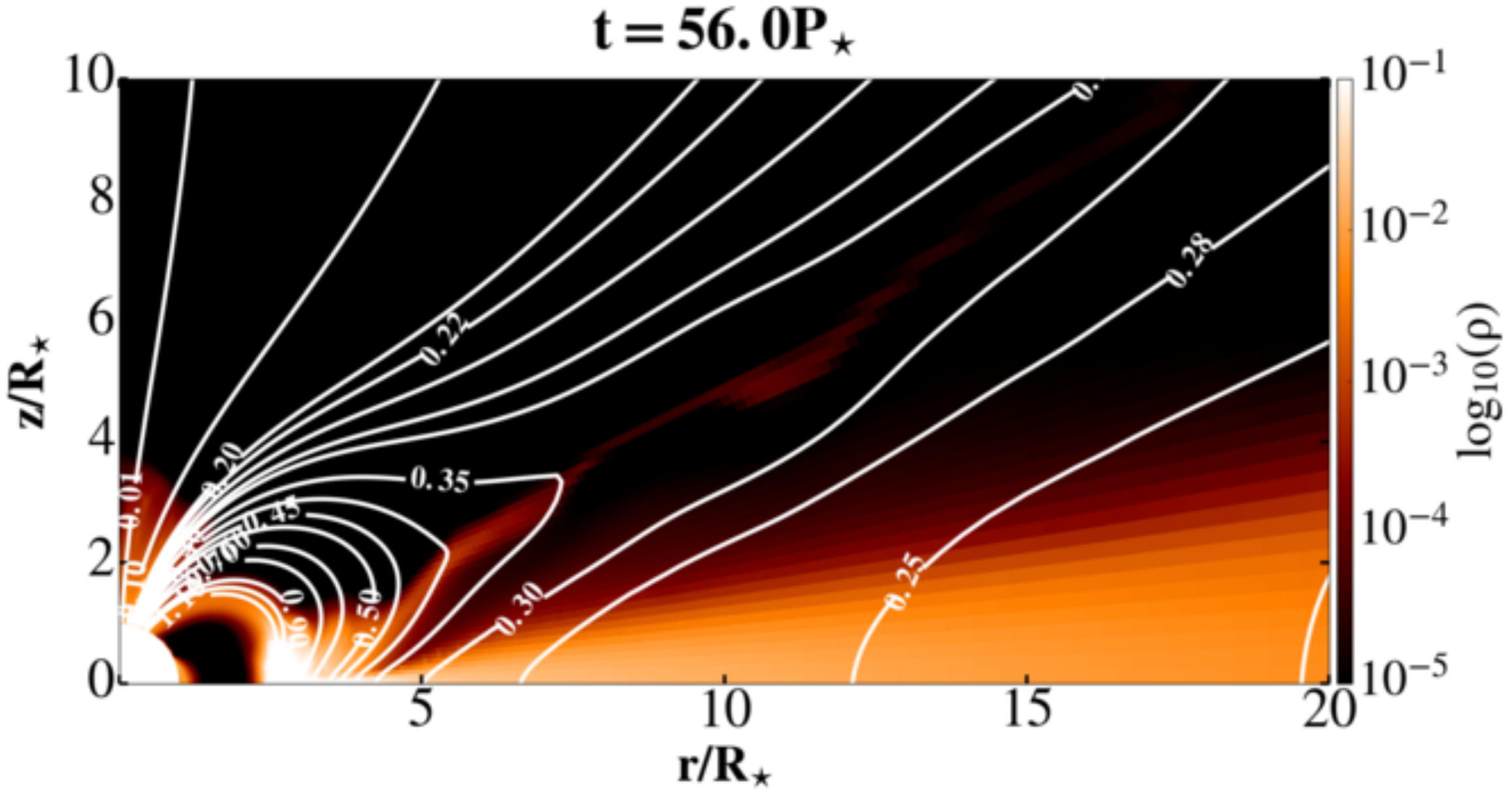}
\includegraphics[width=\columnwidth]{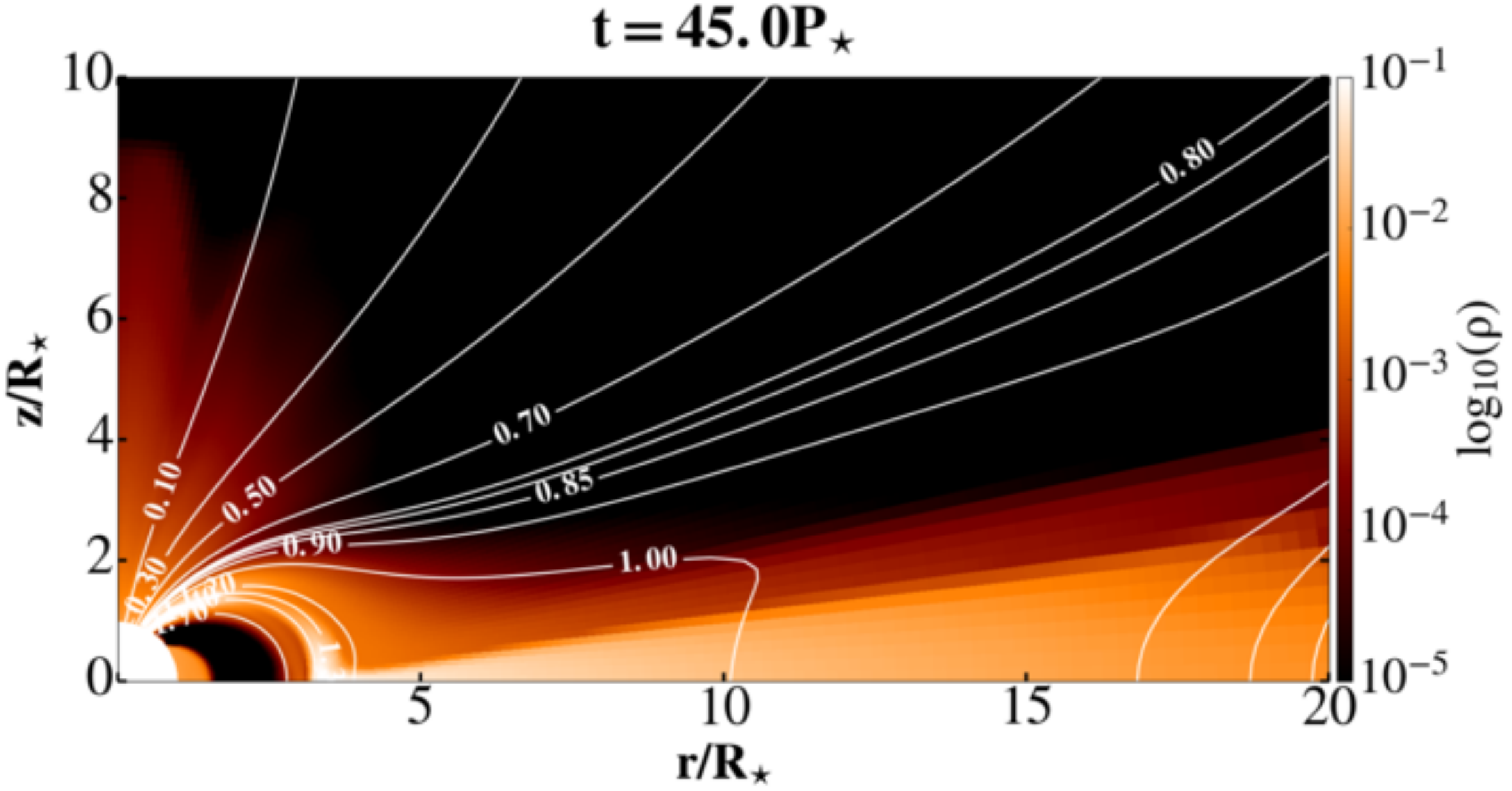}
\includegraphics[width=\columnwidth]{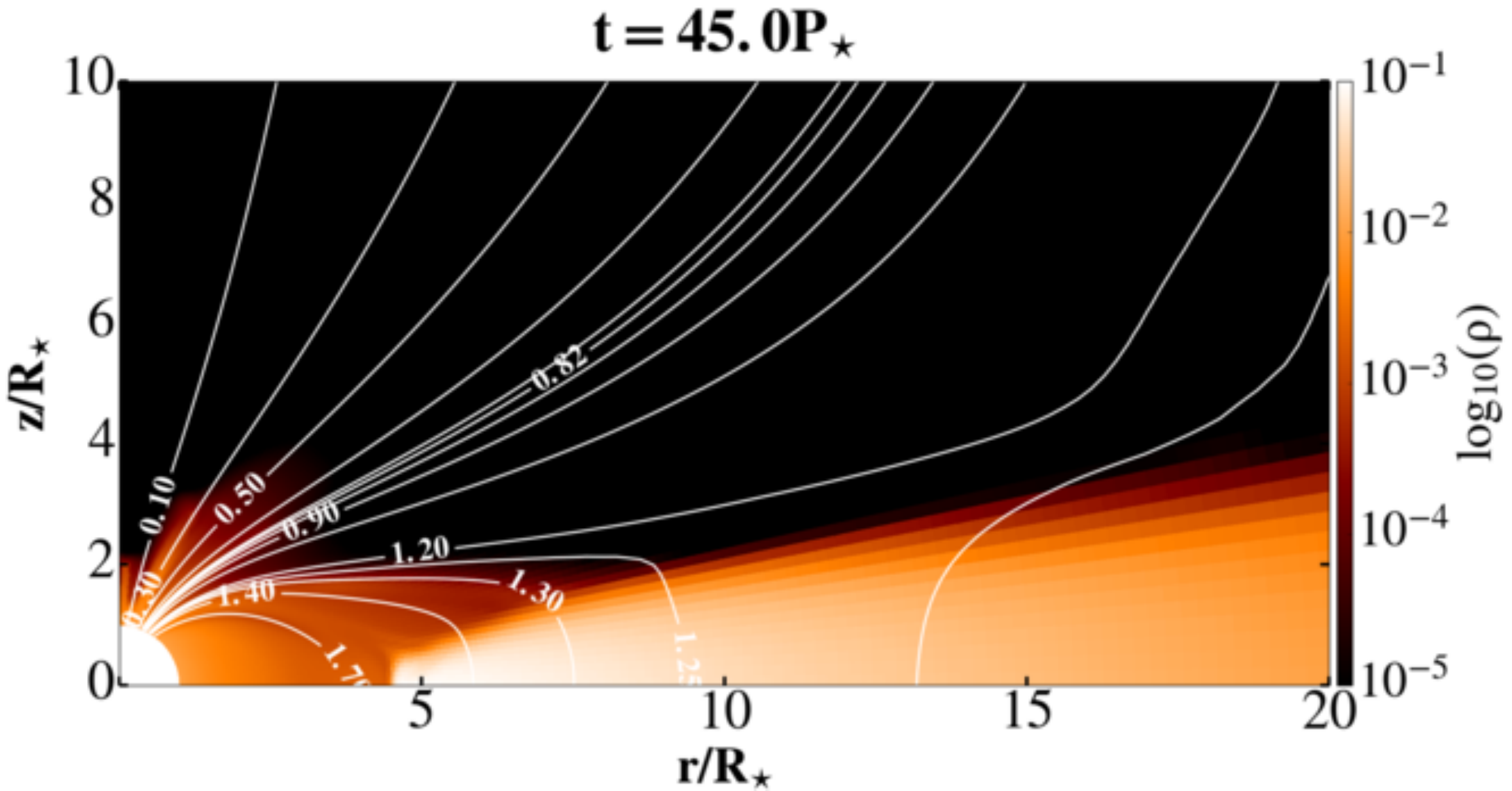}
\includegraphics[width=\columnwidth]{{plotmu1.4Av1Am1om.1}.pdf} 
\caption{Matter density and poloidal magnetic field distribution in
the quasi-stationary interval in $\mu=1.4$ (1.0 kG) case with
$\Omega_\star$=0.1, with $\alpha_m=0.1$, 0.4, 0.7 and 1.0.}
\end{figure*}
\begin{figure*}
\includegraphics[width=\columnwidth]{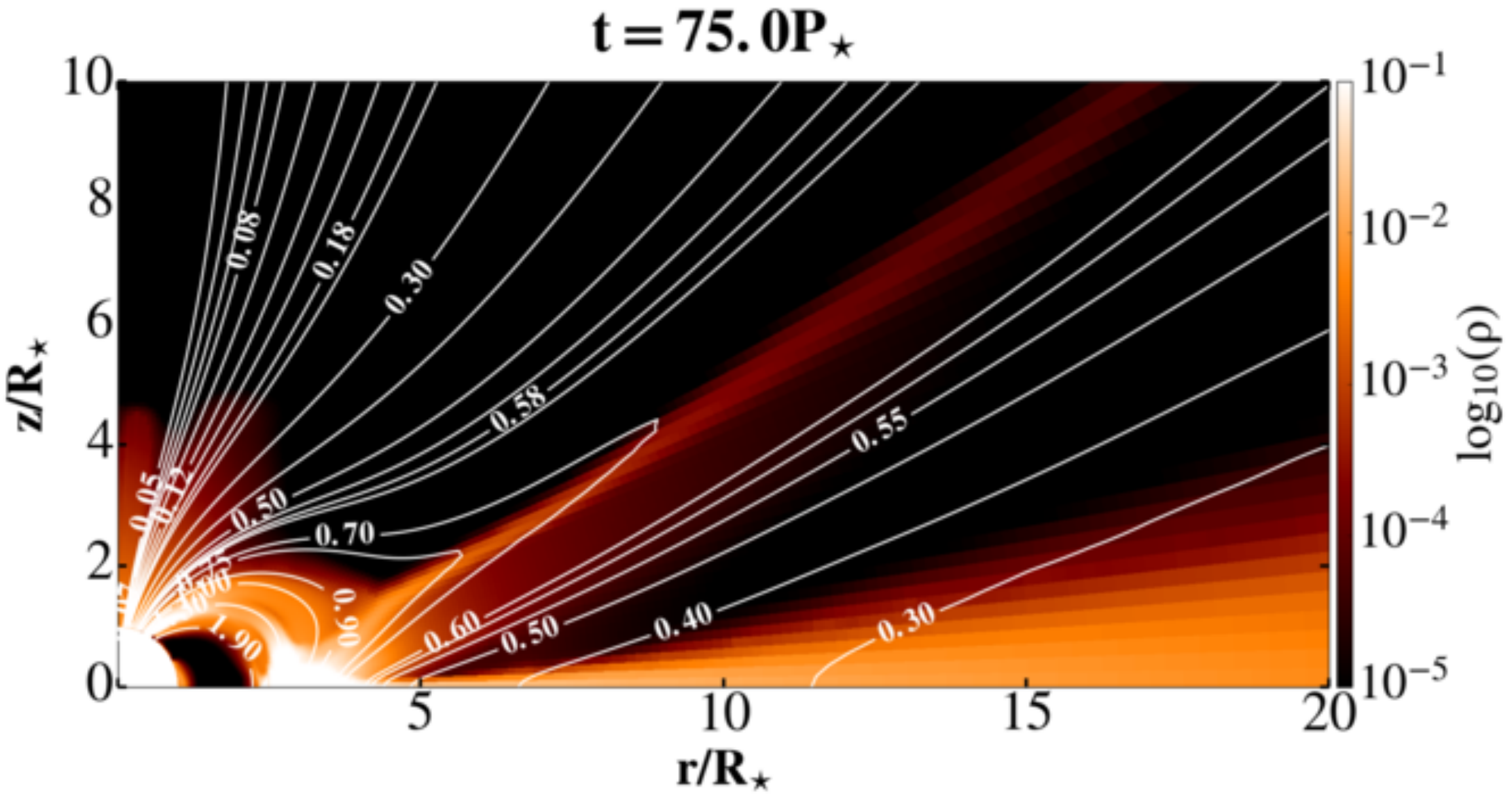}
\includegraphics[width=\columnwidth]{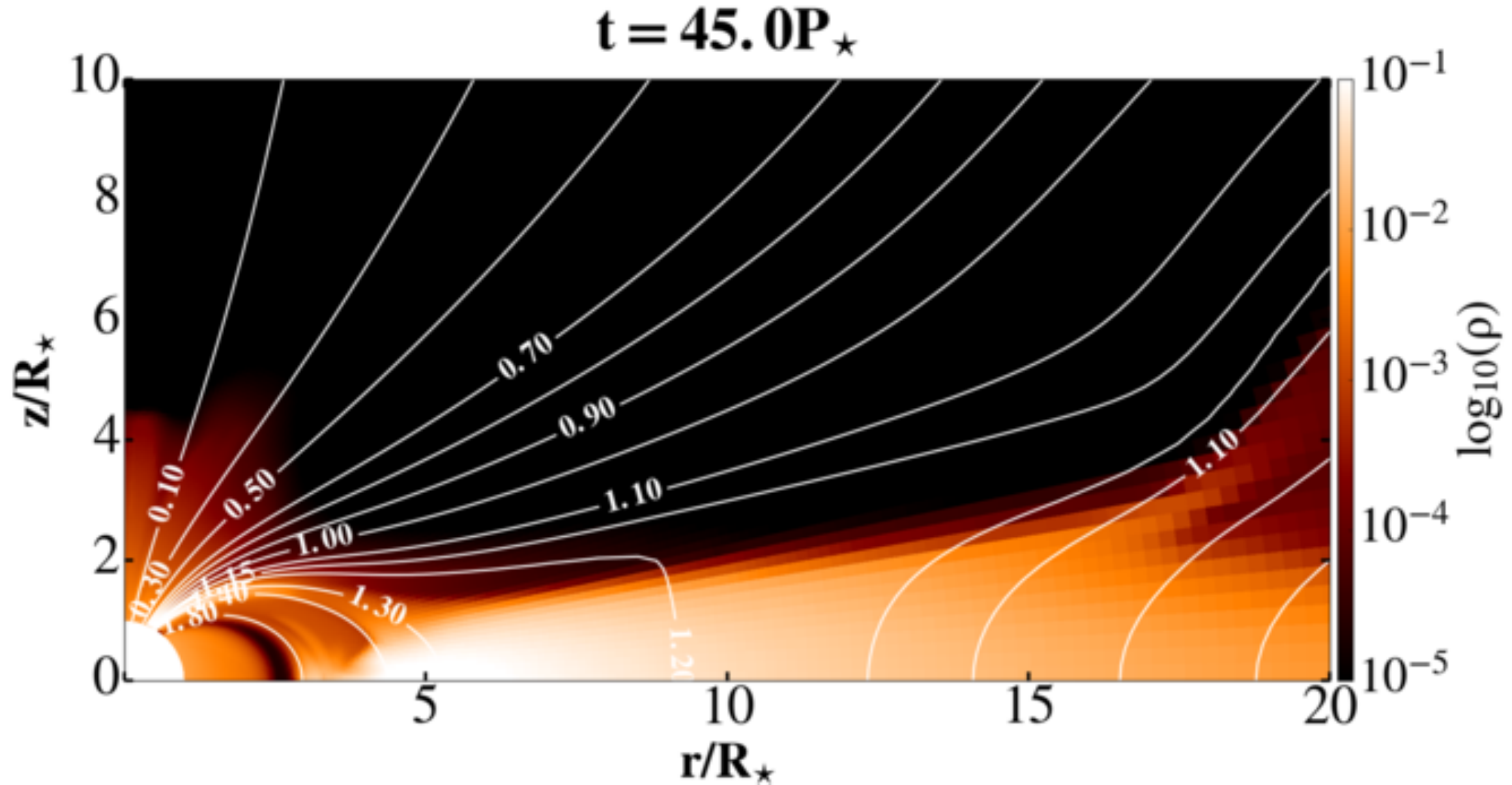}
\includegraphics[width=\columnwidth]{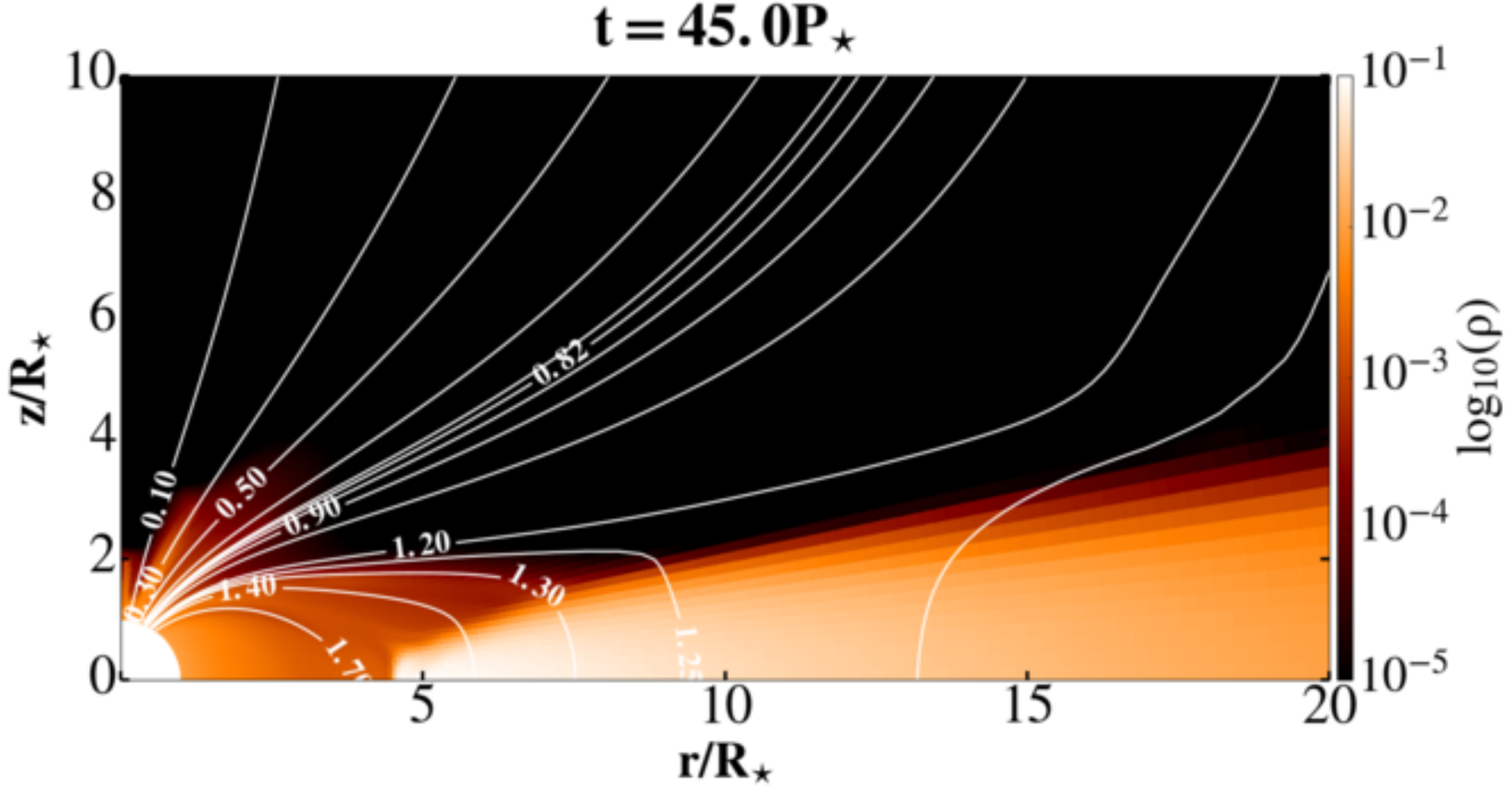}
\includegraphics[width=\columnwidth]{{plotmu1.4Av1Am1om.15}.pdf} 
\caption{Matter density and poloidal magnetic field distribution in
the quasi-stationary interval in $\mu=1.4$ (1.0 kG) case with
$\Omega_\star$=0.15, with $\alpha_m=0.1$, 0.4, 0.7 and 1.0.}
\end{figure*}
\begin{figure*}
\includegraphics[width=\columnwidth]{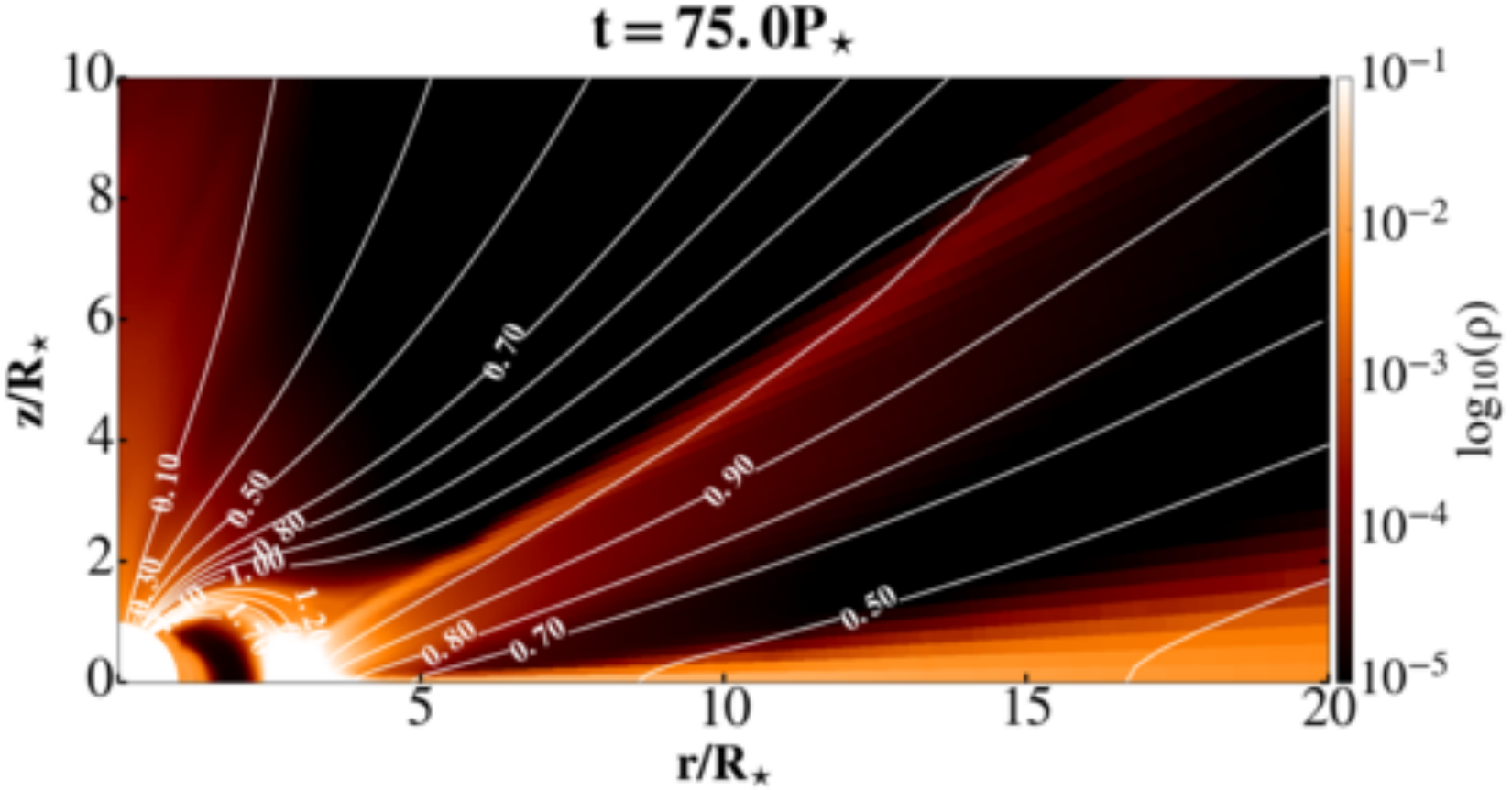}
\includegraphics[width=\columnwidth]{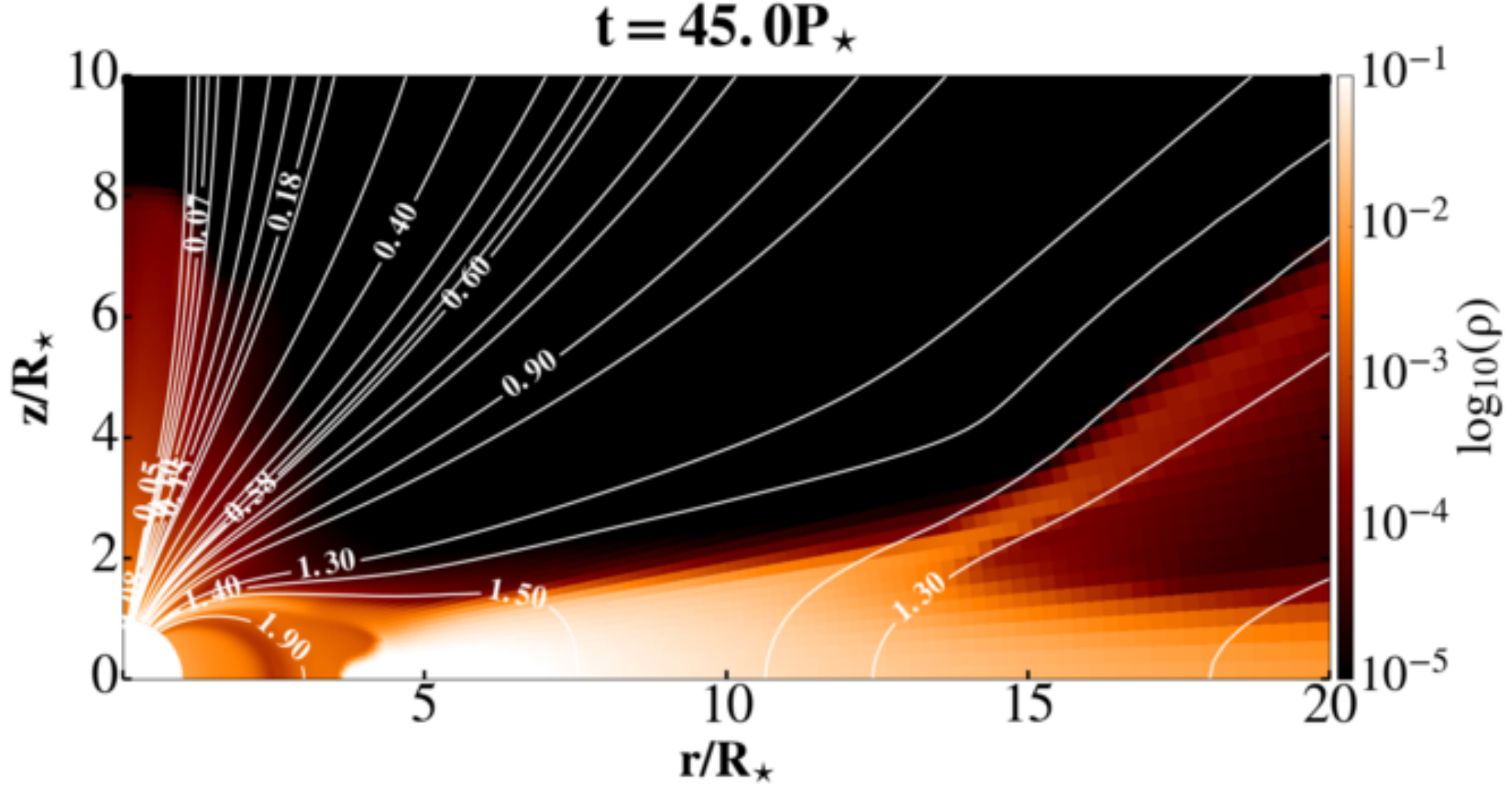}
\includegraphics[width=\columnwidth]{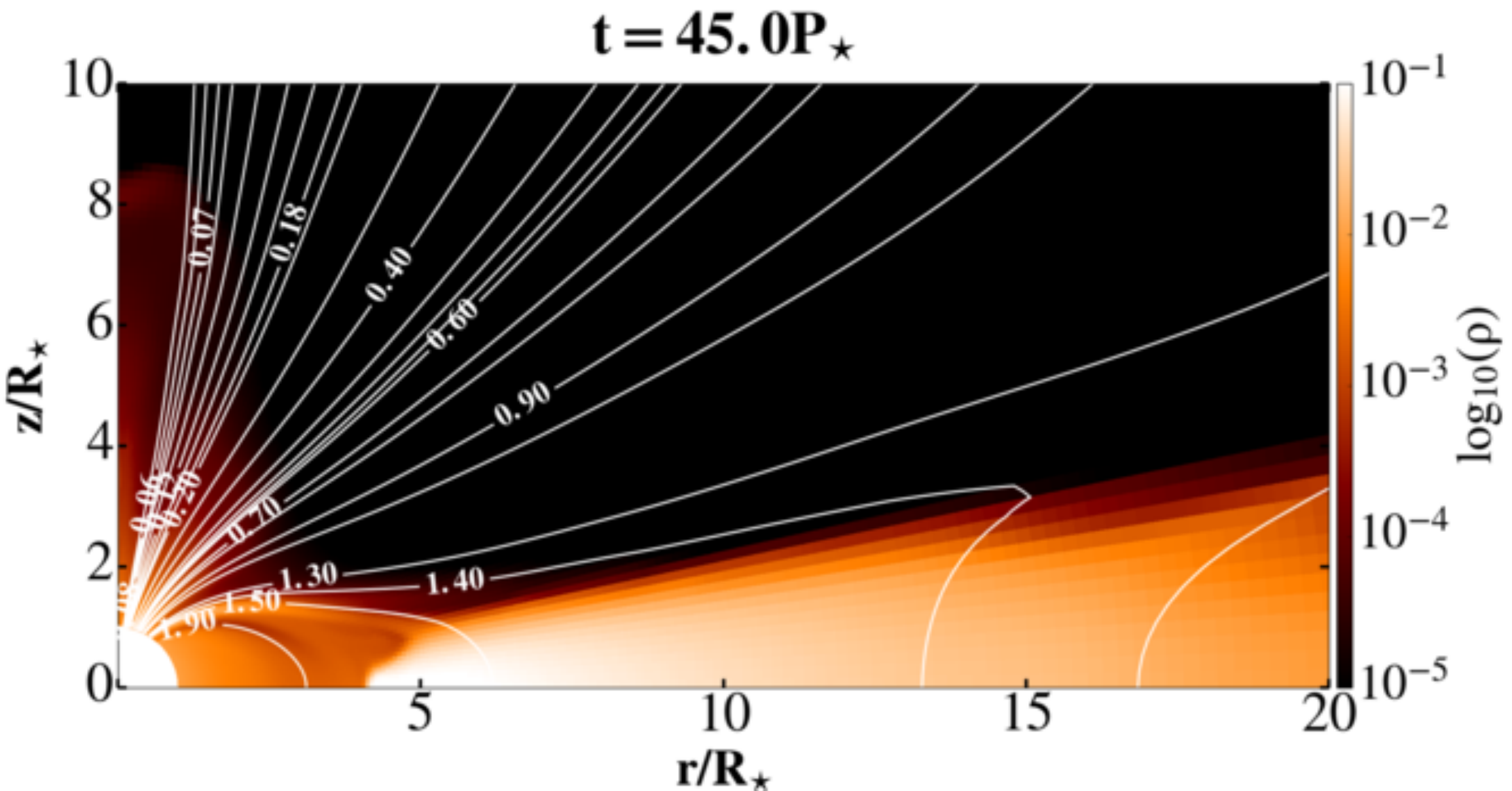}
\includegraphics[width=\columnwidth]{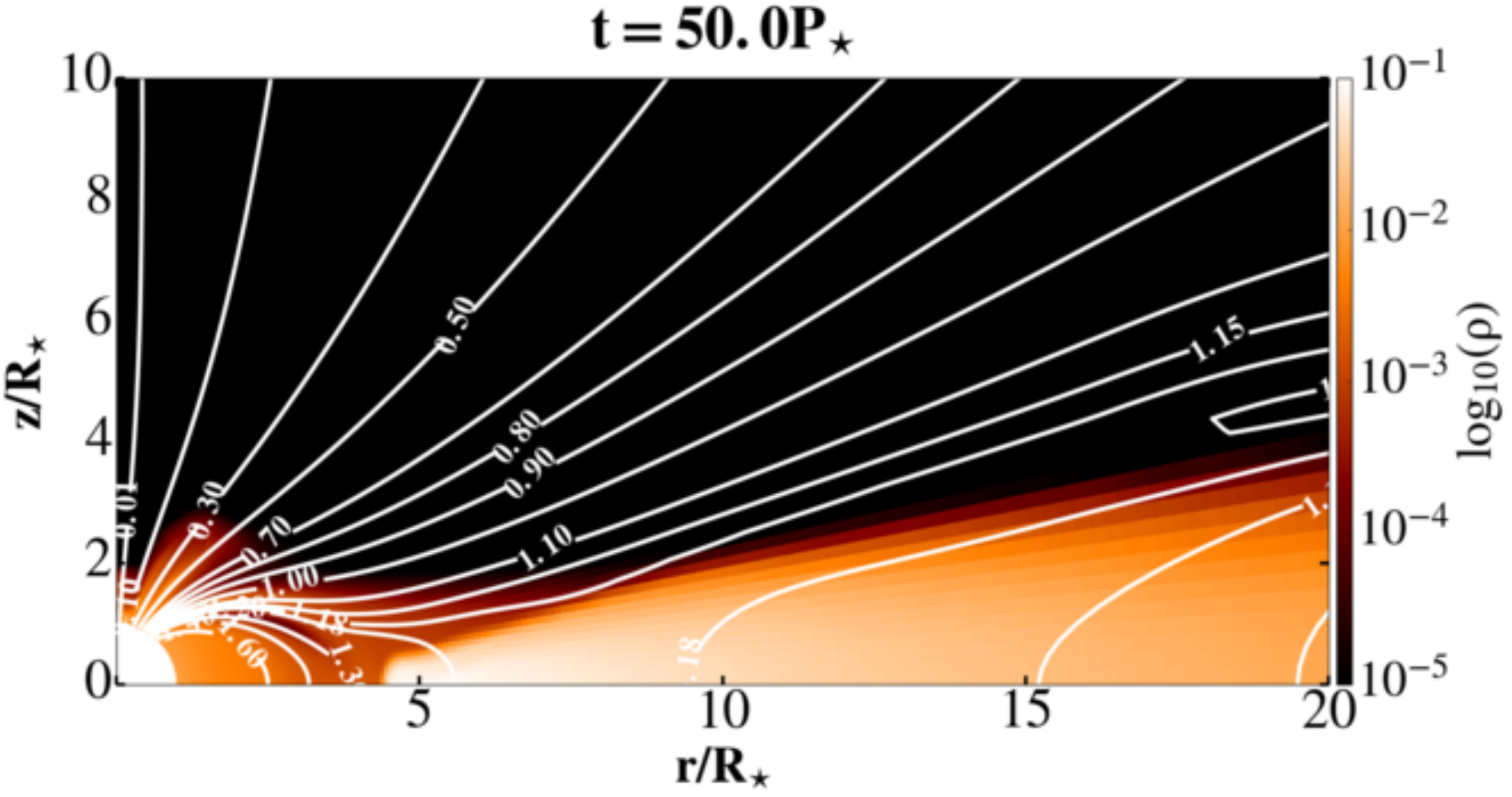} 
\caption{Matter density and poloidal magnetic field distribution in
the quasi-stationary interval in $\mu=1.4$ (1.0 kG) case with
$\Omega_\star$=0.2, with $\alpha_m=0.1$, 0.4, 0.7 and 1.0.}
\end{figure*}
\clearpage


\end{document}